\documentclass[a4paper,fleqn,usenatbib]{mnras}

\usepackage{newtxtext,newtxmath}

\usepackage[T1]{fontenc}
\usepackage{ae,aecompl}

\usepackage{graphicx}	
\usepackage{amsmath}	
\usepackage{amssymb}	
\usepackage{xcolor}

\defcitealias{2011MNRAS.415.3308G}{G11}
\defcitealias{2017MNRAS.467..179G}{G17}
\bibpunct[; ]{(}{)}{,}{a}{}{;}
\usepackage{siunitx}
\usepackage{graphicx}

\title[Properties of barred galaxies from the Auriga simulations]{Structural and photometric properties of barred galaxies from the Auriga cosmological simulations}
\author[G. Bl{\'a}zquez-Calero et al.]{
Guillermo Bl{\'a}zquez-Calero,$^{1}$\thanks{E-mail: gblazquez@ugr.es}
Estrella Florido,$^{1,2}$
Isabel P{\'e}rez,$^{1,2}$
Almudena Zurita,$^{1,2}$ 
\newauthor
Robert J. J. Grand,$^{3}$
Francesca Fragkoudi,$^{3}$
Facundo A. G{\'o}mez,$^{4,5}$
Federico Marinacci,$^{6}$
\newauthor
and R{\"u}diger Pakmor$^{3}$
\\
$^{1}$Departamento de F{\'i}sica Te{\'o}rica y del Cosmos, Universidad de Granada, Campus de Fuentenueva, E-18071 Granada, Spain\\
$^{2}$Instituto Carlos I de F{\'i}sica Te{\'o}rica y Computacional, Universidad de Granada, E-18071 Granada, Spain\\
$^{3}$Max-Planck-Institut f{\"u}r Astrophysik, Karl-Schwarzschild-Str. 1, D-85748, Garching, Germany\\
$^{4}$Instituto de Investigaci{\'o}n Multidisciplinar en Ciencia y Tecnolog{\'i}a, Universidad de La Serena, Ra{\'u}l Biltr{\'a}n 1305, La Serena, Chile\\
$^{5}$Departamento de F{\'i}sica y Astronom{\'i}a, Universidad de La Serena, Av. Juan Cisternas 1200 Norte, La Serena, Chile\\
$^{6}$Department of Physics {\&} Astronomy, University of Bologna, via Gobetti 932/2, 40129 Bologna, Italy
}
\date{Accepted XXX. Received YYY; in original form ZZZ}

\pubyear{2015}

\begin{document}
\label{firstpage}
\pagerange{\pageref{firstpage}--\pageref{lastpage}}
\maketitle

\begin{abstract}

In this work we analyse the structural and photometric properties of 21 barred simulated galaxies from the Auriga Project. These consist of Milky Way-mass magneto-hydrodynamical simulations in a $\Lambda$CDM cosmological context. In order to compare with observations, we generate synthetic SDSS-like  broad-band images from the numerical data at $z = 0$ with different inclinations (from face-on to edge-on). Ellipse fits are used to determine the bar lengths, and 2D bulge/disc/bar decompositions with {\sc galfit} are also performed, modelling the bar component with the modified Ferrer profile. We find a wide range of bar sizes and luminosities in the sample, and their structural parameters are in good agreement with the observations. All bulges present low S{\'e}rsic indexes, and are classified as pseudobulges. In regard to the discs, the same breaks in the surface brightness profiles observed in real galaxies are found, and the radii at which these take place are in agreement with the observations.  Also, from edge-on unsharp-masked images at $z=0$, boxy or peanut-shaped (B/P) structures are clearly identified in the inner part of 4 bars, and also 2 more bars are found in buckling phase. The sizes of the B/P match fairly well with those obtained from observations. We thus conclude that the observed photometric and structural properties of galaxies with bars, which are the main drivers of secular evolution, can be developed in present state-of-the-art $\Lambda$CDM cosmological simulations.

\end{abstract}




\section{Introduction}

According to observations, roughly two thirds of disc galaxies in the local Universe host bars \citep[e.g.][]{2007ApJ...657..790M}. They play an important role in secular galaxy evolution, since they redistribute angular momentum to the disc, bulge, and the dark matter halo component \citep{2003MNRAS.341.1179A, 2004ARA&A..42..603K, 2005CeMDA..91....9A}. In order to understand the secular evolution of disc galaxies, numerous dynamical simulations have been carried out. Bars appeared even in the earliest N-body simulations of galaxies  \citep{1970ApJ...161..903M}, and subsequent simulations succeeded in reproducing some of the observed bar properties \citep{1990A&A...233...82C, 2000ApJ...543..704D, 2002MNRAS.330...35A}. They provided a description of bar evolution, how they can grow in relation with the transferred angular momentum, and showed that bars can buildup pseudobulges by redistributing mass within the disc \citep{2013seg..book....1K}. Also, N-body simulations offered insight into the formation of boxy/peanut structure (hereafter B/P) in the inner regions of a bar through buckling instability \citep[e.g.][]{1994ApJ...425..551M}.

Although non-cosmological galaxy simulations shed light on the bar influence in galaxy secular evolution, these assume ad-hoc initial conditions and, in general, they do not take into account the impact of the enviroment \citep[although there are exceptions, e.g.][]{2016ApJ...821...90A}. Thus, idealised dynamical N-body simulations allow us to focus on the effect of specific properties and mechanisms in galaxy evolution, but cannot provide an overall evolutionary picture within a $\Lambda$CDM cosmology. In order to fully understand the properties of bars in the observed galaxies, they must be reproduced in cosmological simulations, where the key processes responsible for galaxy formation and evolution are included \citep{2012MNRAS.425L..10S}.

The simulation of realistic barred galaxies in a cosmological context is a relatively new achievement. Early $\Lambda$CDM cosmological simulations did not succeed in producing disc galaxy morphologies in accordance with observations. Simulated galaxies tended to have too small discs, shortened by angular momentum loss due to dynamical friction \citep[e.g.][]{2001MNRAS.321..471B}, and unrealistic massive bulge components \citep[e.g.][]{2000ApJ...538..477N}. This prominent spheroidal concentration of mass present in the central regions of simulated galaxies was the result of the commonly known as `overcooling' problem: gas cools too quickly and becomes too dense at the inner regions of the halos \citep{2001MNRAS.326.1228B}. $\Lambda$CDM cosmological simulations were at the time far from generating the morphology of Milky Way-like galaxies. 

Cosmological simulations have deeply changed since then. Current hydrodynamical simulations take into account mechanisms such as stellar and AGN feedback, together with a high numerical resolution, and therefore the simulation of Milky Way-mass galaxies with small bulges and extended discs have finally been accomplished \citep{2011ApJ...728...51B,2013MNRAS.428..129S,2014MNRAS.437.1750M}. Then, it is reasonable to check if bar properties in relation to their host galaxies in current $\Lambda$CDM cosmological simulations are in agreement with observations. 

Studies of bars in realistic spiral galaxies from fully cosmological simulations have recently been carried out, and they have been compared with previous dynamical simulations with idealised initial conditions, as well as with observational results. For instance, \citet{2012MNRAS.425L..10S} studied the properties of two bars of Milky Way-mass  galaxies at $z = 0$, a subsample of the simulations described in \citet{2009MNRAS.396..696S}. Subsequent work developed by \citet{2015MNRAS.447.1774G}, \citet{2015PASJ...67...63O}, and \citet{2017MNRAS.465.3729S}, focused on the evolution of bar properties for one or two galaxies. Although the aforementioned works were generally in good agreement with both previous idealised simulations and observations, they lacked of a statistically representative sample to verify if these cosmological simulations could reproduce the statistical properties of barred galaxies. In this direction, \citet{2012ApJ...757...60K} obtained a bar fraction evolution with redshift for 33 simulated galaxies that are consistent with observations. Also, \citet{2017MNRAS.469.1054A} concludes that the population of bar fractions and lengths for EAGLE cosmological simulations are in agreement with observational constraints; although they find that their bar patterns speeds are too slow. Other recent work performed by \citet{2019MNRAS.483.2721P}, studied  tidally induced bars formed from flyby interactions in Illustris simulations.

In this paper, the bar structural and photometric properties of the Auriga galaxy simulation suite at $z = 0$ are presented. The Auriga project represents one of the largest sample of cosmological zoom-in simulations so far, performed with high resolution and state-of-the-art galaxy formation model. These are extensively described in \citet[][hereafter \citetalias{2017MNRAS.467..179G}]{2017MNRAS.467..179G}. Although \citet{2012ApJ...757...60K} and \citet{2017MNRAS.469.1054A} have already studied the properties of a significant amount of bars from cosmological galaxy simulations, a different approach is presented here. In this work, we analyse the simulated barred galaxies following the same treatment usually performed over real astronomical images. To achieve this purpose, synthetic images from the cosmological simulated galaxies are created. The photometric data was already modeled by \citetalias{2017MNRAS.467..179G}, and {\it g}, {\it r} and {\it i} broad-bands are used for the synthetic image creation. Our results are mainly compared with a Sloan Digital Sky Survey (SDSS) sample of 291 barred galaxies found in the local Universe, analysed by \citet[][ hereafter \citetalias{2011MNRAS.415.3308G}]{2011MNRAS.415.3308G}. Often, the morphological decomposition of simulated galaxies is performed taking into account stellar mass density, and 1D or kinematic decomposition is preferred over 2D decompositions. In this work, a 2D bulge/disc/bar morphological decomposition from photometric synthetic images \citep[as in][]{2010MNRAS.407L..41S} is performed using {\sc galfit}. 

This paper is structured as follows. In Section \ref{sec:simulation}, the Auriga galaxy simulations and the observational sample for comparison are briefly described. In Section \ref{sec:methods}, we explain how synthetic images are created from the simulations and the methodology followed in our analysis. In Section \ref{sec:results}, the main results of the bar, bulge, disc and B/P characterisation of the Auriga galaxies is presented and compared with observations. Finally, in Section \ref{sec:conclusions}, the main results and conclusions from this work are summarised.

\section{The simulations and observational data}\label{sec:simulation}

In this section, we shortly mention the main characteristics of the Auriga galaxy simulations, which are extensively described in \citetalias{2017MNRAS.467..179G}. Also, we present a brief description of the galaxy sample used for comparison with the Auriga galaxies.

\subsection{The Auriga cosmological simulations}
The Auriga $\Lambda$CDM cosmological magneto-hydrodynamical simulation suite comprises 30 high-resolution Milky Way-mass dark haloes. The simulations are carried out using the zoom-in technique, and ran with the moving mesh code {\sc AREPO} \citep{2010ARA&A..48..391S}. The simulations include a state-of-the-art galaxy formation model that is capable of reproducing realistic properties of disc galaxies \citep{2013MNRAS.436.3031V, 2014MNRAS.437.1750M, 2016MNRAS.459..199G}. The numerical data used in this work corresponds to the level 4 resolution simulations at $z=0$, in which the baryonic resolution and dark matter typical particle mass is $\sim 5\times 10^4~\text{M}_\odot$ and $\sim 3\times 10^5~\text{M}_\odot$, respectively. The physical softening length for star particles grows with redshift until $z=1$ is reached, for lower redshifts it is fixed to $369$ pc. Every star particle in the simulation is assumed to represent a single stellar population characterised by a given age and metallicity. The photometric data is determined by stellar population synthesis model \citep{2003MNRAS.344.1000B}, and luminosities in {\it U}, {\it V}, {\it B}, {\it K}, {\it g}, {\it r}, {\it i}, and {\it z} bands are available, without modelling the effects of dust attenuation.

The simulations successfully exhibit the characteristic morphology from early-type to late-type disc galaxies, with the presence not only of bars, but also of spiral arms, pseudobulges, rings, and B/P bulges when seen from an edge-on perspective (their morphology can be inspected from the RGB images in Fig. B1, included in online material). This allows us to characterise the bars with respect to their host galaxy sizes and luminosities, and also to check if the observed trends of the bar properties with the galaxy morphology are reproduced. Bars are visually detected at $z=0$ in 21 galaxies, and these form the subsample analysed in this paper. 

\subsection{The observational sample for comparison}

In order to check if bar properties from the Auriga simulations are in accordance with the observations, the galaxy sample presented by \citetalias{2011MNRAS.415.3308G} has been chosen for comparison. It comprises 291 Milky Way-mass like  galaxies with bars from the SDSS, selected from the parent sample described in \citet{2009MNRAS.393.1531G}. The latter paper presents a 2D bulge/disc/bar decomposition in {\it g}, {\it r}, and {\it i} bands, of 946 barred and non-barred galaxies representative of the population found in the local Universe (within a redshift range of $0.02\leq z\leq 0.07$). Only close to face-on galaxies ($b/a \geq 0.9$) are included; consequently, dust and projection effects are minimized. These characteristics enable us to compare \citetalias{2011MNRAS.415.3308G} sample with the barred galaxies from the Auriga simulations. In \citetalias{2011MNRAS.415.3308G}, the 2D multicomponent decomposition is carried out with the software {\sc budda}, which uses a S{\'e}rsic function for the bar model, providing bar lengths and bar-to-total luminosity ratios. These are the main parameters that are compared with the Auriga simulations in this work.

\section{Methodology}\label{sec:methods}

In this section, we describe the methodology followed to characterise the Auriga barred galaxies. First, synthetic images are created from the simulations. Then, isophotal ellipse fits are used to determine the bar length lower and upper limits, as well as the disc profiles. Later, 2D bulge/disc/bar decompositions provide us the bar length relative to the disc scalelength, and the bar-to-total luminosity ratio. Finally, we describe the method used for the detection of B/P bulges, and how their sizes are measured.

\begin{figure*}
	\centering
	\includegraphics[width=0.8\textwidth]{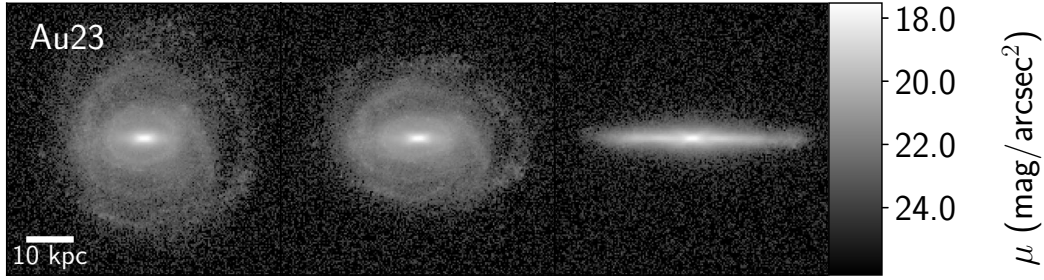}
    \caption{Synthetic images based on SDSS images are created for different inclination angles $i$. On the left, $i=0\degr$; center, $i=45\degr$; and right, $i=90\degr$. From an edge-on perspective, when the bar axis is contained in the plane of the sky, X-shaped bulges can be observed in some galaxies (see Section \ref{sec:bp_method}).}
    \label{fig:mock_images}
\end{figure*}

\subsection{Synthetic image creation}\label{sec:synthetic}

In order to characterise the bars of the Auriga simulations from an observational perspective, the first step is to create synthetic images, so as to treat them as observational ones. Every synthetic image shows a galaxy centered at zero gravitational potential, and is created from the simulation numerical data at $z=0$, for different inclination angles, as depicted in Fig. \ref{fig:mock_images}. The synthetic images are obtained by building a 2D grid where each bin represents a pixel with a given flux, with the following procedure:

\begin{itemize}
\item Since star particles do not represent individual stars, but a single stellar population (see Section \ref{sec:simulation}), a Triangular Shape Cloud (TSC) interpolation method \citep{1988csup.book.....H} is performed over the particles flux. This renders a more realistic image than a pure 2D-histogram, keeping the total flux conserved, and avoiding empty pixels that might be populated by stars. 

\item The image characteristics are chosen in order to resemble an astronomical image from SDSS. Then, in the synthetic images flux is given in nMgy units, defined as 
\begin{equation}
m = 22.5 - 2.5\log_{10}f[\text{nMgy}],
\end{equation}
where the value $22.5$ represents the photometric zeropoint of the magnitude scale. 

\item Although numerical data at $z=0$ is used, we assume the SDSS pixel scale ($0.396\, \text{arcsec}\,\text{pixel}^{-1}$) and an intermediate galactic redshift $z=0.04$, that is within the redshift range of the galaxy sample from \citetalias{2011MNRAS.415.3308G}. An image resolution of $0.340\,\text{kpc}\,\text{pixel}^{-1}$ is obtained, considering a Hubble constant of $H_0 = 100h~\text{km}~\text{s}^{-1}~\text{Mpc}^{-1}$, where $h=0.6777$ \citep{2014A&A...571A..16P}.

\item Gaussian noise is added to images, based on the one measured on SDSS sky subtracted images with $0.02$, $0.03$, and $0.04$ nMgy standard deviation for {\it i}, {\it r}, and {\it g} bands, respectively. This emulates the signal-to-noise ratio that would be measured in astronomical images. It is needed to estimate uncertainties and indicators of goodness of fits ($\chi^2_\nu$), and to determine the limit at which the noise level starts to be significant in the surface brightness profiles.

\end{itemize}

In addition, edge-on synthetic images with the bar contained in the plane of the sky allows us to characterise B/P bulges. In order to improve the B/P detection and size measurement, the resolution of these edge-on synthetic images is increased to $0.139$ kpc/pixel (assuming a galactic redshift $z=0.01635$).

\subsection{Isophotal ellipse fitting}\label{sec:length}

\subsubsection{Bar length and ellipticity}
In this work, the method presented in \citet{2003ApJS..146..299E} is followed in order to determine the bar length and bar ellipticity. Using {\sc ellipse} task from {\sc iraf}, isophotal ellipse fitting is performed to the synthetic images. It provides the variation of ellipticity ($\epsilon$) and position angle (PA) as a function of the semimajor-axis (SMA). This method allows us to define lower and upper limits of the bar semimajor-axis length (or simply, bar length):
\begin{itemize}
\item The lower bar length limit, $a_\text{max}$, is defined as the SMA at which maximum ellipticity in the bar region is reached.
\item The upper bar length limit, $L_\text{bar}$, is defined by the minimum of two other distances, $L_\text{bar}=\text{min}(a_\text{min}, a_{10})$. We denote $a_\text{min}$ as the point at which the first minimum in ellipticity takes place, and $a_{10}$ is defined as the radial distance where the fitted ellipses differ at least by  $10\degr$ from the PA of the isophote at $a_\text{max}$.
\end{itemize}

The ellipticity of the bar is a measure of its strength, or of how strong are the tangential forces induced by bars. Although there is no standard way to define the bar strength, we use the parametrisation present in \citet{2002MNRAS.336.1281W},

\begin{equation}\label{eq:strength}
f_\text{bar} = \frac{2}{\pi}\left[\arctan(1 - \epsilon_\text{max})^{-1/2} - \arctan(1 - \epsilon_\text{max})^{1/2}\right],
\end{equation}
where $\epsilon_\text{max}$ is the ellipticity at $a_\text{max}$, and $f_\text{bar}\in [0,1]$. When the galaxy is unbarred $f_\text{bar}=0$ is obtained, and if $f_\text{bar}=1$ the galaxy is considered to be strongly barred.

\subsubsection{Disc breaks}

From isophotal ellipse fits we have obtained the surface brightness profiles of the Auriga galaxies (these are shown in Fig. 1C, included in the online material). During the ellipse fitting, the inner isophotes were fitted allowing the PA and $\epsilon$ to vary, but for the regions beyond the bar where isophotes start to have $\epsilon \sim 0.1$ and the PA abruptly varies, the fitting were manually stopped; the PA and $\epsilon$ of the isophotes beyond this radius were fixed to the ones of the last iteration. It is well known that most of galaxy discs do not follow a pure exponential profile \citep[e.g.,][]{1979A&AS...38...15V, 2004ASSL..319..713P, 2004A&A...427L..17P, 2006A&A...455..475F}, and for the Auriga galaxies we find the same deviations found in real galaxies. In this work we adopt the same basic classification scheme as \citet{2006A&A...454..759P}, that distinguishes discs of type I, pure exponentials; Type II, downward breaks; and Type III, upward breaks.

In order to characterise the disc breaks, if any, we have modeled the surface brightness profiles obtained by isophotal ellipse fits beyond $L_\text{bar}$, using the 1D broken-exponential function proposed by \citet{2008AJ....135...20E}:
\begin{equation}
I(r) = S I_0 e^\frac{-r}{r_\text{s,1}} \left[1+e^{\alpha\left(r - r_\text{br}\right)}\right]^{\frac{1}{\alpha}\left(\frac{1}{r_\text{s,1}}-\frac{1}{r_\text{s, 2}}\right)},
\end{equation}
where $I_0$ is the central intensity of the inner exponential, $r_\text{s,1}$ and $r_\text{s,2}$ are the inner and outer disc scalelength, respectively, $r_\text{br}$ is the radius at which the break takes place, $\alpha$ is a parametrisation of the break sharpness, and the scaling factor $S$ is given by
\begin{equation}
S = \left(1+e^{-\alpha R_\text{br}}\right)^{\frac{1}{\alpha}\left(\frac{1}{r_\text{s,1}}-\frac{1}{r_\text{s, 2}}\right)}.
\end{equation}
Since discs can display several breaks along their profiles, and in simulations there is no restriction in characterising them at low surface brightness, we have fitted the profiles down to $\mu_\text{crit}$. Beyond this point, the fits performed over the synthetic images are no longer reliable due to the noise level. Our definition of $\mu_\text{crit}$ is based on the one used in observations: the limit at which the standard error of the azimuthally averaged flux of isophotal ellipses yields an uncertainty of $0.2~\text{mag}\,\text{arcsec}^{-2}$ \citep{2006A&A...454..759P, 2008AJ....135...20E}. For our synthetic images, we obtain reliable fits down to $\sim 26~\text{mag}\,\text{arcsec}^{-2}$. The results from the isophotal ellipse fitting over {\it r} band synthetic images of face-on galaxies from the numerical data at $z=0$ are presented in Table \ref{tab:ellipse_results}.

\subsection{2D bulge/disc/bar decomposition}\label{sec:decomposition}

The 2D photometric decomposition of galaxy synthetic images is performed by fitting the bulge, the disc, and the bar component. The software used for this fitting is {\sc galfit} \citep{2011ascl.soft04010P}. Next, we describe the fitting function used for each component.

\subsubsection{Bulge component}
The S{\'e}rsic profile is used for fitting the surface brightness of the bulge. In magnitude units, the radial profile is given by
\begin{equation}
\mu_\text{bulge} = \mu_e + 1.086b_n\left[\left(\frac{r}{r_e}\right)^{1/n}-1\right],
\end{equation}
where $\mu_e$ is the surface brightness at effective radius $r_e$; $n$ is the S{\'e}rsic index, which measures the brightness central concentration; and $b_n$ is a variable coupled to $n$ by $b_n \approx (1.9992n-0.3271)$ \citep{2005PASA...22..118G}. In the particular case of $n=4$, the Vaucouleurs function is obtained, and $n=1$ gives the exponential function. 

\subsubsection{Disc component}
The surface brightness radial profile of the disc is modeled with an exponential function, that in magnitude units is given by
\begin{equation}
\mu_\text{disc} = \mu_0 + 1.086 r/r_s,
\end{equation}
where $\mu_0$ is the central bar surface brightness, and $r_s$ is the disc scalelength. Although most of the Auriga galaxies show disc breaks, which are typically modeled in 2D decompositions with two exponential components with different scalelengths \citep[e.g.,][]{2017A&A...598A..32M}, we have limited our decomposition to only one disc component, since we wanted to prevent our solutions from degeneracy.

\subsubsection{Bar component}
For the bar component, the most commonly used models are Ferrer and S{\'e}rsic profiles. In \citet{2015ApJ...799...99K}, a comparison between both models is presented. In this paper, the modified Ferrer profile is chosen over S{\'e}rsic, since the former allows us to define a truncation radius $r_\text{bar}$, that can be regarded as an additional bar length  measurement independent to the method presented in Section \ref{sec:length}. The modified Ferrer function is given by
\begin{equation}
\mu (r) = \mu_0 \left[1-\left(\frac{r}{r_\text{bar}}\right)^{2-\beta}\right]^\alpha,
\end{equation}
where $\mu_0$ is the central surface brightness; $r_\text{bar}$ is the radius beyond which the function has a value of $0$; $\alpha$ determines how sharply the bar profile decreases near $r_\text{bar}$; and $\beta$ controls the central slope. As it has been previously reported by \citet{2017ApJ...845..114G}, $r_\text{bar}$ is correlated with $\alpha$: as $\alpha$ increases  $r_\text{bar}$ grows. Since $a_\text{max}$ and $L_\text{bar}$ are usually considered as a lower and upper boundaries for bar length, respectively, these are used as constraints of $r_\text{bar}$ for {\sc galfit} fitting if $\alpha$ or $r_\text{bar}$ increases up to unrealistic values; this ensures that $r_\text{bar}$ can be regarded as a measurement of the bar truncation radius. Moreover, when we treat $\alpha$ and $\beta$ as free parameters, {\sc galfit} could not always converge to a solution. This has also been reported by \citet{2015ApJ...799...99K}, where only $75$ per cent of bars from the Spitzer Survey could be characterised by letting $\alpha$ and $\beta$ to vary. For this reason, we decided to fix $\alpha$ and let $\beta$ to be a free parameter, in order to obtain meaningful solutions for all our sample by applying the same fitting method, even at the expenses of more accurate results. We could also fix $\beta$ and let $\alpha$ be a free parameter, but as shown in \citet{2015ApJ...799...99K}, $\beta$ seems to show a stronger correlation with bar flatness than $\alpha$. Thus, we fixed $\alpha=1.5$, since for this value $r_\text{bar}$ usually falls within bar length boundaries $a_\text{max}$ and $L_\text{bar}$, and we obtain meaningful solutions for all the Auriga galaxies.

The fitting strategy followed in this work is similar to the one described in \citet{2008ASPC..393..279W}. The complexity of the model is built up gradually, starting by fitting the whole galaxy with a single S{\'e}rsic function. Then, a second model is fitted with two components, taking into account the previous results for S{\'e}rsic function for the bulge as initial parameters, and using an exponential function to model the disc. The PA and $q$ (the axis ratio) are fixed to the values of the outermost isophote as determined by the ellipse task in {\sc iraf}. Finally, from the residuals it is possible to guess some initial parameters for the bar fitting with a modified Ferrer function. Since the morphological components of some galaxies display asymmetric features, the center of its bar and bulge models are let to vary, but the center of the disc component is kept fixed. 

The 2D disc/bulge/bar decomposition results are presented in Table \ref{tab:decomposition_results}. The decomposition is performed over {\it r} band synthetic images of face-on galaxies from the numerical data at $z=0$. In Fig. \ref{fig:decomposition_1-5}, the synthetic images, 2D multicomponent decomposition models, together with the residuals, are presented for Au1, Au2, and Au5 (for the remaining galaxies, these are included in Fig. \ref{fig:decomposition_6-9}). The indicator of the fitting goodness is given by $\chi^2$, that is of the order of the decomposition of real images \citep{2010AJ....139.2097P}.

\subsection{B/P identification and size measurement}\label{sec:bp_method}

\begin{figure*}
	\vspace{-1.6em}
	\includegraphics[height=0.97\textheight]{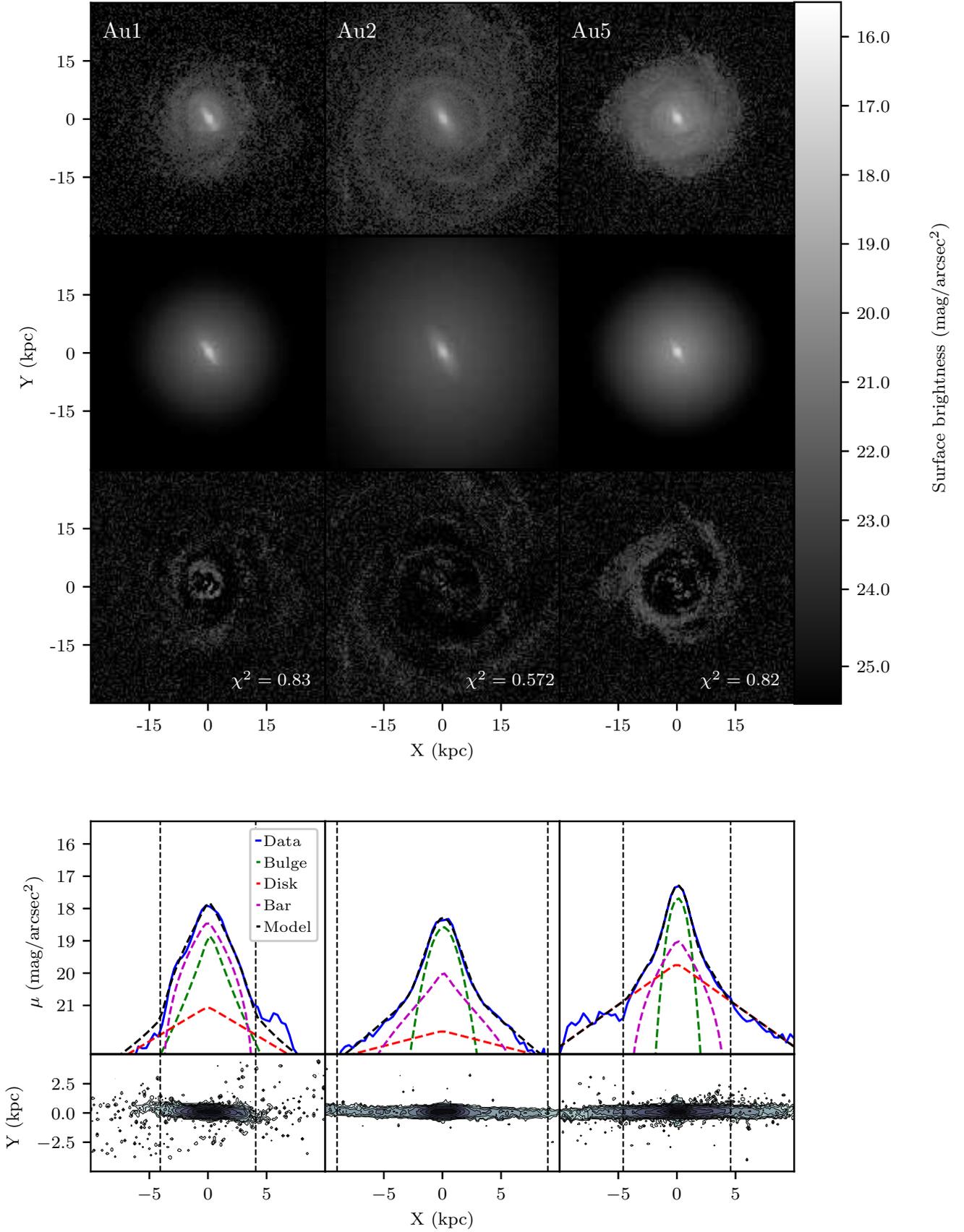}
    \caption{Upper panel: Top: synthetic face-on images in {\it r} band. Center: the model performed by {\sc galfit}. Bottom: the residuals. Bottom panel: {\it r} band surface brightness profiles along bar axis for the different components, including at the bottom the {\it z} band unsharp-masked edge-on images with intensity contours overlaid. Dashed vertical lines indicate the bar length upper limits, $L_\text{bar}$.  For the rest of the galaxies these plots are included in Fig. \ref{fig:decomposition_6-9}. }
    \label{fig:decomposition_1-5}
\end{figure*}

Boxy-peanut or X-shaped bulges (both referred as B/P) are detected by performing an unsharp-mask over the edge-on synthetic images, where the bar major axis is contained in the plane of the sky. These images are shown below the lower panels of Figures \ref{fig:decomposition_1-5} and \ref{fig:decomposition_6-9}. The unsharp-masked images have been obtained by subtracting a median filtered image from the original ones, as described by \citet{2006MNRAS.370..753B}. A squared window was chosen for the median filter, and its size varies for each galaxy, depending on how magnified the X-shape structure was rendered after subtracting the median filter. The window sizes ranges from 15 to 23 pixels.

Several methods to measure the B/P strength (or B/P scaleheight) and size have been proposed, not necessarily from edge-on unsharp-masked images \citep[e.g.][]{2013MNRAS.431.3060E, 2015MNRAS.450..229F, 2016MNRAS.459.1276C, 2017MNRAS.471.3261S}. The method followed in this paper is based on \citet{2017A&A...598A..10L}. The tips of the X-shape branches are determined visually on the unsharp-masked images, and these measurements were repeated three times for different window sizes. Then, the semimajor and semiminor axis of the box that encloses the X-shape were determined, denoted by $a_X$ and $b_X$, respectively. 

We remind to the reader that these methods, both for B/P detection and size measurement, are based on the ones used for astronomical images. Nevertheless, other methods usually employed in simulations \citep[e.g.][]{2008IAUS..245..103M, 2017A&A...606A..47F}, could result in a finer B/P detection and more accurate B/P size measurements. Since we are more interested in the direct comparison with observations, rather than a detailed description of each galaxy simulation, we limit our study to the method described here.

\section{Results}\label{sec:results}
In this section, we present the main results for the characterisation of the barred galaxies from Auriga simulations. First, the properties of the bar, the bulge, and the disc component, are presented and compared with observations. The last part of this section is dedicated to the B/P bulges and buckling bars present in the Auriga galaxies, where also their sizes and frequencies are compared with observations.

\begin{figure*}
    \centering
    \begin{minipage}[t]{0.45\textwidth}
        \centering
        \includegraphics[width=\textwidth]{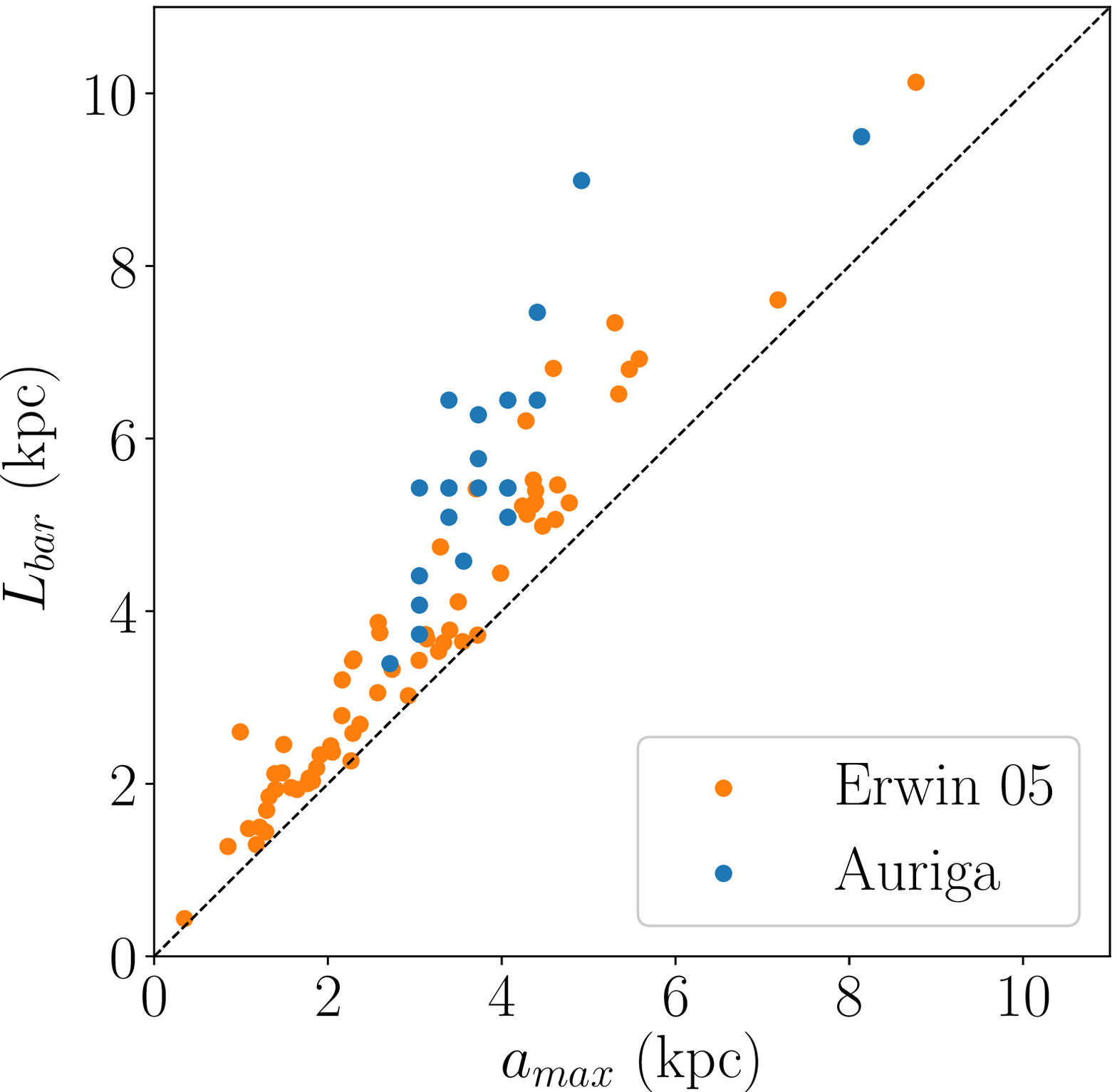}
        \caption{Correlation between the different bar length measurements, $a_\text{max}$ and $L_\text{bar}$, compared with the observational results obtained by \citet{2005MNRAS.364..283E}.  $L_\text{bar}$ and $a_\text{max}$ can be regarded as an upper and lower bound for the bar length, respectively. The dashed line indicates the perfect correspondence.}
        \label{fig:lcorrelation}
    \end{minipage}%
    \hspace{2em}
    \begin{minipage}[t]{0.45\textwidth}
        \centering
        \includegraphics[width=\textwidth]{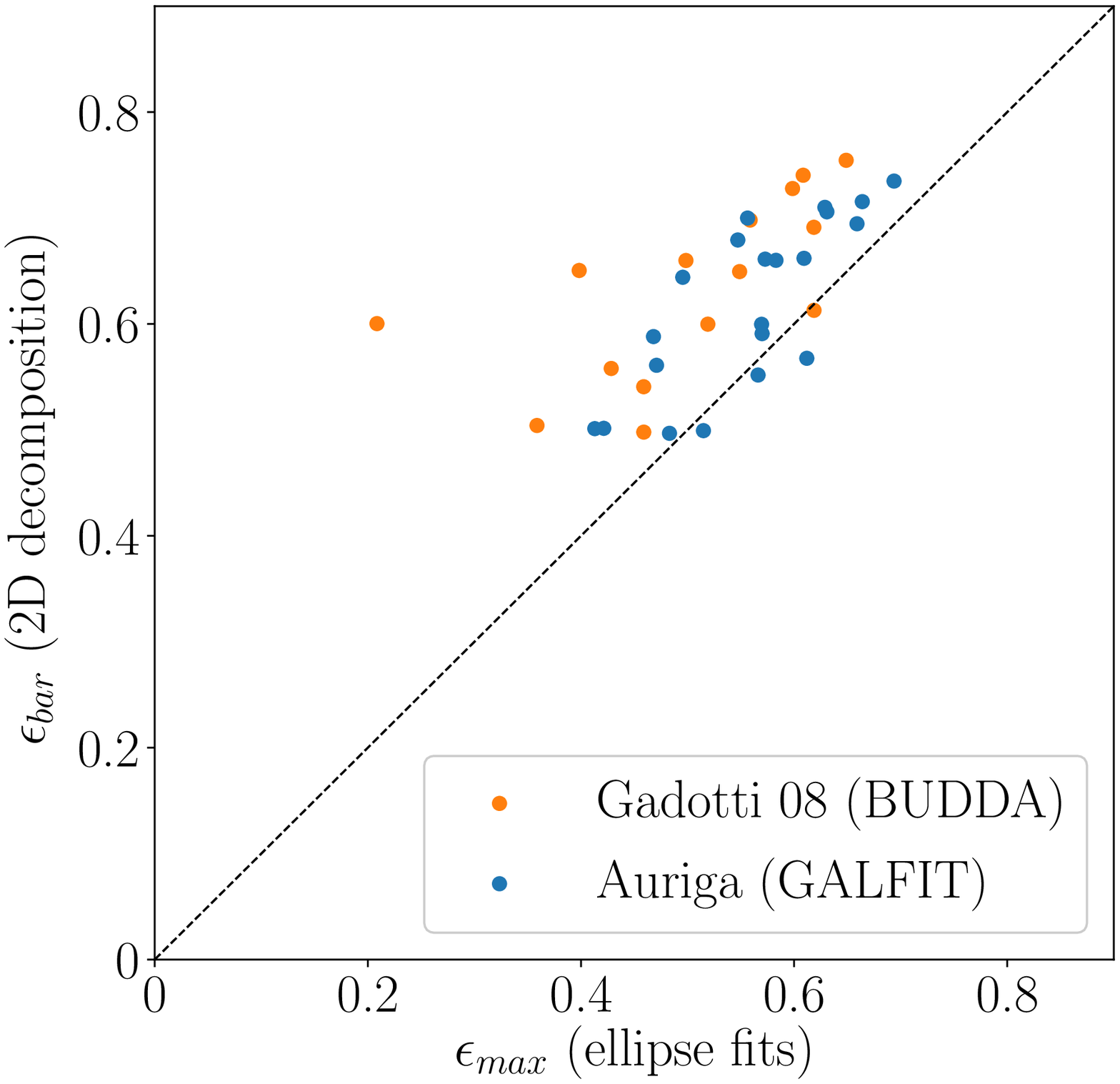}
    	\caption{Correlation between $\epsilon_\text{max}$, the bar ellipticity defined throught ellipse fits; and $\epsilon_\text{bar}$, the bar ellipticity obtained from the 2D decomposition results. For both results, using {\sc galfit} for the Auriga galaxies (blue points) and {\sc budda} for the sample presented in \citet{2008MNRAS.384..420G} (yellow points), $\epsilon_\text{max}$ is lower than $\epsilon_\text{bar}$. The dashed line indicates the perfect correspondence.}
    	\label{fig:epsilon_correlation}
    \end{minipage}
\end{figure*}

\begin{table*}
	\centering
	\caption{Bar and disc morphological parameters obtained from ellipse fits, using {\it r} band synthetic images. The columns show: (1) the galaxy simulation name; (2) and (3), the lower and upper limit of bar lengths from ellipse fits, respectively; (4) the maximum ellipticity that defines $a_\text{max}$; (5) the bar strength; (6) the disc break type; (7) the inner disc scalelength; (8) the outer disc scalelength; (9) the break radius; (10) the surface brightness limit at which the isophotal ellipses yield an uncertainty of $0.2\, \text{mag}\, \text{arcsec}^{-2}$; (11)-(12) stellar masses and gas fraction within $0.1$ times the virial radius, taken from \citetalias{2017MNRAS.467..179G}; and (13), the morphological classification, performed visually. Galaxies with B/P bulges are indicated by $^*$, and buckling bars by $^\dagger$.}
	\label{tab:ellipse_results}
	\begin{tabular}{l c c c c c S[table-format=1.1] c S[table-format=1.1] c c c l}
		\hline
		Galaxy & $a_\text{max}$ & $L_\text{bar}$ & $\epsilon_\text{max}$ & $f_\text{bar}$ & Disc Type & $r_\text{s,1}$ & $r_\text{s,2}$ & $r_\text{br}$ & $\mu_\text{crit}$ & $\log_{10}\left(\frac{M_*}{\text{M}_\odot}\right)$ & $f_\text{gas}$ & Morph. Type \\
		~ & (kpc) & (kpc) &  &  &  & {(kpc)} & (kpc) & {(kpc)} & $\left(\text{mag}\, \text{arcsec}^{-2}\right)$ & & & \\
		(1) & (2) & (3) & (4) & (5) & (6) & {(7)} & (8) & {(9)} & (10) & (11) & (12) & (13) \\
		\hline
		Au1  & 3.05 & 4.07 & 0.56 & 0.57 & I & 5.69 & 0.00 & 0.00 & 26.05 & 10.44 & 0.41 & SB(r)b \\

		Au2  & 4.92 & 8.99 & 0.57 & 0.59 & II & 16.45 & 6.76 & 38.93 & 26.41 & 10.85 & 0.17 & SB(r)c \\

		Au5  & 3.56 & 4.58 & 0.50 & 0.50 & II & 6.28 & 2.86 & 16.05 & 26.02 & 10.83 & 0.27 & SAB(r)b \\

		Au6  & 4.07 & 5.43 & 0.42 & 0.42 & II & 8.48 & 2.05 & 30.92 & 26.16 & 10.68 & 0.19 & SAB(r?)bc \\

		Au7  & 3.05 & 5.43 & 0.47 & 0.47 & II & 7.48 & 5.14 & 12.75 & 26.39 & 10.69 & 0.39 & SAB(r?)b \\

		Au9  & 4.41 & 6.45 & 0.66 & 0.76 & II & 6.37 & 2.98 & 16.72 & 25.94 & 10.79 & 0.21 & SB(r)b \\

		Au10  & 3.39 & 6.45 & 0.66 & 0.75 & III & 2.29 & 4.66 & 14.99 & 26.08 & 10.77 & 0.24 & SB(s)a \\

		Au12  & 3.05 & 3.73 & 0.57 & 0.60 & II & 8.51 & 3.22 & 8.38 & 26.66 & 10.78 & 0.31 & SAB(s)ab \\

		Au13$^*$  & 3.73 & 5.43 & 0.57 & 0.59 & III & 3.38 & 8.70 & 16.11 & 26.08 & 10.79 & 0.17 & SB(r)0/a \\

		Au14  & 4.07 & 5.09 & 0.61 & 0.65 & II & 6.96 & 4.45 & 12.56 & 26.41 & 11.02 & 0.28 & SB(r?)b \\

		Au17$^\dagger$  & 4.07 & 5.43 & 0.57 & 0.59 & II & 6.67 & 3.49 & 8.91 & 26.38 & 10.88 & 0.19 & SB(r)a \\

		Au18$^*$  & 4.07 & 6.45 & 0.58 & 0.61 & II & 5.65 & 3.31 & 15.64 & 26.21 & 10.91 & 0.13 & SB(rs)b \\

		Au20  & 3.39 & 5.43 & 0.61 & 0.66 & I & 9.31 & 0.00 & 0.00 & 25.77 & 10.68 & 0.33 & SB(rs)c \\

		Au21  & 2.71 & 3.39 & 0.41 & 0.41 & II & 7.31 & 4.41 & 17.38 & 26.38 & 10.89 & 0.28 & SAB(r?)b \\

		Au22$^*$  & 3.73 & 6.28 & 0.55 & 0.56 & I & 2.24 & 0.00 & 0.00 & 25.98 & 10.78 & 0.11 & SB(r)a \\

		Au23$^*$  & 8.14 & 9.50 & 0.63 & 0.69 & II & 6.31 & 4.57 & 23.97 & 26.53 & 10.96 & 0.20 & SB(r)bc \\

		Au24  & 3.39 & 5.09 & 0.51 & 0.52 & II & 12.63 & 4.23 & 31.93 & 26.51 & 10.82 & 0.16 & SB(r)c \\

		Au25  & 3.05 & 4.41 & 0.47 & 0.47 & I & 7.27 & 0.00 & 0.00 & 26.33 & 10.50 & 0.34 & SAB(s)b \\

		Au26$^\dagger$  & 3.39 & 5.43 & 0.48 & 0.48 & I & 3.81 & 0.00 & 0.00 & 26.39 & 11.04 & 0.12 & SB0/a \\

		Au27  & 3.73 & 5.77 & 0.63 & 0.69 & II & 6.21 & 3.49 & 24.96 & 26.28 & 10.98 &  0.22 & SB(r)bc \\

		Au28  & 4.41 & 7.46 & 0.69 & 0.82 & III & 2.32 & 4.70 & 13.81 & 26.28 & 11.02 & 0.14 & SB(r)a \\
   		\hline
	\end{tabular}
\end{table*}

\subsection{Bars}

\subsubsection{Bar profiles}
In the lower panels of Figures \ref{fig:decomposition_1-5} and \ref{fig:decomposition_6-9}, the surface brightness profiles along the bar axis, averaged over a width of 2 pixels ($\sim 0.7$ kpc), are presented. Photometrical studies on bars report the existence of different types of bar profiles \citep{1985ApJ...288..438E, 1990ApJ...357...71O, 1996AJ....111.2233E, 1998MNRAS.299..672S}, that ranges from flat to exponential. Looking at the region where the bar component dominates in the surface brightness profiles along the bar axis, it can be concluded that all bars seem to be exponential. This is reflected also in the Ferrer profile $\beta$ parameter, since for our decomposition $\beta>1$ is obtained for most of galaxies (see Table \ref{tab:decomposition_results}). As shown in \citet{2015ApJ...799...99K}, these values are typical of exponential or intermediate bar profiles. Nevertheless, it is worth noting that $\beta$ was the parameter with higher uncertainties determined by {\sc galfit} (the median of the relative errors is $\sim 60$ per cent), and its value drastically changes in some cases when compared to $\beta$ obtained from noiseless images (see Table D2 from the online supplementary material), affecting also to the bar central surface brightness, $\mu_{0,\text{bar}}$. Thus, neither the central slope of the bar, nor its central surface brightness are well determined.

It was claimed by \citet{1996AJ....111.2233E} that early-type galaxies tend to host flat bar profiles, whereas bars from late-type galaxies are preferentially exponential. For the Auriga galaxies, no correlation of the bar profile nature with Hubble type is found, but we might also lack the statistics needed to show this trend.

\begin{figure*}
	\includegraphics[width=0.45\textwidth]{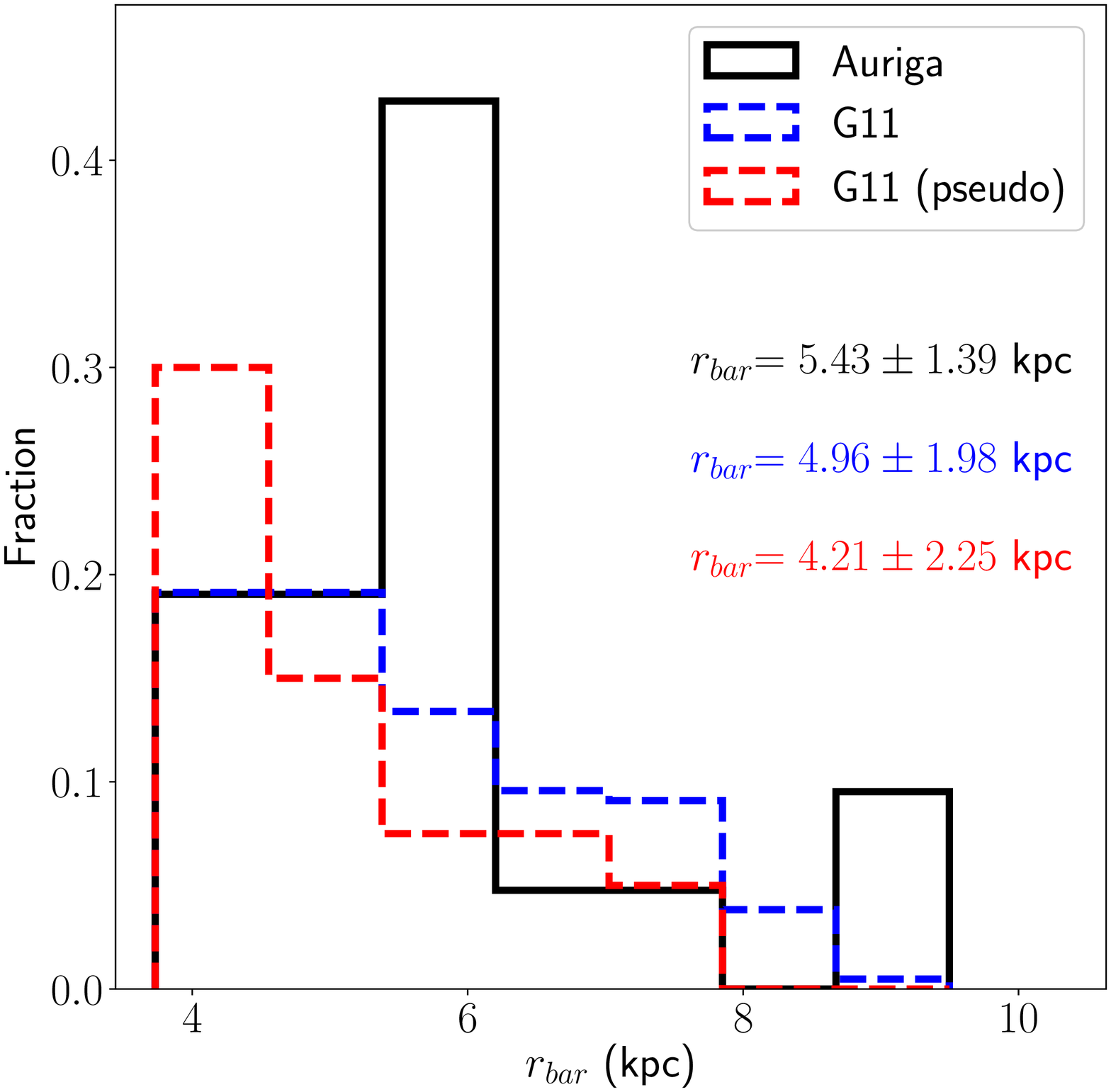}
	\includegraphics[width=0.45\textwidth]{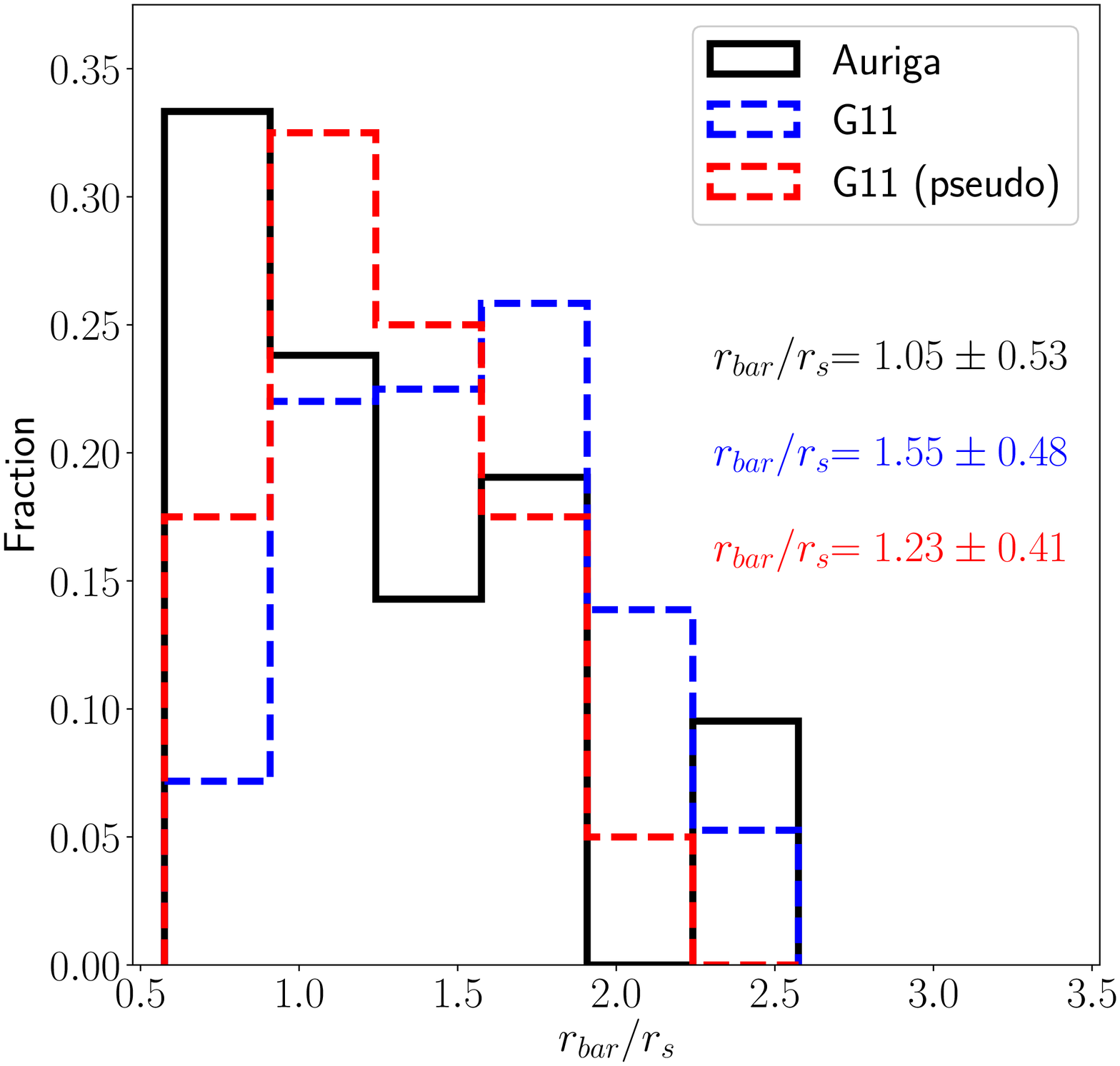}
    \caption{Histograms of the bar length (left panel) and the bar length relative to the disc scalelength (right panel) for the Auriga galaxies in comparison to the sample studied in \citetalias{2011MNRAS.415.3308G}, for both pseudobulges and classical bulges (dashed blue lines), and only pseudobulges (dashed red lines). Only galaxies of \citetalias{2011MNRAS.415.3308G} with mass within the Auriga galaxies mass range $2\times 10^{10}\leq M \leq 1.1\times 10^{11} \text{M}_\odot$ are included. The median values and the standard deviation of the population are indicated on the right side. Bin sizes are $\sim 3\bar{\sigma}$ for $r_\text{bar}$, and $\sim 1.5\bar{\sigma}$ for $r_\text{bar}/r_s$, where $\bar{\sigma}$ is the average of standard errors of individual measurements for the Auriga galaxies.}
    \label{fig:bar_length_histograms}
\end{figure*}

\begin{table*}
	\centering
	\caption{{\sc galfit} 2D multicomponent decomposition results for the face-on {\it r} band synthetic images of the 21 Auriga barred galaxies. Disc, bulge, and bar components are fitted with an exponential, S{\'e}rsic, and modified Ferrer profile, respectively. The columns show: (1) the galaxy simulation name; (2)-(4) the central surface brightness for each component; (5) the bulge effective radius; (6) the disc scalelength; (7) the bar radius; (8) the bulge S{\'e}rsic index; (9)-(10) the Ferrer function $\alpha$ and $\beta$ parameters, respectively; (11)-(12) the axis ratio for the outermost ellipse for bulge and bar component; and (13)-(15) the luminosity ratio for bulge, disc, and bar, respectively. Galaxies with B/P bulges are indicated by $^*$, and buckling bars by $^\dagger$.}
	\label{tab:decomposition_results}
	\begin{tabular}{lcccccccccccccc} 
		\hline
		Galaxy & $\mu_\text{e,B}$ & $\mu_{0,\text{D}}$ & $\mu_{0,\text{bar}}$ & $r_\text{e,B}$ & $r_\text{s,D}$ & $r_\text{bar}$ & $n$ & $\alpha$ & $\beta$ & $q_\text{B}$ & $q_\text{bar}$ & $B/T$ & $D/T$ & ${\rm Bar}/T$\\
		~ & $\left(\frac{\text{mag}}{\text{arcsec}^2}\right)$ & $\left(\frac{\text{mag}}{\text{arcsec}^2}\right)$ & $\left(\frac{\text{mag}}{\text{arcsec}^2}\right)$ & (kpc) & (kpc) & (kpc) &  &  &  &  &  &  &  & \\
		(1) & (2) & (3) & (4) & (5) & (6) & {(7)} & (8) & {(9)} & (10) & (11) & (12) & (13) & (14) & (15)\\
		\hline
		Au1  & 20.58 & 21.03 & 17.71 & 2.03 & 5.01 & 4.19 & 1.45 & 1.50 & 1.49 & 0.51 & 0.30 & 0.19 & 0.65 & 0.16 \\
		Au2  & 19.33 & 21.79 & 19.12 & 1.22 & 11.72 & 8.95 & 0.56 & 1.50 & 1.72 & 0.49 & 0.40 & 0.07 & 0.86 & 0.07 \\
		Au5  & 18.26 & 19.69 & 18.47 & 0.77 & 4.27 & 4.46 & 0.46 & 1.50 & 1.45 & 0.70 & 0.36 & 0.10 & 0.84 & 0.06 \\
		Au6  & 20.12 & 21.07 & 20.18 & 0.98 & 6.55 & 5.43 & 0.41 & 1.50 & 1.40 & 0.61 & 0.50 & 0.04 & 0.91 & 0.05 \\
		Au7  & 18.52 & 19.91 & 19.64 & 0.96 & 5.02 & 5.67 & 0.44 & 1.50 & 1.21 & 0.52 & 0.44 & 0.09 & 0.85 & 0.06 \\
		Au9  & 18.22 & 20.27 & 18.87 & 0.98 & 4.40 & 6.10 & 0.51 & 1.50 & 1.11 & 0.41 & 0.28 & 0.14 & 0.71 & 0.15 \\
		Au10  & 17.10 & 19.33 & 17.04 & 0.85 & 2.89 & 5.43 & 0.31 & 1.50 & 1.35 & 0.46 & 0.31 & 0.19 & 0.48 & 0.33 \\
		Au12  & 18.21 & 19.18 & 19.22 & 0.85 & 3.70 & 3.73 & 0.27 & 1.50 & 0.01 & 0.46 & 0.34 & 0.06 & 0.88 & 0.06 \\
		Au13$^*$  & 16.93 & 19.62 & 18.14 & 0.97 & 2.88 & 5.43 & 0.47 & 1.50 & 0.99 & 0.43 & 0.45 & 0.32 & 0.39 & 0.29 \\
		Au14  & 17.26 & 19.38 & 17.47 & 0.81 & 4.96 & 5.09 & 0.34 & 1.50 & 1.53 & 0.53 & 0.34 & 0.11 & 0.81 & 0.09 \\
		Au17$^\dagger$  & 17.53 & 20.39 & 19.26 & 1.26 & 4.17 & 5.43 & 0.48 & 1.50 & 0.01 & 0.41 & 0.41 & 0.37 & 0.44 & 0.19 \\
		Au18$^*$  & 18.24 & 20.01 & 19.02 & 1.11 & 4.48 & 6.45 & 0.47 & 1.50 & 1.22 & 0.48 & 0.34 & 0.16 & 0.72 & 0.12 \\
		Au20  & 18.78 & 21.27 & 19.39 & 0.99 & 8.17 & 5.43 & 0.51 & 1.50 & 1.13 & 0.42 & 0.43 & 0.08 & 0.82 & 0.10 \\
		Au21  & 18.63 & 19.82 & 20.19 & 1.02 & 5.32 & 4.81 & 0.36 & 1.50 & 0.70 & 0.56 & 0.50 & 0.07 & 0.89 & 0.04 \\
		Au22$^*$  & 18.61 & 19.19 & 17.20 & 0.80 & 2.44 & 5.76 & 0.64 & 1.50 & 1.68 & 0.71 & 0.32 & 0.14 & 0.63 & 0.23 \\
		Au23$^*$  & 18.60 & 20.44 & 18.07 & 1.31 & 6.03 & 9.50 & 0.57 & 1.50 & 1.77 & 0.43 & 0.29 & 0.13 & 0.76 & 0.11 \\
		Au24  & 18.16 & 21.53 & 20.95 & 1.10 & 8.84 & 5.09 & 0.58 & 1.50 & 0.15 & 0.43 & 0.50 & 0.18 & 0.78 & 0.04 \\
		Au25  & 20.81 & 21.26 & 20.40 & 1.01 & 6.00 & 4.41 & 0.32 & 1.50 & 1.27 & 0.42 & 0.41 & 0.02 & 0.93 & 0.04 \\
		Au26$^\dagger$  & 16.33 & 19.60 & 17.62 & 1.18 & 3.31 & 5.22 & 0.50 & 1.50 & 1.24 & 0.41 & 0.50 & 0.48 & 0.31 & 0.22 \\
		Au27  & 18.45 & 19.81 & 18.11 & 0.87 & 5.42 & 5.52 & 0.36 & 1.50 & 1.49 & 0.49 & 0.29 & 0.05 & 0.87 & 0.08 \\
		Au28  & 17.20 & 18.87 & 16.55 & 0.90 & 2.96 & 7.46 & 0.38 & 1.50 & 1.55 & 0.45 & 0.27 & 0.13 & 0.51 & 0.36 \\
        \hline
	\end{tabular}
\end{table*}

\begin{figure*}
	\includegraphics[width=0.33\textwidth]{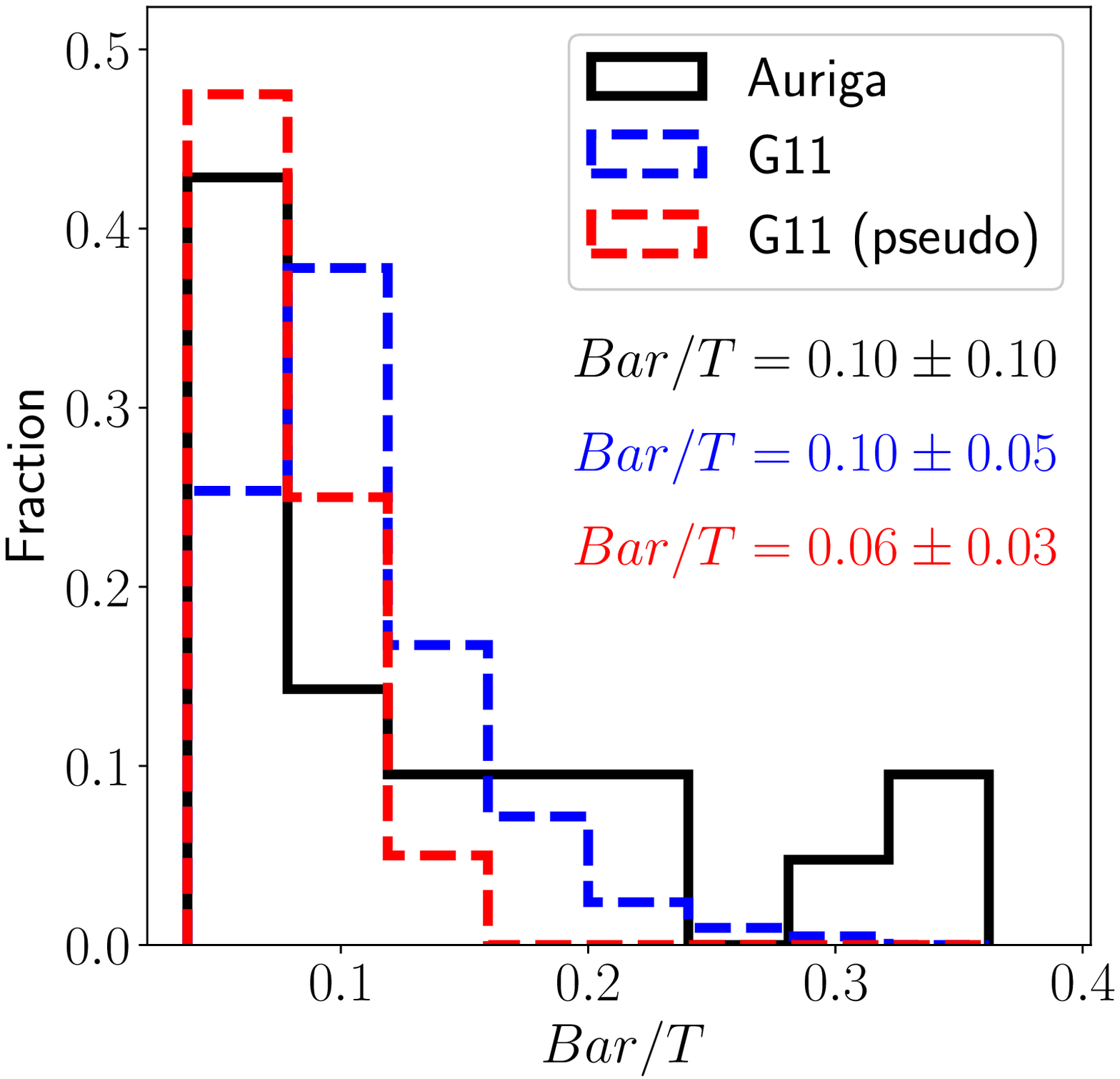}
	\includegraphics[width=0.33\textwidth]{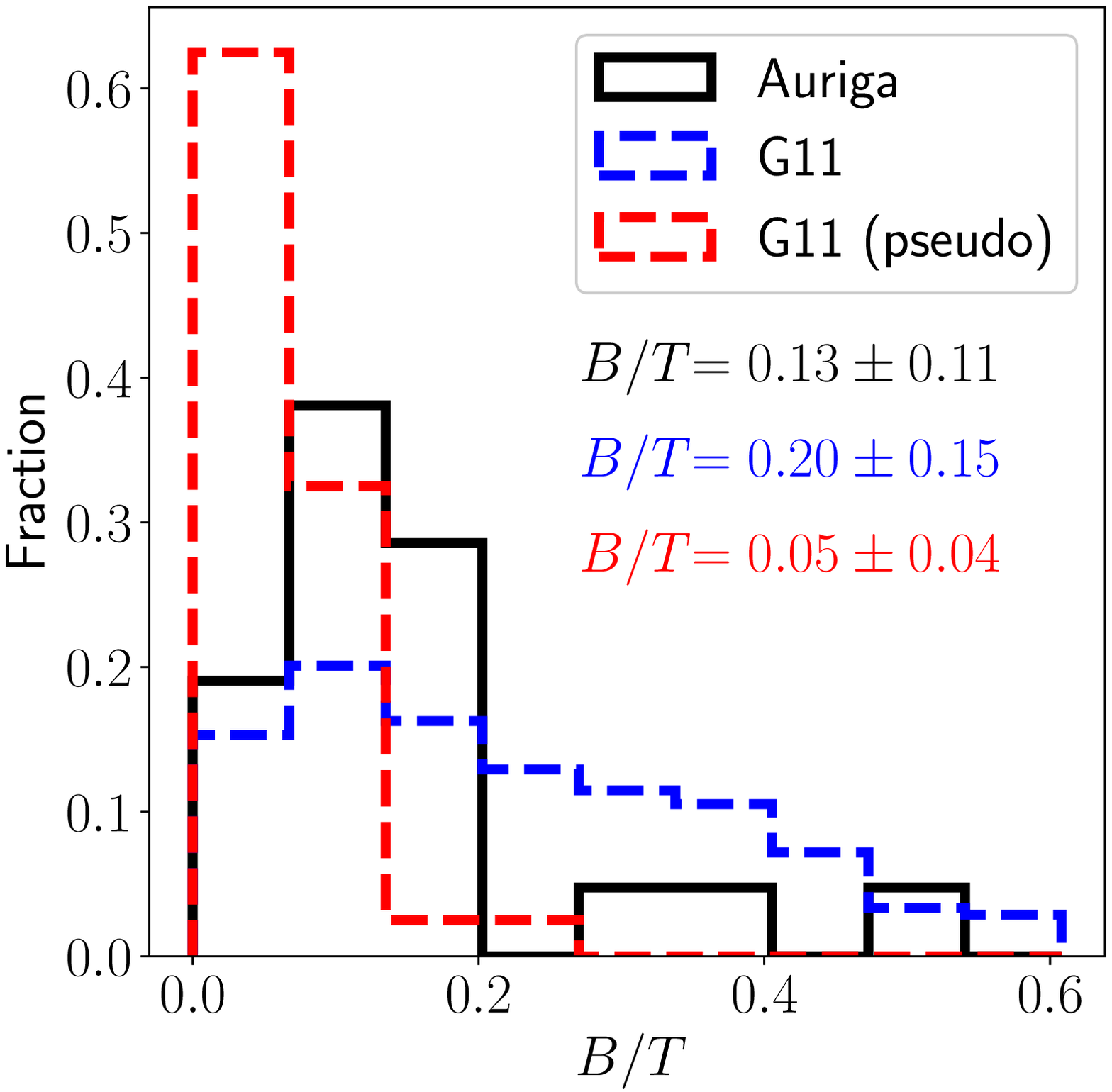}
	\includegraphics[width=0.33\textwidth]{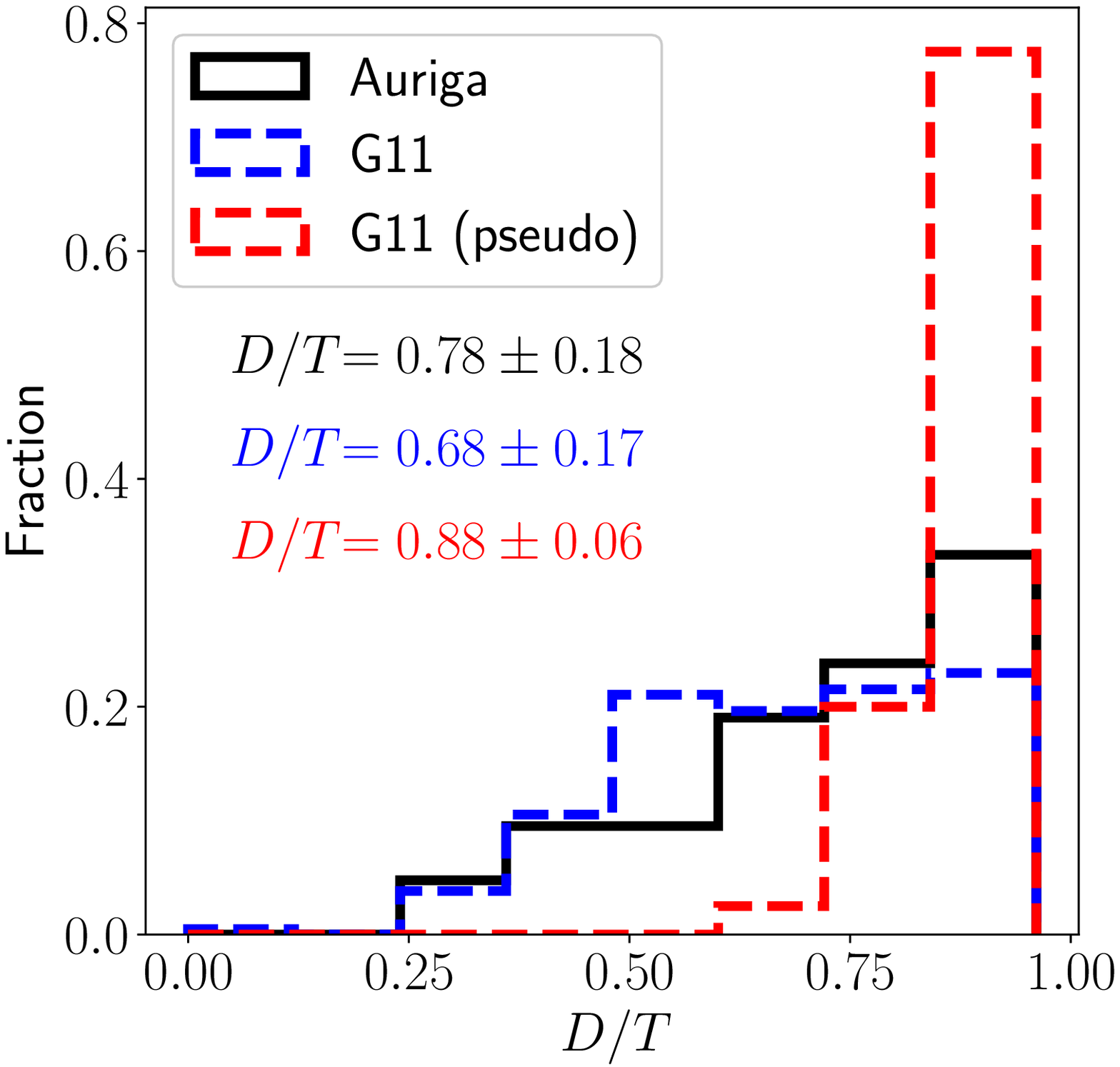}
    \caption{Histograms of the bar-to-total (left panel), bulge-to-total (center panel), and the disc-to-total (right panel) luminosity ratio for the Auriga galaxies in comparison to the sample studied in \citetalias{2011MNRAS.415.3308G}, for both pseudobulges and classical bulges (dashed blue lines), and only pseudobulges (dashed red lines). Only galaxies of \citetalias{2011MNRAS.415.3308G} with mass within the Auriga galaxies mass range $2\times 10^{10}\leq M \leq 1.1\times 10^{11} \text{M}_\odot$ are included. The median values and the standard deviation of the population are indicated in the figures. Bin sizes are $\sim 1\bar{\sigma}$ for ${\rm Bar}/T$, $\sim 1.5\bar{\sigma}$ for $B/T$, and $\sim 5\bar{\sigma}$ for $D/T$, where $\bar{\sigma}$ is the average of standard errors of individual measurements for the Auriga galaxies.}
    \label{fig:comp_to_total_histograms}
\end{figure*}

\subsubsection{Bar length and ellipticity}
In Table \ref{tab:ellipse_results}, the main results from isophotal ellipse fitting are presented for {\it r} band synthetic images. The correlation between parameters $a_\text{max}$ and $L_\text{bar}$ is presented in Fig. \ref{fig:lcorrelation}, and can be compared with the same bar length bounds obtained from real images by \citet{2005MNRAS.364..283E}. The mean ratio of both determinations for our synthetic images is $L_\text{bar}/ a_\text{max}=1.49$, whereas \citet{2005MNRAS.364..283E} obtains $L_\text{bar}/a_\text{max}=1.25$. $L_\text{bar}$ and $a_\text{max}$ can be regarded as a lower and upper limits of bar length, respectively, although it seems that in our measurements we obtain larger ratios between these two parameters than \citet{2005MNRAS.364..283E}.

Also, the bar ellipticity obtained by isophotal ellipse fitting, $\epsilon_\text{max}$, can be compared to the bar ellipticity obtained by 2D multicomponent decomposition, $\epsilon_\text{bar}$, presented in Table \ref{tab:decomposition_results}. From Fig. \ref{fig:epsilon_correlation}, we conclude that $\epsilon_\text{max}$ is in general lower than $\epsilon_\text{bar}$. It is worth mentioning that this occurs not only using {\sc galfit} (the software used in this work) for 2D decomposition, but also \citet{2008MNRAS.384..420G} concludes the same using {\sc budda}. Thus, the bar strength $f_\text{bar}$, defined by Equation \ref{eq:strength}, yields lower values when ellipticities obtained from 2D decompositions are employed.

In Fig. \ref{fig:bar_length_histograms}, the histograms for the bar length $r_\text{bar}$, obtained from the 2D decomposition (left panel), and bar length relative to the disc scalelength (right panel) are presented. Results from the decomposition performed by \citetalias{2011MNRAS.415.3308G} are also shown for comparison. To avoid any correlation of these quantities with the stellar mass of the galaxies, only those whose masses are within the range of the Auriga simulations, from $2\times 10^{10}$ to $1.1\times 10^{11} \text{M}_\odot$, have been taken into account (209 from 291 barred galaxies from the \citetalias{2011MNRAS.415.3308G} sample). Bar lengths are all realistic, since for any of the Auriga simulation we can always find a galaxy in \citetalias{2011MNRAS.415.3308G} sample with a similar bar length.  Since galaxies with pseudobulges have in general shorter and less massive bars than galaxies with classical bulges, the population of galaxies with any kind of bulge (in blue), and the ones with pseudobulges (in red), are represented separately in Fig. \ref{fig:bar_length_histograms}. The median values and standard deviation of the distribution of bar lengths seem to be higher than the ones obtained from observations (these are indicated on the right side of the plots in Fig. \ref{fig:bar_length_histograms}). Nevertheless, bar length is strongly correlated with disc scalelength \citep[e.g.][]{2005MNRAS.364..283E}. Therefore, any different distribution of galaxy sizes would produce different distributions in bar length, provided that bar and disc sizes correlate in the same way in simulated an real galaxies. We have inspected both disc scalelength distributions, and we find that the median values for the sample from \citetalias{2011MNRAS.415.3308G} is $\sim 1.6$ kpc shorter in comparison to the Auriga barred galaxies. Consequently, although we find that all bars are realistic, the bars for Auriga galaxies are expected to be longer, and this is what we obtain. If we aim to compare the bar length of the Auriga galaxies in relation to its host galaxies, we must consider its bar length relative to the disc scalelength. We conclude from the right panel of Fig. \ref{fig:bar_length_histograms} that the relative bar lengths of Auriga galaxies are also realistic, and the median values are closer to the galaxies with pseudobulges. In order to show if these distributions could be drawn from the same parent population, we have performed Anderson-Darling two-sample tests, recommended for small samples. The Auriga galaxy relative bar length distribution is independent of the galaxies with any kind of bulge from \citetalias{2011MNRAS.415.3308G} under a confidence level of $>99\%$ ($p=0.0015$); but when compared to the sample with only pseudobulges, 
the test does not find evidence for differences (we obtain $p=0.15$).

The deficit of short bars reported by \citet{2005MNRAS.364..283E} in non-cosmological simulations, is neither found in the Auriga galaxies nor in the cosmological simulations from EAGLE project \citep{2017MNRAS.469.1054A}. Indeed, it seems that the opposite might be observed: there could be an excess of short bars (in relation to the disc scalelength) for the Auriga galaxies. In principle, a significant amount of short bars could have been undetectable for the sample studied by \citetalias{2011MNRAS.415.3308G}, where the physical spatial resolution is $1.5$ kpc (taking into account the FWHM of the PSF typical of SDSS images), and likely missed most bars shorter than $r_\text{bar}\approx 2-3$ kpc. Nevertheless, since even the shortest bars of the Auriga galaxies have $r_\text{bar} > 3$ kpc, these bars should have been resolved if the image resolution is assumed to be the same.

\subsubsection{Bar-to-total luminosity ratio}

In Fig. \ref{fig:comp_to_total_histograms} (left panel), the histogram of bar-to-total luminosity ratios in the  {\it r} band for the Auriga galaxies is presented, together with the results from \citetalias{2011MNRAS.415.3308G} barred galaxies with masses within the range of $2\times 10^{10}$ to $1.1\times 10^{11} \text{M}_\odot$. The median values of the bar-to-total luminosity ratios, indicated on the plots of Fig. \ref{fig:comp_to_total_histograms}, seem to agree well with observations.  In order to compare the distributions with observations, we have performed Anderson-Darling tests, which reject the null hypothesis that the Auriga galaxy distribution of bar-to-total luminosity ratios comes from the same parent distribution that the one of galaxies with any kind of bulge, with a confidence level of $>95\%$ ($p=0.033$). The null hypothesis is rejected with a higher confidence level, $>99\%$, when compared with the distribution of galaxies with pseudobulges ($p=6.7\times 10^{-4}$). From Fig. \ref{fig:comp_to_total_histograms} an excess of luminous bars in the Auriga galaxies can be found in comparison to \citetalias{2011MNRAS.415.3308G} sample, where only 1 out of 291 barred galaxies has $\text{Bar}/\text{T}>0.3$. Nevertheless, this abundance of luminous bars in the Auriga galaxies does not mean that these are not realistic. The Auriga galaxies with higher bar-to-total luminosity ratio are Au28, Au10, Au13, and Au22 (see Table \ref{tab:decomposition_results}). These are all early-type galaxies (SB0/a or SBa), and real examples of galaxies of this kind can be found with similar bar-to-total ratios. For instance, from the sample studied in \citetalias{2011MNRAS.415.3308G}, most of the few galaxies with $\text{Bar}/\text{T} \sim 0.3$, (NGC 4314, J100008.1+024555, and J22344737-0952545)  are early-type galaxies. Also, from multicomponent decomposition of S$^4$G galaxies \citep{2014ApJ...782...64K,2015ApJS..219....4S}, barred galaxies such as NGC 7552, NGC 5728, and NGC 7582 have morphology and bar-to-total ratios comparable to those Auriga galaxies with the most luminous bars. Other morphological decompositions performed by \citet{2009ApJ...696..411W} or \citet{2018MNRAS.473.4731K} reports also several early-type galaxies with $\text{Bar}/\text{T}>0.3$. 

\begin{figure}
	\centering
	\includegraphics[width=0.45\textwidth]{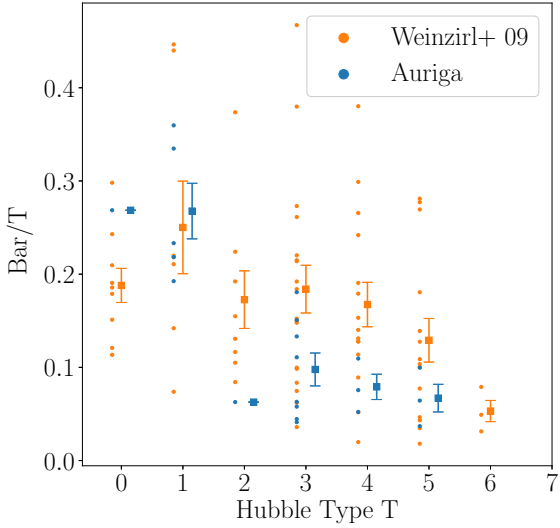}
    \caption{Bar-to-total luminosity ratio plotted against Hubble type for the Auriga galaxies (blue points) and observational results from \citet{2009ApJ...696..411W} (orange points). The square points represent the mean value for each Hubble Type T, the numerical labeling equivalent to the morphological classification presented in Table \ref{tab:ellipse_results}. Error bars width represents the standard deviation of the mean.}
    \label{fig:bar_to_total_weinzirl}
\end{figure}

In Fig. \ref{fig:bar_to_total_weinzirl}, the bar-to-total luminosity ratios in {\it r} band are plotted against the corresponding Hubble type value of the morphological classification presented in Table \ref{tab:ellipse_results}, from S0/a (T=0) to Sd (T=7). The results in {\it H} band obtained by \citet{2009ApJ...696..411W} are also included. Although we lack of statistical significance to confirm any clear trend of bar-to-total ratios as a function of the Hubble type, it seems that, as observations indicate, bar-to-total ratios in early-type galaxies are larger than in late-type galaxies.

\subsubsection{Bar lengths for different image resolutions}

\begin{figure}
	\centering
	\centering
	\includegraphics[width=0.45\textwidth]{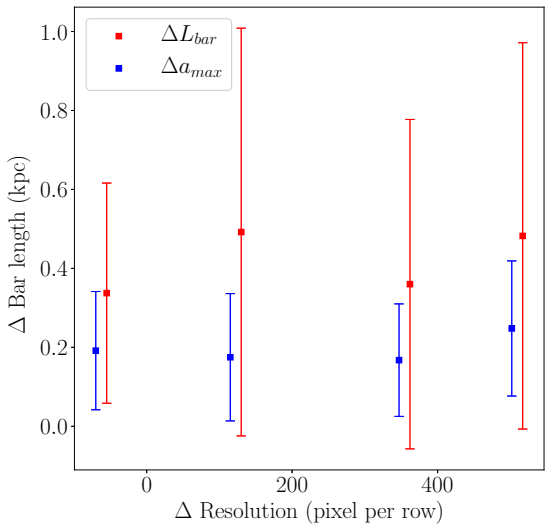}
    \caption{Mean bar length differences as a function of the variation in image resolution with respect to the results showed in Table \ref{tab:ellipse_results}, where synthetic images with 348 pixel per row were used. The error bar represents the standard deviation of the different estimates. Differences in $L_\text{bar}$ are larger and more scattered than differences in the $a_\text{max}$ measurements.}
    \label{fig:all_res}
\end{figure}

In order to study how sensitive are bar length estimates, $L_\text{bar}$ and $a_\text{max}$, to the image resolution, we have also performed ellipse fits over synthetic images of the Auriga barred galaxies with 278, 463, 695, and 850 pixel per row (and per column). In order to create these images, the SDSS pixel scale ($0.396\, \text{arcsec}\,\text{pixel}^{-1}$) was assumed, and we adopted the redshifts of $0.05$, $0.03$, $0.02$, and $0.01635$, respectively\footnote{These correspond to $\sim 0.425$, $\sim 0.255$, $\sim 0.170$, and $\sim 0.139\,{\rm kpc\, pixel}^{-1}$, respectively.}. In order to compare these results with the ones presented in Table \ref{tab:ellipse_results}, we have plotted the averaged differences (over the 21 barred galaxies) in  $L_\text{bar}$ and $a_\text{max}$ as a function of the variation in resolution, taking as reference 348 pixel per row. This is presented in Fig. \ref{fig:all_res}, where error bars indicate the standard deviation of the distribution. Neither the averaged difference in  $L_\text{bar}$ nor in $a_\text{max}$ seem to be significantly affected when resolution is changed. Nevertheless, we observe that the difference in $L_\text{bar}$ measurements are larger and significantly more scattered that $a_\text{max}$. Then, we  conclude that, although $L_\text{bar}$ could represent a closer measurement of the actual bar length, it is in fact a less precise measurement than $a_\text{max}$. In order to show that this is not due to the noise level of the images, we have performed the same to noiseless images, reaching the same conclusions (this shown in Fig. D1 included in the online material).

\subsection{Bulges}
For a detailed study on all bulges of the Auriga galaxies we refer the reader to \citet{2019MNRAS.489.5742G}, that includes an extensive characterisation of their structural properties. Here, we only comment the main results obtained from the 2D multicomponent decomposition of the subsample of galaxies studied in this paper.

Figures \ref{fig:decomposition_1-5} and \ref{fig:decomposition_6-9} show that the S{\'e}rsic models for the bulges are rather flat. This fact is reflected as well in the low $n$ S{\'e}rsic indices (see Table \ref{tab:decomposition_results}), that turned out to be lower than 2 for all the Auriga barred galaxies. This is regarded as an approximated method to discriminate pseudobulges from classical bulges \citep{2008AJ....136..773F}. Moreover, it has been checked that all bulges satisfy the following inequality:
\begin{equation}
\left\langle \mu_e \right\rangle > 13.95 + 1.75 \log r_e,
\end{equation}
where $\left\langle\mu_e\right\rangle$ is the mean surface brightness within the effective radius $r_e$ (we refer the reader to \citet{2005PASA...22..118G} for more information about this quantity), being measured at {\it i} band, and with $r_e$ in parsec. This is the condition proposed by \citet{2009MNRAS.393.1531G}, that a bulge must fulfill in order to be classified as a pseudobulge. These results are in agreement with the bulge characterisation performed by \citet{2019MNRAS.489.5742G}, that concludes that none of the bulges of the Auriga galaxies (barred and non-barred) can be classified as a classical bulge.

In Fig. \ref{fig:comp_to_total_histograms} (center panel), the histogram of the bulge-to-total luminosity ratio for the Auriga barred galaxies is presented, where results from \citetalias{2011MNRAS.415.3308G} barred galaxies with masses within the range of $2\times 10^{10}$ to $1.1\times 10^{11} \text{M}_\odot$ are included for comparison. The median values of the bulge-to-total luminosity ratio compare well with observations, which is between the sample of galaxies with any kind of bulge (pseudo- or classical bulge), and the subsample of galaxies with pseudobulges. The distribution of the Auriga galaxies seems to be closer to galaxies with any kind of bulge, since Anderson-Darling tests yield $p=0.030$ for both kinds of bulges, and $p=1.6\times 10^{-4}$ for only pseudobulges. It is worth noting that the Auriga galaxies are successful in reproducing galaxies with low $B/T$, as Au25, Au6, and Au7, whose bulge components account for $\lesssim 5$ per cent of the total luminosity at {\it r} band. 

Our 2D model yields a mean value and standard deviation of the bulge effective radius $\sim 20 \pm 30$ per cent shorter with respect to the 1D bulge/disc decomposition presented in \citetalias{2017MNRAS.467..179G}. Also, our bulge-to-total luminosity ratios are $\sim 60 \pm 20$ per cent smaller that the bulge-to-total mass ratios from \citetalias{2017MNRAS.467..179G}. These are the typical consequences if bar component is not taken into account, which is already reported comparing different decomposition methods using observational data \citep[e.g.,][]{2006AJ....132.2634L, 2008MNRAS.384..420G}. We can also compare our results of the bulge-to-total ratio with the ones obtained from kinematic decomposition performed by \citetalias{2017MNRAS.467..179G}, or with the combination of spatial and kinematic definition of a bulge proposed by \citet{2019MNRAS.489.5742G}. Both disc-to-total and bulge-to-total ratios are expected to change when a kinematic decomposition is performed instead of a photometric one \citep[e.g.,][]{2010MNRAS.407L..41S}. We conclude that our bulge-to-total ratios are more severely altered than disc-to-total when compared with the kinematic decomposition presented in \citetalias{2017MNRAS.467..179G} and \citet{2019MNRAS.489.5742G}. This is mainly due to the fact that, when a kinematic decomposition is considered, most of the particles that belong to the bar are regarded as part of the spheroidal component. 

\begin{figure}
\centering
	\includegraphics[width=0.45\textwidth]{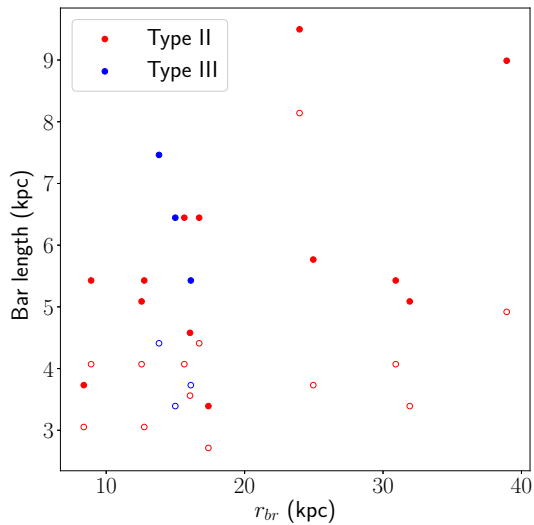}
    \caption{$L_\text{bar}$ (filled circles) and $a_\text{max}$ (empty circles) as a function of the radius at which the disc break takes place, for both Type II (in red), and III (in blue). No correlation is found between the bar length and the disc break radius.}
    \label{fig:rbreak_lbar_correlation}
\end{figure}

\subsection{Discs}

In Fig. \ref{fig:comp_to_total_histograms} (right panel), the histogram for the disc-to-total luminosity ratio for the Auriga barred galaxies is presented, including results from \citetalias{2011MNRAS.415.3308G} for barred galaxies with masses within the range of $2\times 10^{10}$ to $1.1\times 10^{11} \text{M}_\odot$ for comparison. The median values of the disc-to-total luminosity ratio compare fairly well with observations, which is between the median values obtained when taking into account all galaxies from \citetalias{2011MNRAS.415.3308G} and only the subsample of galaxies with pseudobulges. An Anderson-Darling test finds no statistical difference between the Auriga galaxy disc-to-total distribution and the observations for galaxies with any kind of bulges ($p=0.32$); when only pseudobulges from \citetalias{2011MNRAS.415.3308G} are considered, the distributions are statistically different (with $p=1.7\times 10^{-4}$). Then, as well as for rest of total luminosity ratios shown in Fig. \ref{fig:comp_to_total_histograms}, the disc-to-total ratio seem to be closer to the distribution of galaxies with any kind of bulge, even if we do not classify any of Auriga bulges as classical. We nevertheless note that both disc-to-total and bulge-to-total are somewhere in between both distributions from \citetalias{2011MNRAS.415.3308G}, and we consider that a larger sample is needed in order to check if this trend is real.

The 1D bulge/disc decompositions from \citetalias{2017MNRAS.467..179G} yield, in general, shorter values of the disc scalelengths $r_s$ with respect to our 2D model (we obtain disc scalelengths $\sim 10$ per cent larger than \citetalias{2017MNRAS.467..179G}).

Surface brightness profiles obtained by performing isophotal ellipse fits over the synthetic images show the presence of Type I, II, and III breaks \citep[e.g.][]{2006A&A...454..759P} in the Auriga galaxy discs. When performed over noiseless images, almost in every galaxy a second or a third break after the inner one is found. Indeed, most of the Type II discs display a Type III break in the outermost regions. This feature has also been detected by \citet{2017A&A...608A.126R} in the {\sc RaDES} Milky Way-mass cosmological simulations. Nevertheless, this Type III secondary break is found far beyond $\mu_\text{crit}$, which defines the fitting reliability due to noise level, and only the inner truncation has been taken into account in this work to characterise the disc type. The values for $\mu_\text{crit}$ are around $26\,\text{mag}\,\text{arcsec}^{-2}$, and this allows us to compare our results with observational studies on disc breaks over images with similar depth. From noiseless images, we find beyond $\mu_\text{crit}$ in Au22 and Au26 a type III break, and in Au25 and Au20 a type II break; but these have not been taken into account, since they would not be characterised in observations with the same noise level.

For the 21 barred Auriga galaxies, we find 5 Type I ($\sim 24$ per cent), 13 Type II ($\sim 62$ per cent), and 3 Type III ($\sim 14$ per cent) discs. This can be compared to the frequency found in observations for Sb-Sdm galaxies reported by \citet{2006A&A...454..759P}: $\sim 10$ per cent Type I, $\sim 60$ per cent Type II, and $\sim 30$ per cent Type III. Also, a similar study was performed by \citet{2008AJ....135...20E} for S0-Sb barred galaxies, obtaining a frequency of $27$ per cent, $42$ per cent and $24$ per cent for Type I, II, and III breaks, respectively, and a $6$ per cent of a combination between Type II and Type III. From Table \ref{tab:ellipse_results} we observe that the 3 Type III disc breaks are found in early-type galaxies (SBa/0-SBa), and most of Type II breaks are present in later type galaxies (SBb-SBc). This concurs with the results presented by \citet{2006A&A...454..759P}, which reports the same trend with Hubble type. Both \citet{2008AJ....135...20E} and \citet{2006A&A...454..759P} use 1D models in order to fit the disc, which is consistent with the procedure that we have adopted here. Nevertheless, 2D models usually yields different disc type frequencies \citep{2017A&A...604A...4R}.

The break radius relative to the inner scalelength also compares well with the observations. We obtain mean values and standard deviations of $r_\text{br}/r_\text{s,1} = 2.5\pm 0.9$ for Type II, and $r_\text{br}/r_\text{s,1} = 5.8\pm 0.8$ for Type III; for both kinds of breaks $r_\text{br}/r_\text{s,1} =3.1 \pm 1.5$. For example, \citet{ 2008AJ....135...20E} finds $r_\text{br}/r_\text{s,1} = 2.07 \pm 0.81 $ for Type II, and $r_\text{br}/r_\text{s,1} = 4.4 \pm 0.41$ for type III galaxies (we refer the reader to \citet{2017A&A...604A...4R} for a detailed comparison for different observational results found by several authors). As shown in Fig. \ref{fig:rbreak_lbar_correlation}, no correlation between the break radius and bar length is found.

These results can also be compared to studies on discs of the {\sc RaDES} galaxy cosmological simulations presented in \citet{2016A&A...586A.112R}. A higher amount of type III discs are detected in comparison with the Auriga galaxies, but they have not restricted their fitting to a $\mu_\text{crit}$. Nonetheless, they obtain values for the break radius relative to the disc scalelength $r_\text{br}/r_\text{s,1} = 2.2 \pm 0.4 $ for Type II, and $r_\text{br}/r_\text{s,1} = 6.4 \pm 0.4 $ for Type III, that are also comparable to the ones obtained here for the Auriga galaxies. 

\begin{figure}
	\centering
	\includegraphics[width=0.45\textwidth]{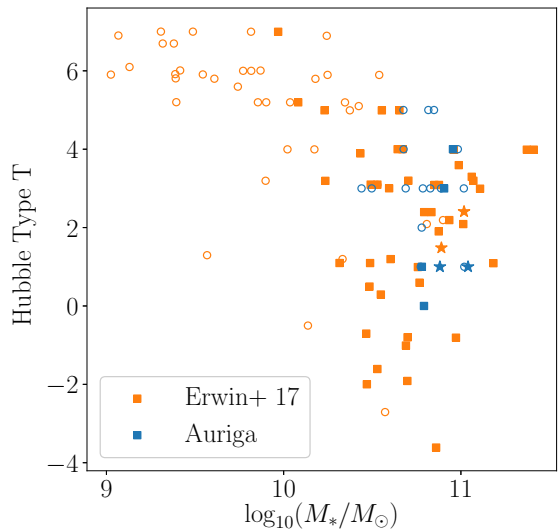}
    \caption{Hubble type T as a function of stellar mass for the Auriga galaxies and for observations \citep{2017MNRAS.468.2058E}, where B/P bulges are represented by filled squares and non-B/P bulges by empty circles. Starred points indicates buckling bars. In the case of the Auriga galaxies, stellar mass is defined as the one enclosed within $0.1$ the virial radius.}
    \label{fig:comparison_erwin}
\end{figure}

\begin{figure}
	\centering
	\includegraphics[width=0.45\textwidth]{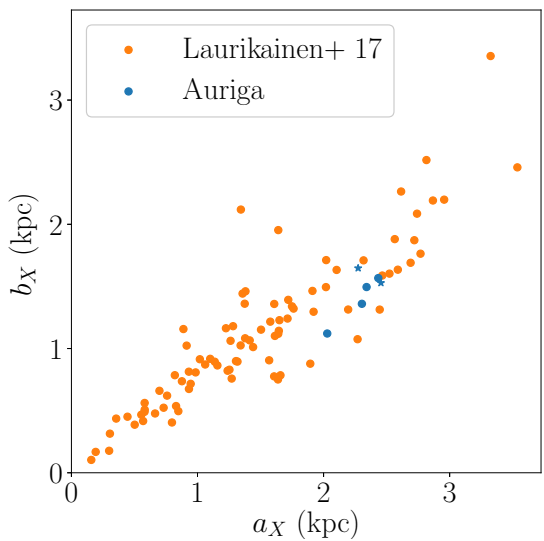}
    \caption{B/P box size measurements compared with the ones obtained by \citet{2017A&A...598A..10L} of edge-on galaxies from S$^4$G and NIRS0S. Starred points indicates bars in buckling, and only the longest semi-minor axis $b_X$ is considered in these cases.}
    \label{fig:comparison_laurikainen}
\end{figure}

\begin{table}
	\centering
	\caption{B/P bulge size measurements: $a_X$ and $b_X$ are the major and minor semiaxis of the box that contains the $X$ shape in unsharp mask images. The X-shape can be offcentered (>-<) or centered (><).} 
	\label{tab:boxy_measures}
	\begin{tabular}{lcccccc} 
		\hline
		Galaxy & $a_X$ & $b_X$ & Buckling & Shape\\
		~ & (kpc) & (kpc) &  & \\
		\hline
		Au13 & 2.341 $\pm$ 0.006 & 1.495 $\pm$ 0.021 &  -  & >-< \\
		Au17 & 2.455 $\pm$ 0.012 & 1.528 $\pm$ 0.012 &  Yes  & >-< \\
		Au18 & 2.434 $\pm$ 0.013 & 1.567 $\pm$ 0.022 &  -  & >< \\
		Au22 & 2.029 $\pm$ 0.009 & 1.121 $\pm$ 0.017 &  -  & >-< \\
		Au23 & 2.304 $\pm$ 0.022 & 1.36 $\pm$ 0.014 &  -  & >-< \\
		Au26 & 2.274 $\pm$ 0.008 & 1.647 $\pm$ 0.016 &  Yes  & >-< \\
        \hline
	\end{tabular}
\end{table}

\subsection{B/P bulge sizes and buckling bars}
The B/P bulges can be examined by unsharp-masked edge-on synthetic images, with the bar major axis horizontally aligned. These edge-on images are shown in the lower part of Figures \ref{fig:decomposition_1-5} and \ref{fig:decomposition_6-9}. The vertical dashed lines indicate $L_\text{bar}$, and the surface brightness step in the contour lines is $0.5\,\text{mag}\,\text{arcsec}^{-2}$. 

The X-shape branches are clearly identified and distinguished from bulges which do not present the B/P structure. Besides, the local density enhancement  that galaxies with B/P bulges usually exhibit on both sides of the boxy region along the bar axis \citep{2003Ap&SS.284..753A}, can also be observed in some of our sample (especially in Au13, Au17, and Au18). Another feature of the Auriga B/P bulges is that the branches of the X-shape usually appear offcentered; that is, they describe a `>-<' shape rather than a `><', which is also seen in observations \citep{2006MNRAS.370..753B, 2016ASSL..418..391A}.

Only 6 out of 21 of the Auriga barred galaxies seem to host B/P bulges \citep[for the most obvious cases, see also][]{2019MNRAS.489.5742G}. This might be a low proportion compared to the observations \citep[e.g.][]{2000A&AS..145..405L, 2015MNRAS.446.3749Y, 2017MNRAS.468.2058E}, if the morphology and masses of the Auriga galaxies are taken into account. For example, \citet{2017MNRAS.468.2058E} studied the presence of B/P bulges in a sample of 84 local barred galaxies that fulfill the following optimal selection criteria for B/P detection: inclination angle within the range of $40-70^\circ$ and relative bar-disc position-angles of $\Delta \text{PA}_\text{bar}\leq 60^\circ$. They argued that this criteria does not show evidences of bias, and they suggested that no significant number of B/P bulges were being missed. From their study it is concluded that B/P frequency for barred galaxies is strongly correlated with stellar mass, and as a side effect, with Hubble type T. This trend is shown in Fig. \ref{fig:comparison_erwin}, where the same parameters for the Auriga galaxies are plotted. It can be observed that from their masses\footnote{The stellar mass of each galaxy the Auriga simulations is the stellar mass enclosed within a radius 0.1 times the virial radius, as presented in \citetalias{2017MNRAS.467..179G}} and Hubble types, more B/P bulges would be expected according to \citet{2017MNRAS.468.2058E}; indeed, they find that 79 per cent of galaxies with masses ${\rm log} (M_*/M_\odot) \gtrsim 10.4$  have B/P bulges. We conclude from Table \ref{tab:ellipse_results} that although none of the galaxies with B/P bulges exceeds $20$ per cent of gas fraction, the low gas content at $z=0$ is not a sufficient condition for the incidence of B/P, in agreement with observations \citep{2017MNRAS.468.2058E}.

Two bars of the Auriga galaxies, those of Au17 and Au26, are clearly undergoing buckling. The fraction of buckling bars observed in local massive barred galaxies is $4.5^{+4.3}_{-2.3}$ per cent \citep{2016ApJ...825L..30E}, that seems to be a low proportion in comparison with our sample.  Nevertheless, in order to check if the Auriga galaxies buckling phase duration and frequency is in accordance to observations and N-body simulations, an evolutionary study at different redshifts should be performed. 

The results of the semimajor ($a_x$) and semiminor axis ($b_x$) of the box that encloses the X-shape feature of B/P, are presented in Table \ref{tab:boxy_measures}. These lengths were measured as described in \citet{2017A&A...598A..10L}, so they can be compared with the results presented therein. These observational measurements of B/P sizes are shown in Fig. \ref{fig:comparison_laurikainen}, where our results from Table \ref{tab:boxy_measures} are also included. Starred points represent buckling bars, and in these cases only the longest $b_X$ length was taken into account. It can be concluded that the Auriga galaxies B/P sizes are in good agreement with observations.

It is worth mentioning that one more method to measure the sizes of B/P bulges has also been tried for the Auriga galaxies. According to \citet{2013MNRAS.431.3060E}, it is not a requisite to have edge-on images and the bar horizontally aligned in order to identify and measure the B/P, but they can be characterised in some cases from images of galaxies with different inclination angles. This is based on the fact that in the inner region of the bar, the B/P bulge should display boxy isophotes, while in the outskirts the isophotes should trace `spurs'. This boxy+spurs morphology was checked also for the isophotes of simulated galaxies by \citet{2013MNRAS.431.3060E}, confirming that the extent of the box major axis would be an estimation of B/P sizes. This is the method they followed to detect the B/P, compared in Fig. \ref{fig:comparison_erwin} with our detections. The proportion of B/P galaxies is in accordance with previous results using edge-on images \citep{2015MNRAS.446.3749Y}. \citet{2013MNRAS.431.3060E} reported that if the inclination angle of a galaxy is in the range of $40-70\degr$ with a relative bar-disc position angle $\leq 60\degr$, the conditions are favorable to characterise B/P. Synthetic images of the Auriga galaxies with inclination angles and relative bar-disc position angle within these ranges have been created in order to check the boxy+spurs morphology. For our synthetic images, we could not find a clear correspondence between the boxy-spur morphology and the presence of B/P observed from edge-on images.

\section{Summary and Conclusions}\label{sec:conclusions}

In this paper, we have presented a characterisation, from an observational perspective, of the structural and photometric properties of bars, bulges, and discs of the barred galaxies from the Auriga cosmological simulations at $z=0$. With this purpose, synthetic images have been created from the simulation photometric data, to mimic SDSS astronomical images of galaxies in the local Universe. Through ellipse fits the bar length upper and lower bounds have been determined. Then, 2D disc/bulge/bar decompositions have been  performed with {\sc galfit}, modelling bars with a modified Ferrer function profile. This enabled us to compare 2D multicomponent decompositions with 1D and kinematical decompositions (presented in \citetalias{2017MNRAS.467..179G}), showing the importance of taking into account the bar component.  

The bar structural parameters are in agreement with those presented by \citetalias{2011MNRAS.415.3308G}, for a sample representative of the galaxy population of the local Universe. Particularly, bar lengths relative to disc scalelengths are within the range found in observations, and bar-to-total luminosity ratios are all realistic. Our results reveal several bars that account for more than $30$ per cent of the whole galaxy luminosity, but some studies report comparable values from real galaxies with similar morphologies. The bar profiles appear exponential rather than flat, and these do not show any trend with Hubble type.

Our 2D bulge/disc/bar decomposition have shown that all bulges of the Auriga galaxies present S{\'e}rsic indexes lower than 2, and they can be considered to be pseudobulges. Their bulge-to-total luminosity ratio at {\it r} band are within the expected range when comparing with galaxies with both kinds of bulges (pseudo- and classical bulges). We report that the Auriga galaxies are successful in reproducing galaxies with low $B/T$; three of them have bulge components that account for $\lesssim 5$ per cent of the total luminosity. 

Regarding the discs, their total luminosity ratio at {\it r} band are found to be within the expected range where galaxies with both kinds of bulges (pseudo- and classical bulges) are considered. From the surface brightness profiles, the Auriga galaxies present the same types of disc breaks observed in real galaxies, (Type II and Type III) and we have shown that the radii at which these take place are in agreement with observations. No correlation between the break radii and bar lengths is found. 

Another feature that the Auriga galaxies successfully simulate is the boxy-peanut structure conspicuous in the inner regions of many of the observed barred galaxies. The sizes of the X-shape bulges have been measured from unsharp-masked images, and they compare remarkably well with the observations. Six out of twenty-one barred galaxy simulations host B/P, two of them being buckling bars. Still, considering the Auriga galaxy Hubble type and masses, a higher fraction of B/P might have been expected to be present according to \citet{2017MNRAS.468.2058E}. Although different B/P detection and size measurement methods which are exclusively used in simulations could be able to characterise them in more detail, we limited our study to observational methods.

We note that no dust obscuration nor reddening have been taking into account in the creation of SDSS synthetic images. Nonetheless, the selection criteria for the galaxies analysed in \citetalias{2011MNRAS.415.3308G} minimize dust effects. As discussed in \citet{2009MNRAS.393.1531G}, there is no substantial difference between the results from their decompositions in different bands. Since dust would have stronger effects on bluer bands, this indicates that their structural and photometric parameters are not significantly perturbed by dust. Also, since in \citetalias{2017MNRAS.467..179G} no dust effects are included, this allows us to compare 1D and 2D bulge/disc/bar decompositions.

Bar formation and evolution have been extensively studied from N-body galaxy simulations, but these assume idealised initial conditions and, in general, isolated systems. Early cosmological galaxy simulations failed to generate galaxies with realistic disc and bulge sizes; thus, parameters as bar length relative to the disc scalelength, or bar-to-total luminosity ratios, could not have been comparable to observations. Nonetheless, this is not the case for present state-of-the-art cosmological simulations. The Auriga cosmological simulations successfully reproduce the bulge and disc structural properties of Milky Way-mass galaxies, allowing their bar lengths with respect to the disc scalelengths, and the bar-to-total luminosity ratios, to be determined and compared with observations. Finally, this work demonstrates that realistic bar features are reproduced in the Auriga cosmological simulations, in agreement with those found in galaxies in the local Universe.

In order to gain insight into bar formation and evolution, the bar characterisation presented in this work could also be performed for synthetic images at different redshifts. Also, future studies on cosmological simulations must pursue statistical significant samples of simulated galaxies with the aim of reproducing the observed statistical properties of bars.

\section*{Acknowledgements}

We acknowledge support from the Spanish Ministry of Science, Innovation and Universities under the grant AYA2014-53506-P, and from the Junta de Andaluc{\'i}a and European Social Fund through the Youth Employment Initiative. FAG acknowledges financial support from CONICYT through the project FONDECYT Regular Nr. 1181264, and funding from the Max Planck Society through a Partner Group grant. This research made use of {\sc Astropy},\footnote{http://www.astropy.org} a community-developed core Python package for Astronomy \citep{2013A&A...558A..33A, 2018AJ....156..123A}; {\sc NumPy} \citep{2011CSE....13b..22V}; {\sc pandas} \citep{mckinney-proc-scipy-2010} and {\sc PyRAF}, a product of the Space Telescope Science Institute, which is operated by AURA for NASA. Plots were generated using {\sc Matplotlib} \citep{2007CSE.....9...90H}





\bibliographystyle{mnras}
\bibliography{bibliography} 




\appendix
\section{2D decompositions and bar profiles}
We present here the images from the 2D decomposition (see Section \ref{sec:decomposition}), analogues to Fig. \ref{fig:decomposition_1-5}, but for the remaining galaxies.


\begin{figure*}
	\includegraphics[height=0.95\textheight]{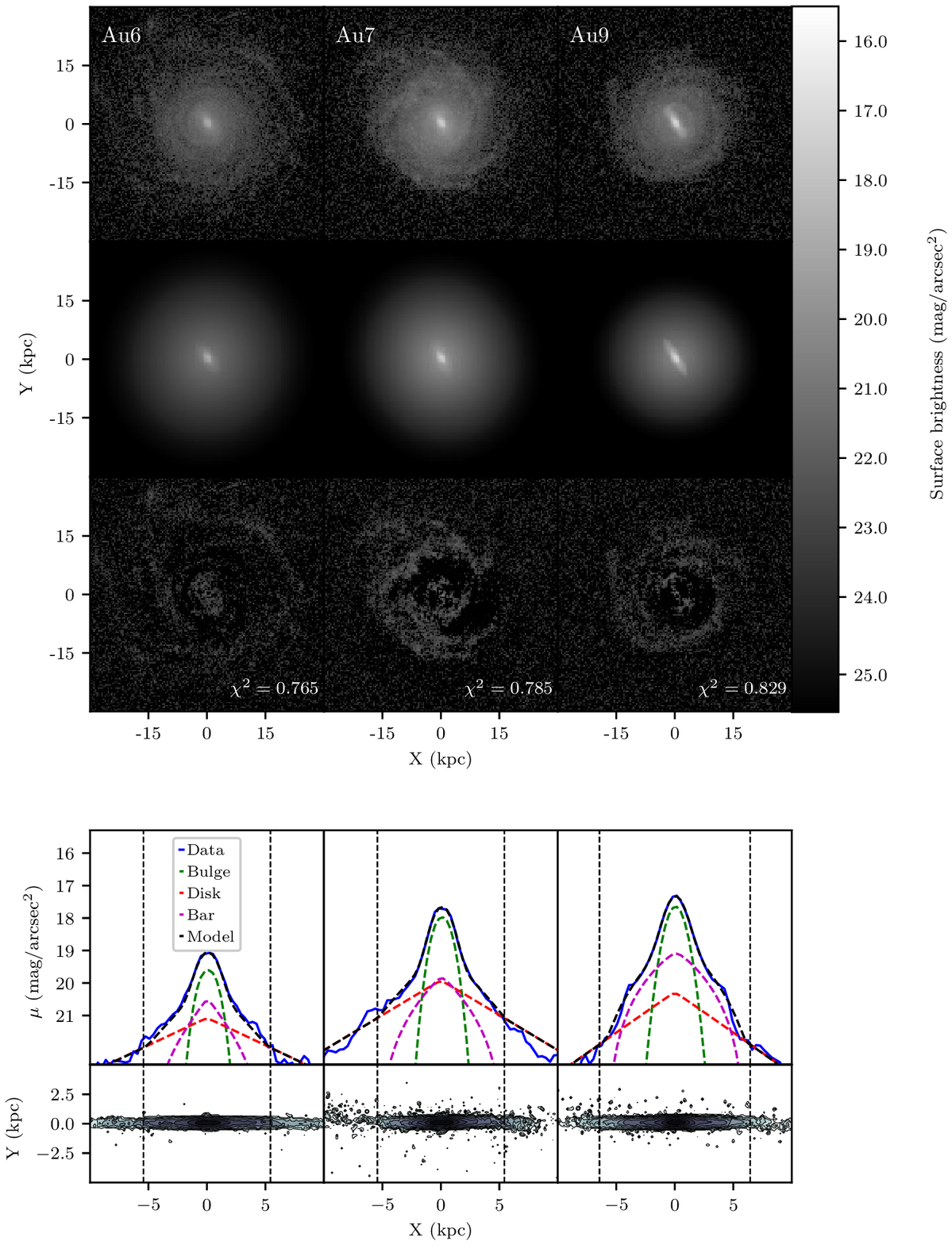}
    \caption{Upper panel: Top: synthetic face-on images in {\it r} band. Center: the model performed by {\sc galfit}. Bottom: the residuals. Bottom panel: {\it r} band surface brightness profiles along bar axis for the different components, including at the bottom the {\it z} band unsharp-masked edge-on images with intensity contours overlaid. Dashed vertical lines indicate the bar length upper limits, $L_\text{bar}$.}
    \label{fig:decomposition_6-9}
\end{figure*}

\begin{figure*}
	\vspace{-1em}
	\includegraphics[height=0.97\textheight]{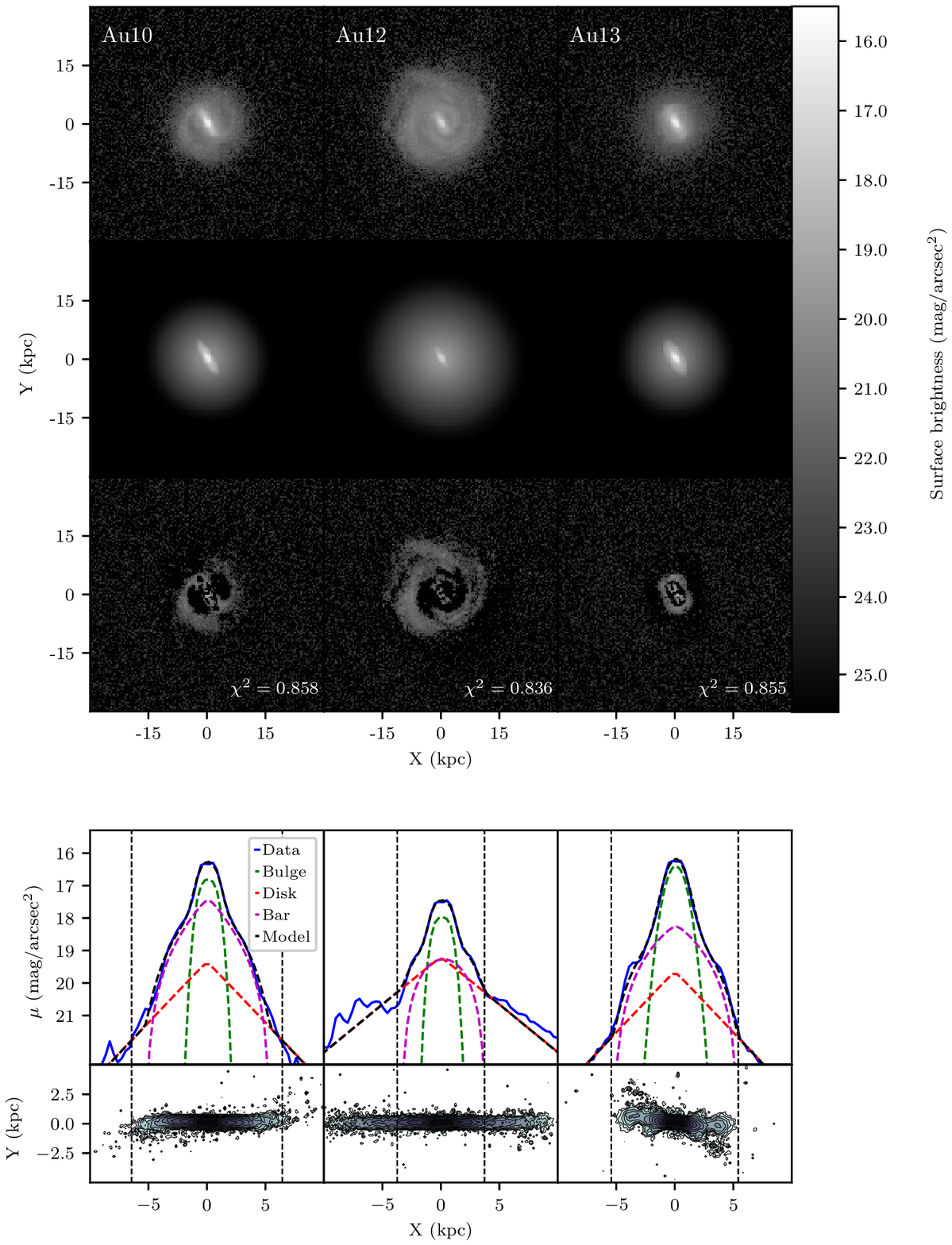}
    \contcaption{}
\end{figure*}

\begin{figure*}
	\vspace{-1em}
	\includegraphics[height=0.97\textheight]{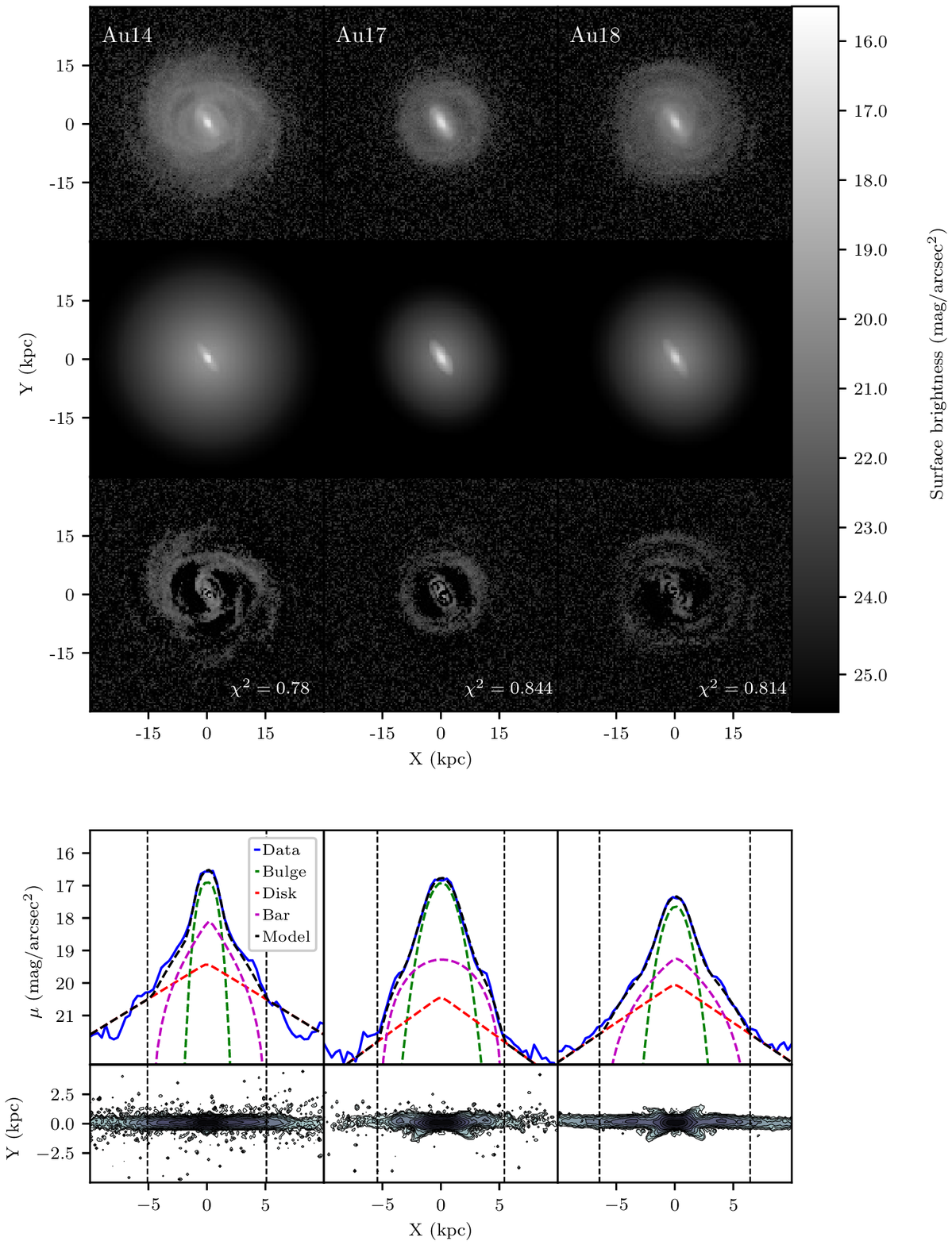}
    \contcaption{}
\end{figure*}

\begin{figure*}
	\vspace{-1em}
	\includegraphics[height=0.97\textheight]{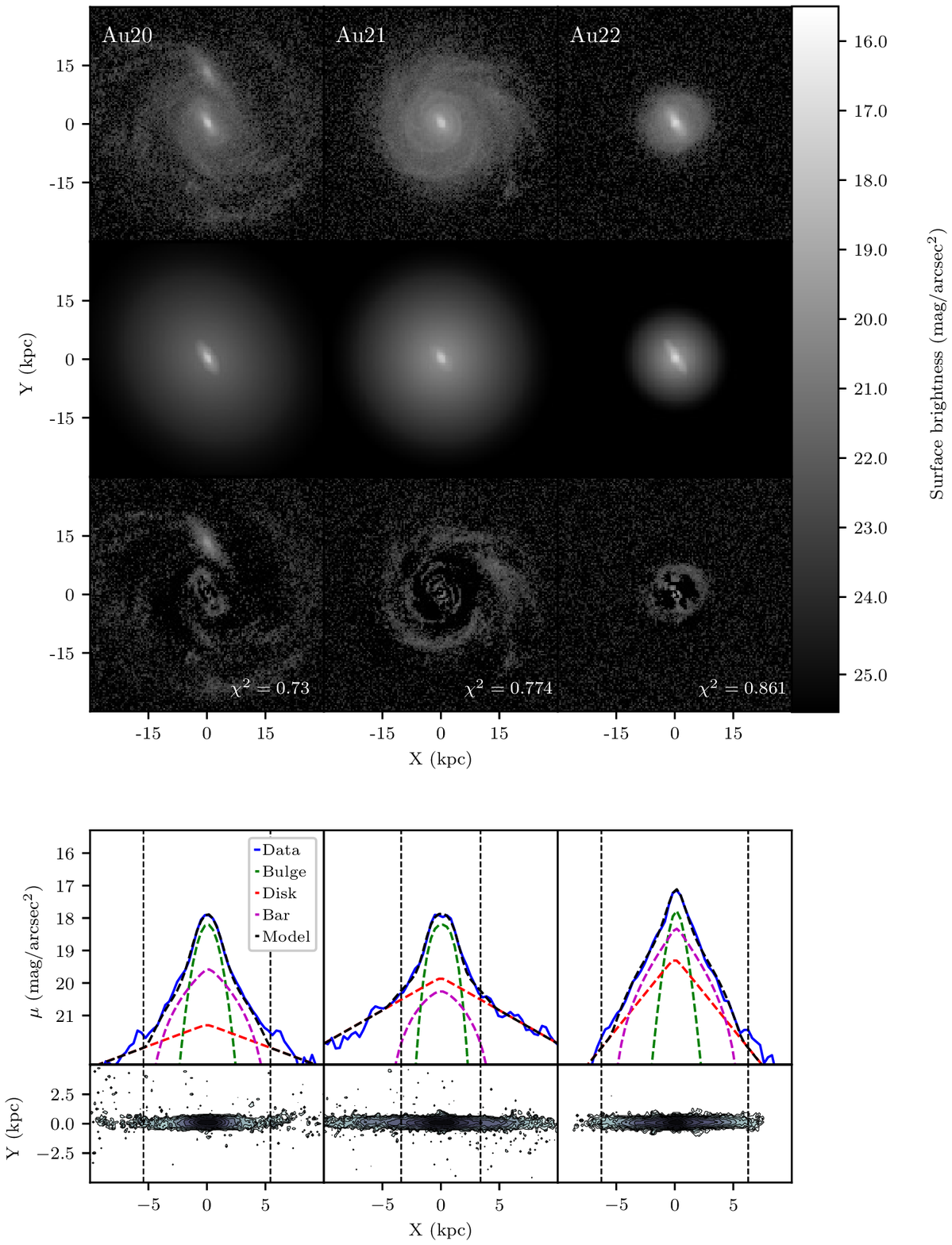}
    \contcaption{}
\end{figure*}

\begin{figure*}
	\vspace{-1em}
	\includegraphics[height=0.97\textheight]{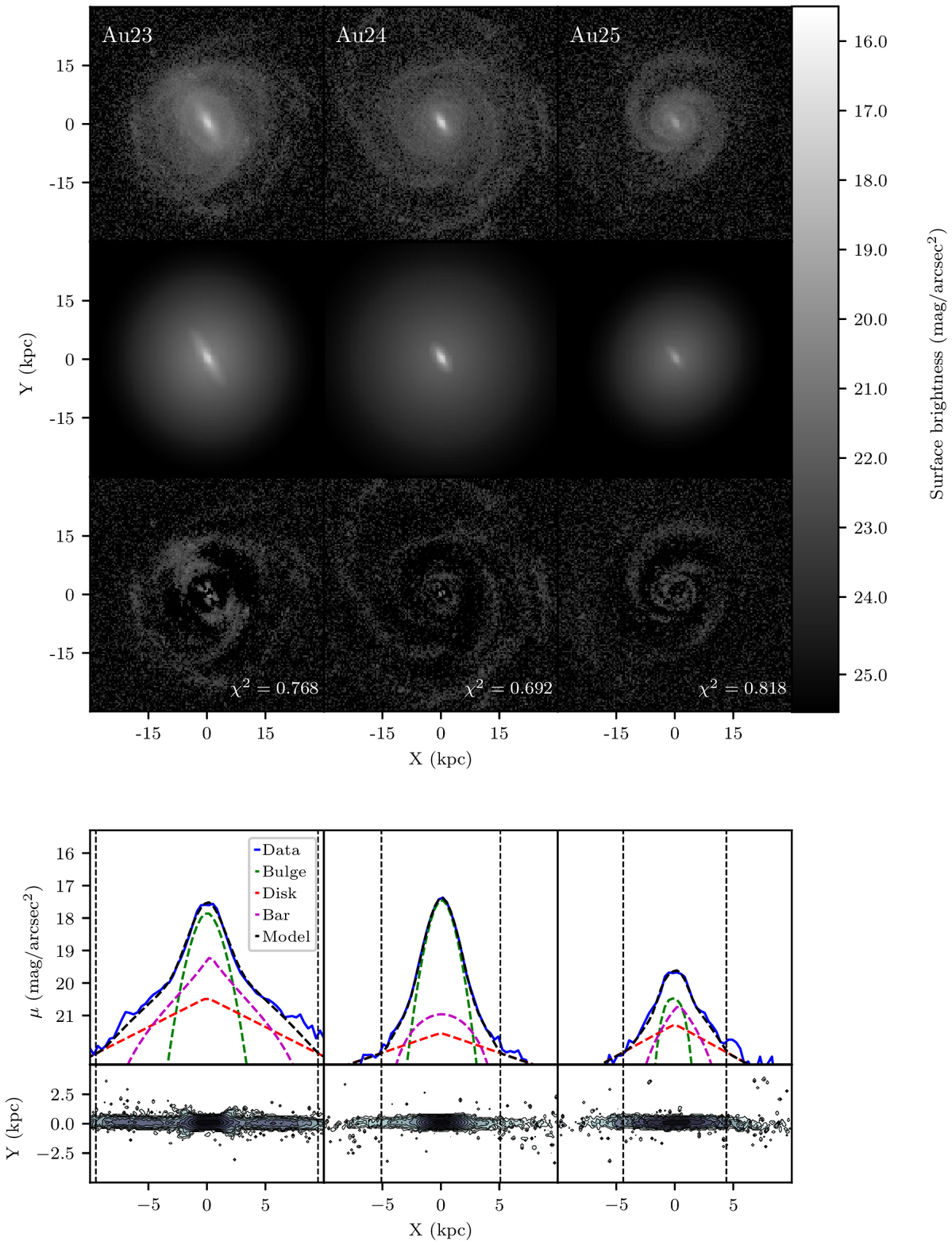}
    \contcaption{}
\end{figure*}

\begin{figure*}
	{\centering
	\vspace{-1em}
	\includegraphics[height=0.87\textheight]{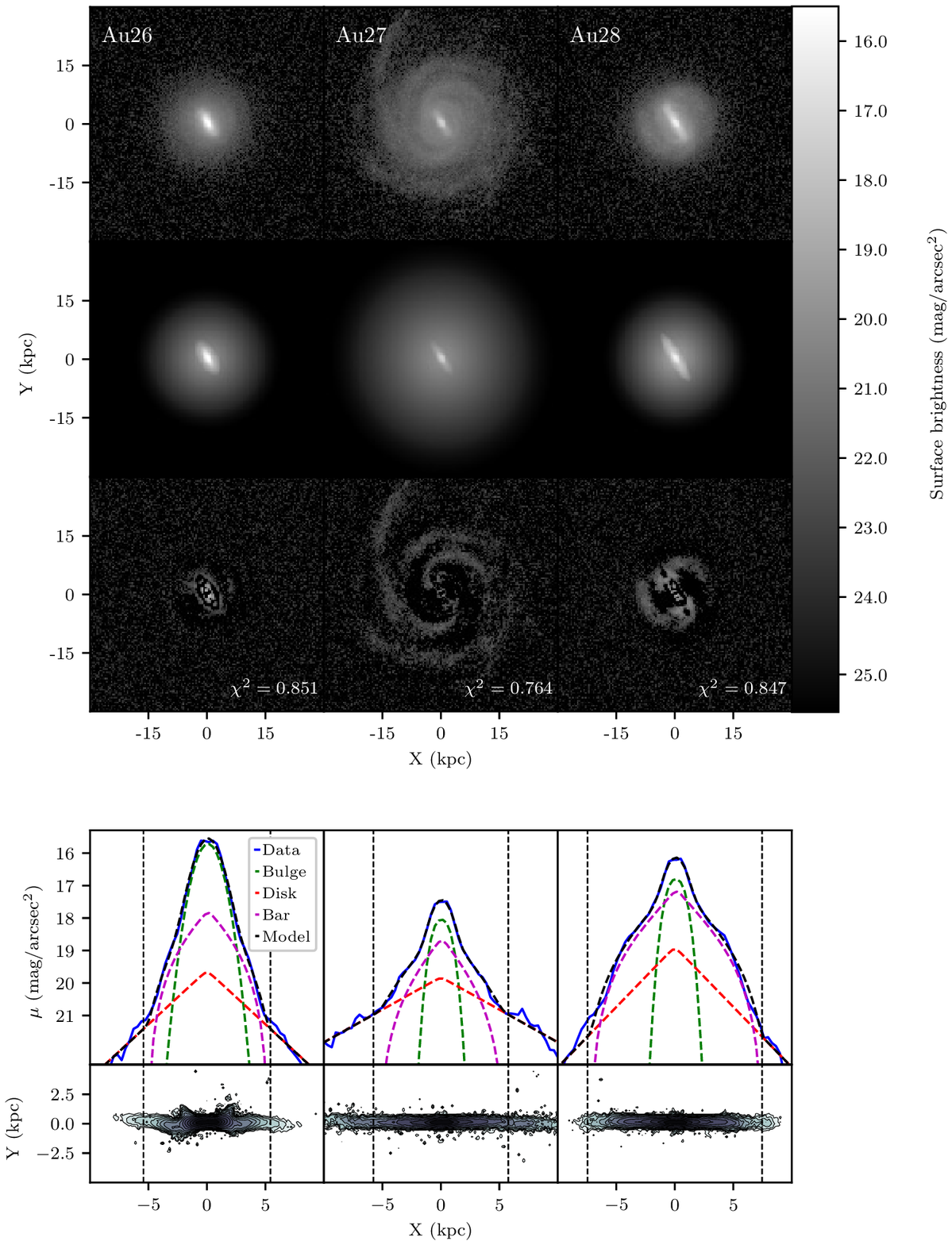}}
    \contcaption{}
\end{figure*}

\bsp	
\label{lastpage}
\end{document}


\appendix
\section{RGB face-on images}

\begin{figure}
	\centering
	\includegraphics[height=0.8\textheight]{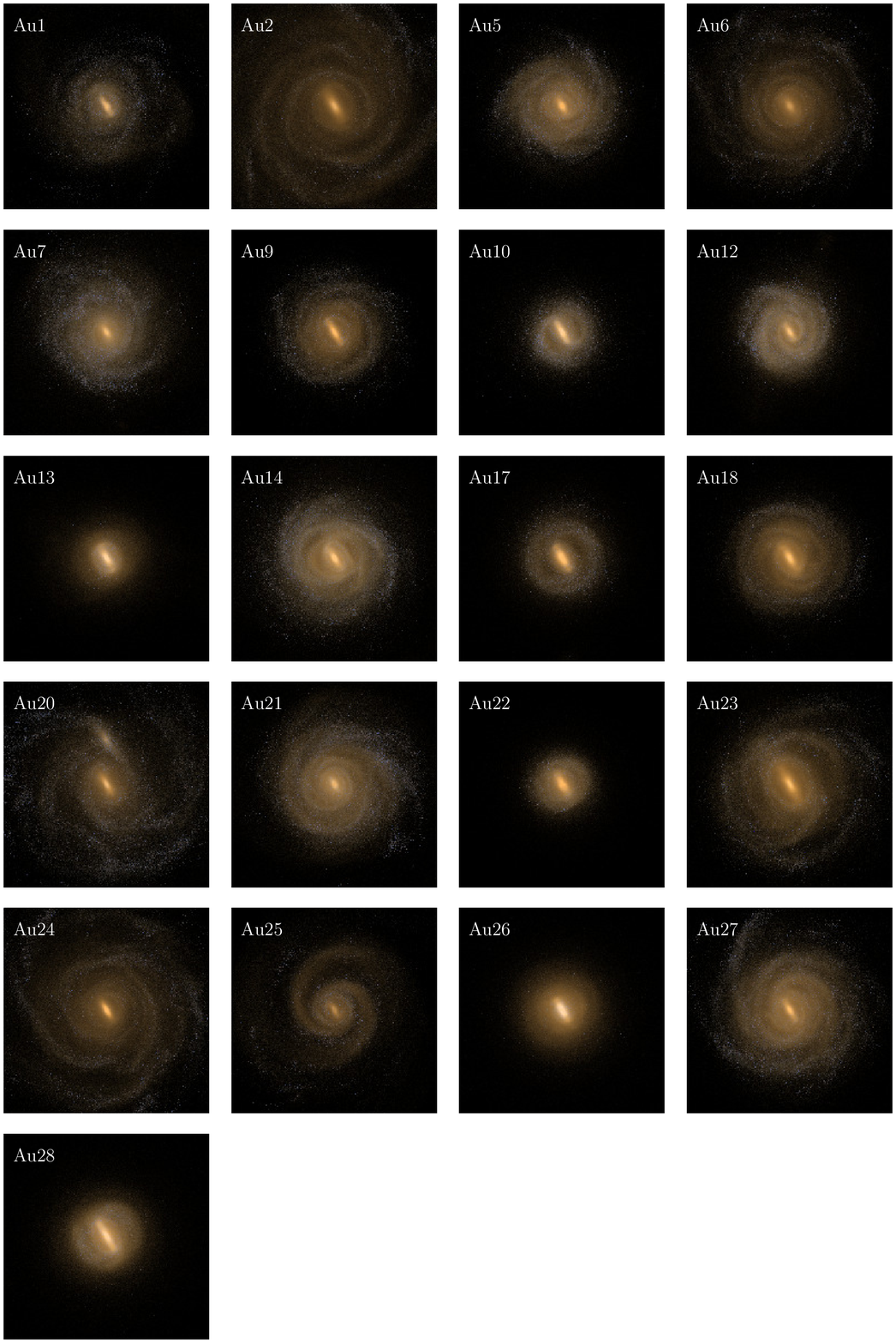}
    \caption{RGB face-on images of the 21 barred galaxies of Auriga simulations, where {\it i}, {\it r}, and {\it g} bands are represented by red, green, and blue colours, respectively. In Table 1 from the paper, the morphological classification of the Auriga galaxies is presented, including the presence of internal rings, bars connected to spiral arms, or a transition between both; denoted by (r), (s) and (rs), respectively.}
    \label{fig:rgb}
\end{figure}

\section{Surface brightness profiles and disc breaks}\label{sec:sbp}

\begin{figure}\label{fig:disk_profiles}
\centering
\begin{tabular}{ccc}
\includegraphics[width=25em]{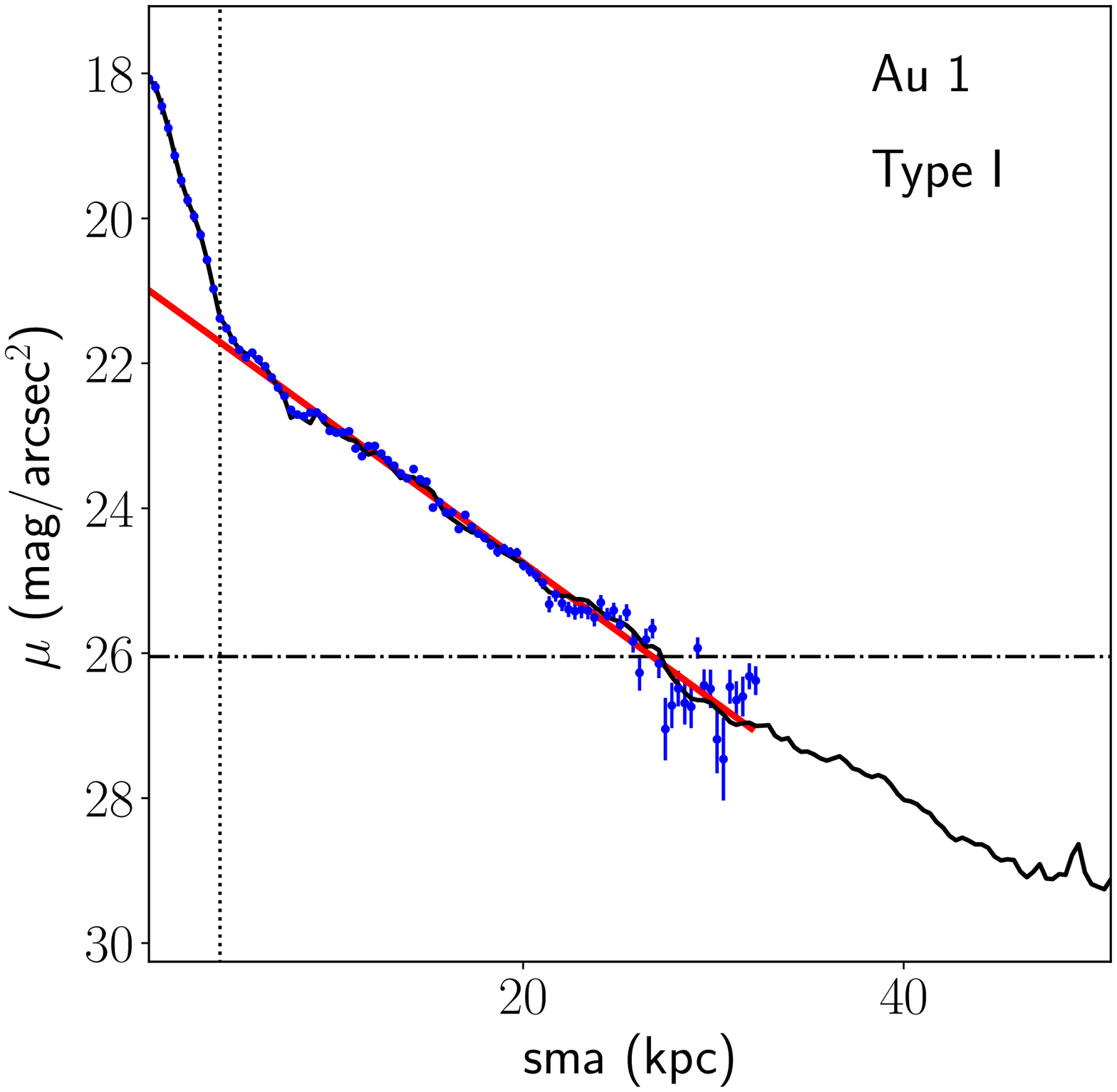} & \includegraphics[width=25em]{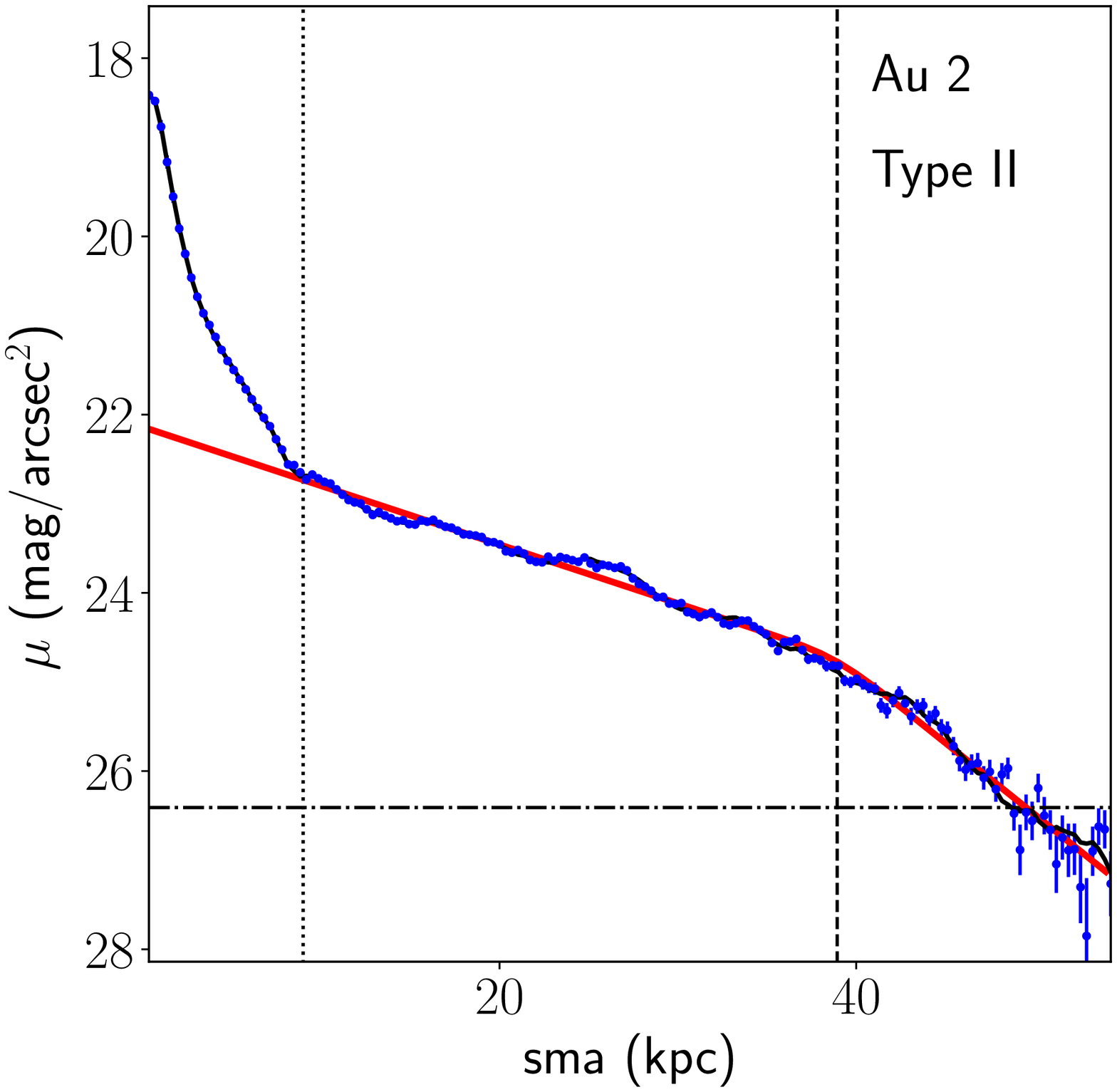} \\
\includegraphics[width=25em]{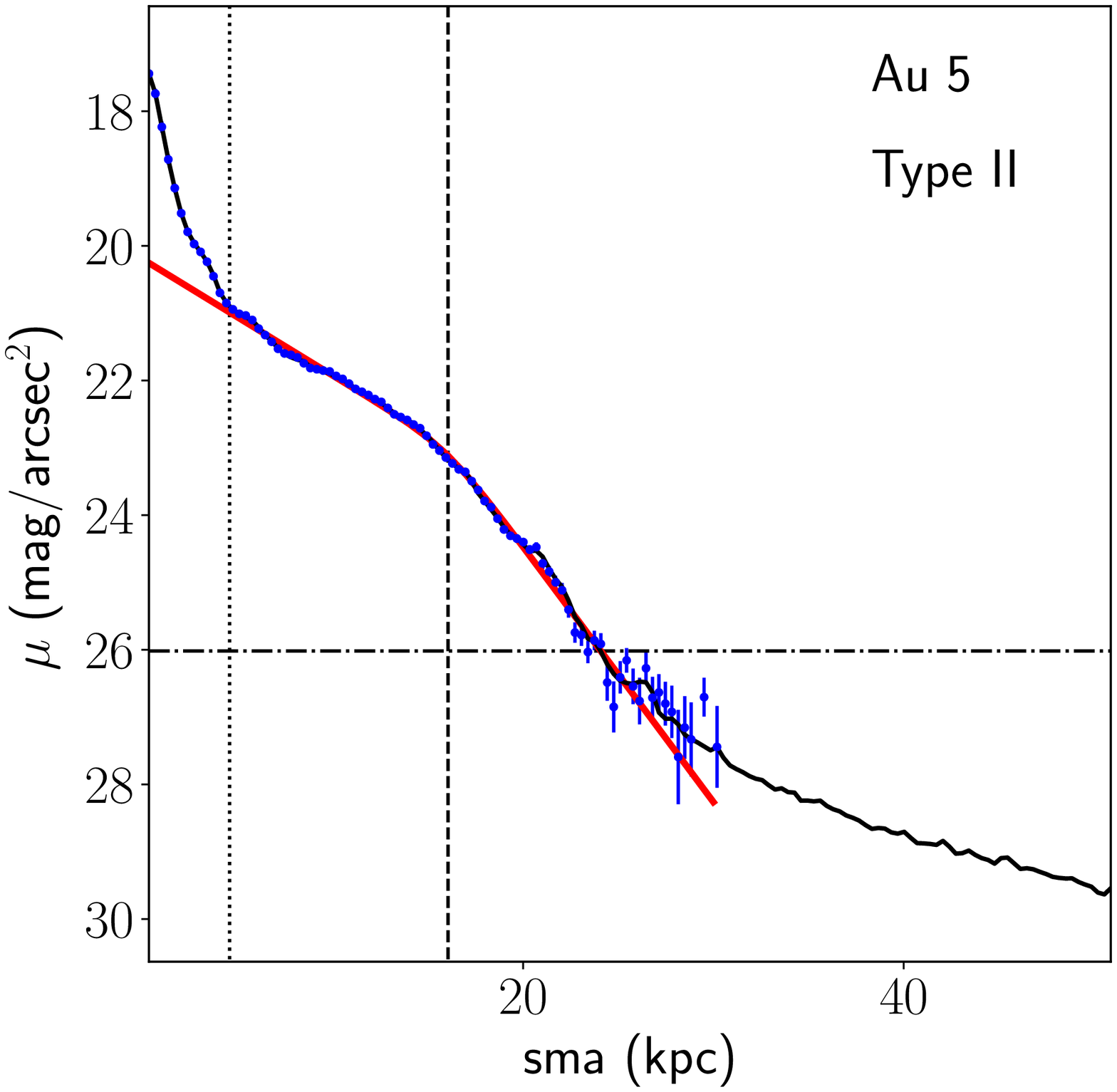} & \includegraphics[width=25em]{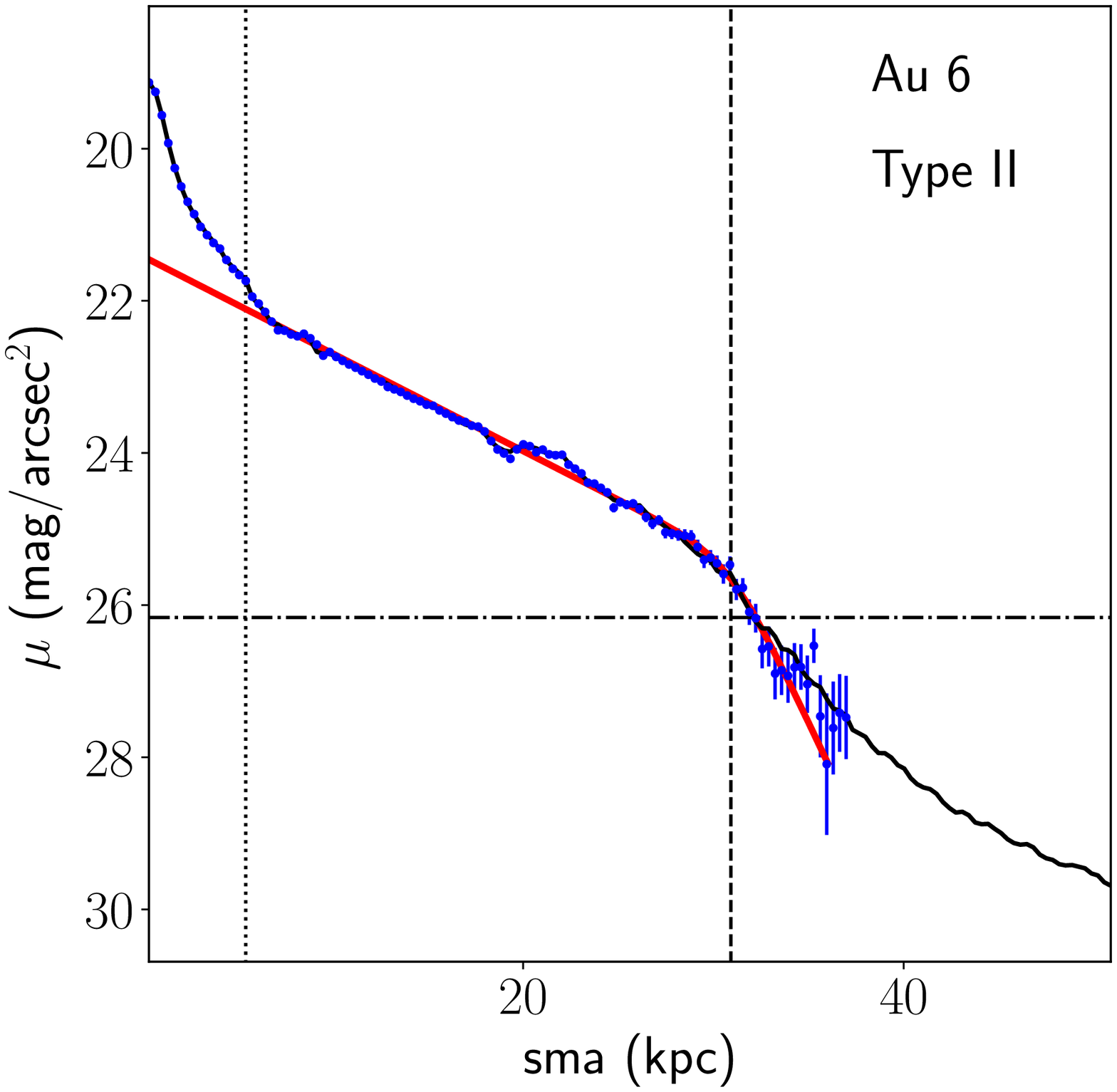} \\
\includegraphics[width=25em]{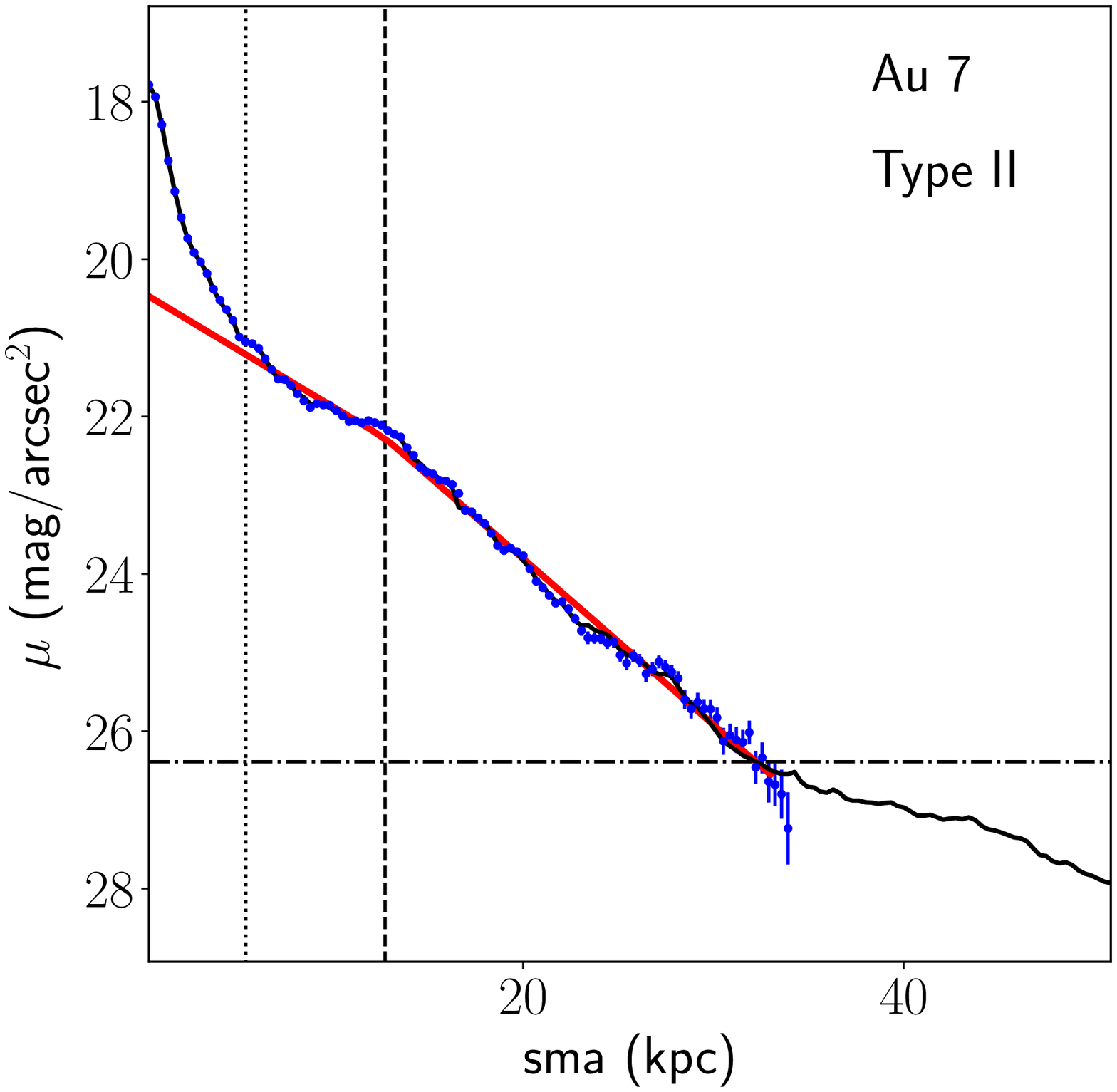} & \includegraphics[width=25em]{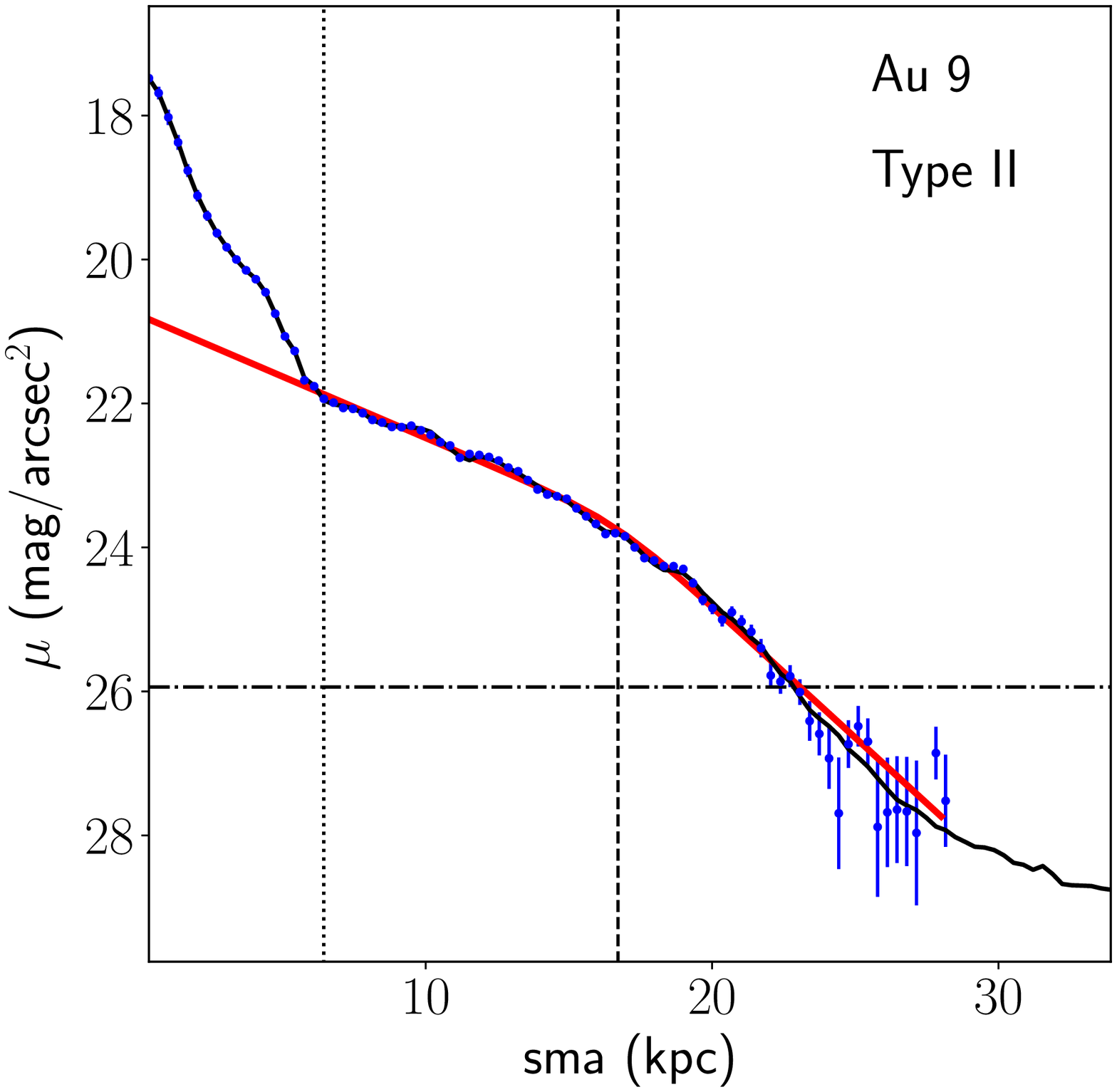} \\
\end{tabular}
\caption{Surface brightness profiles of the synthetic images at $z=0$ using {\it r} band, with noise based on SDSS images (blue points), and without noise (black line). The vertical dotted line indicates the bar length $L_\text{bar}$, distance at which the two-broken exponential fitting (or simple exponential if no break is present) starts. The vertical dashed line indicates the break radius $r_\text{br}$, and the horizontal line is the surface brightness at which the noise level limit is reached, $\mu_\text{crit}$; beyond this point, the fitting is no longer reliable. Red lines are the best fits with the 1D broken-exponential function (Equation 3 in the paper).}
\end{figure}

\begin{figure}
\centering
\begin{tabular}{ccc}
\includegraphics[width=25em]{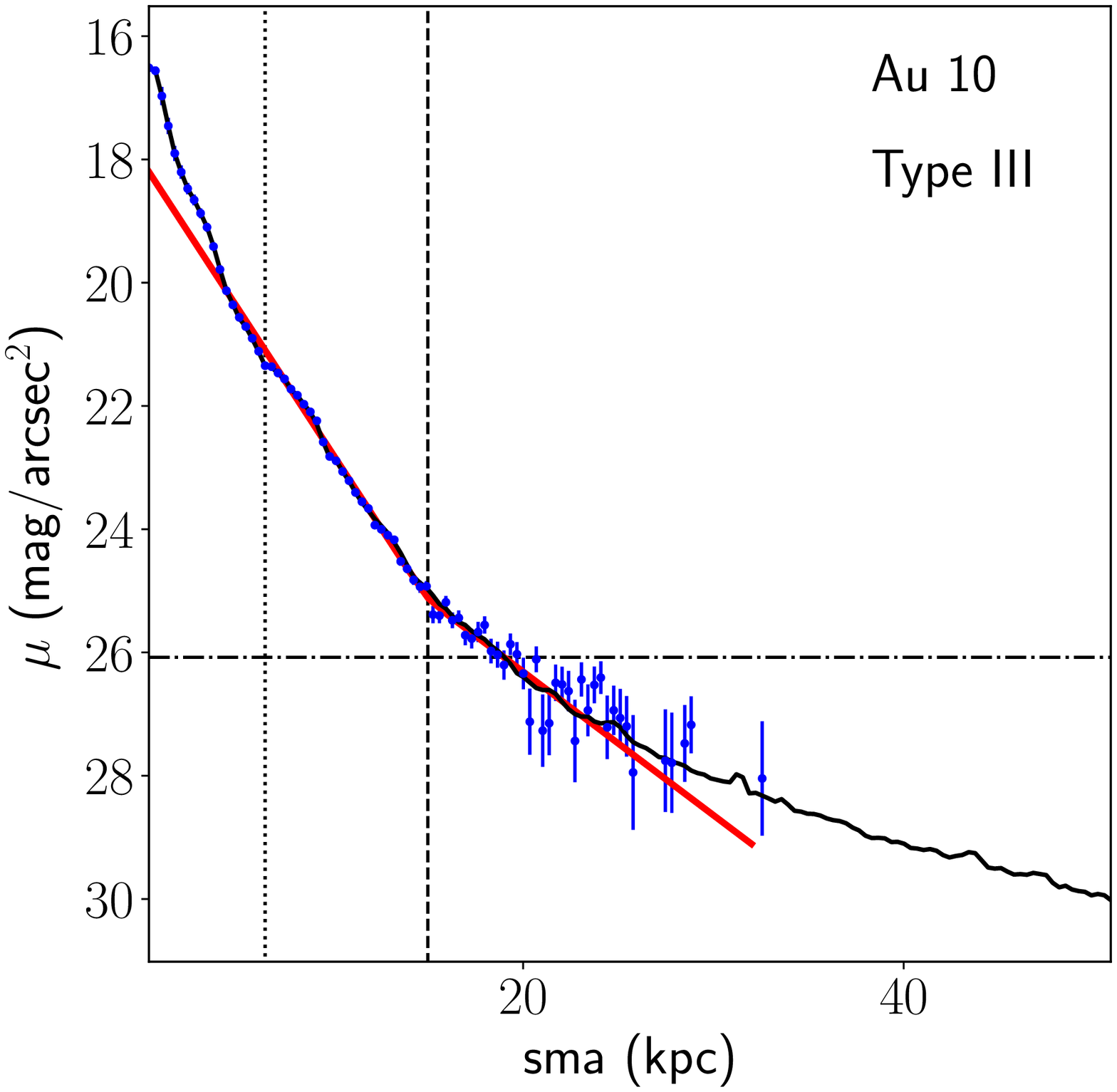} & \includegraphics[width=25em]{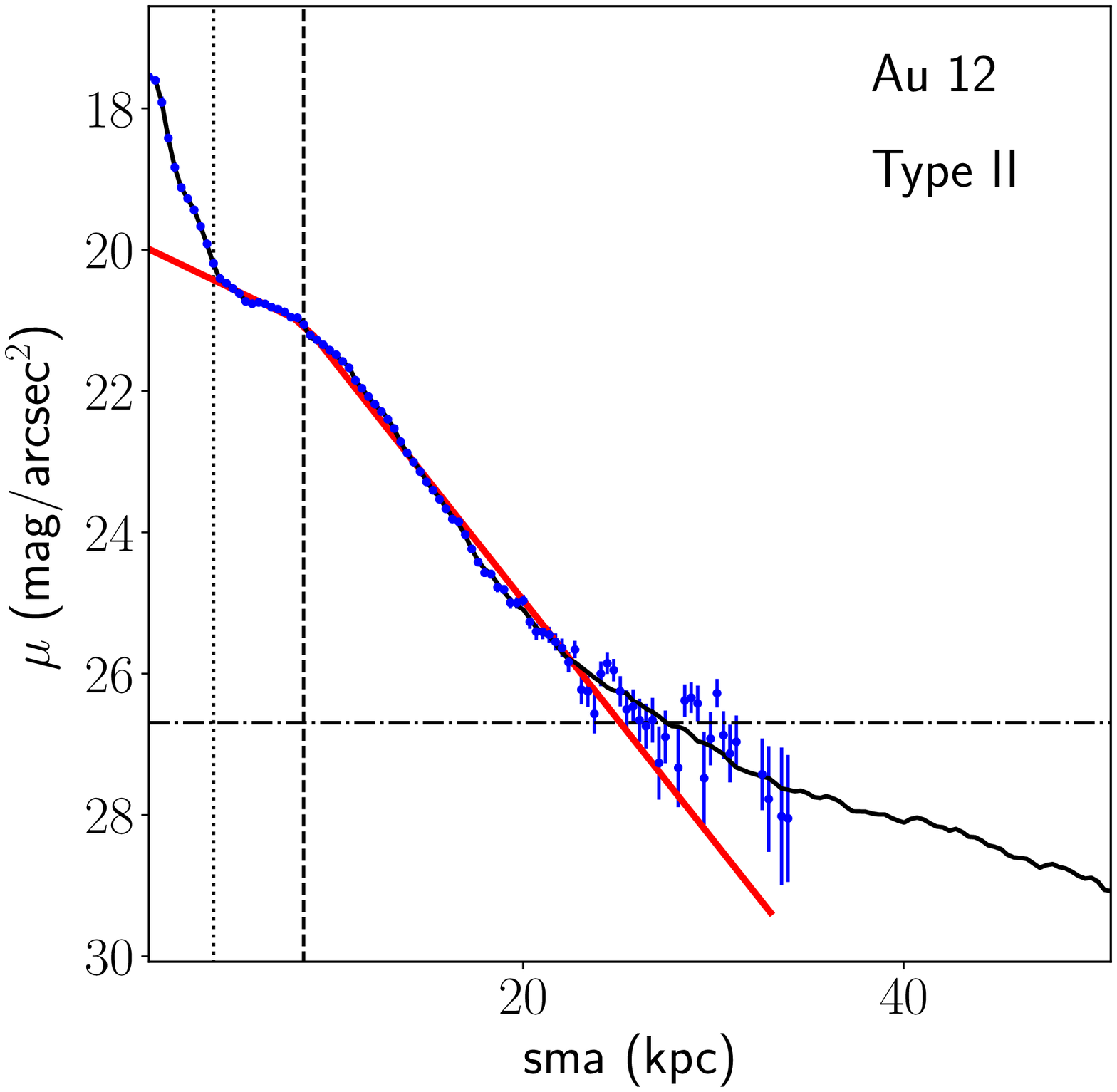} \\
\includegraphics[width=25em]{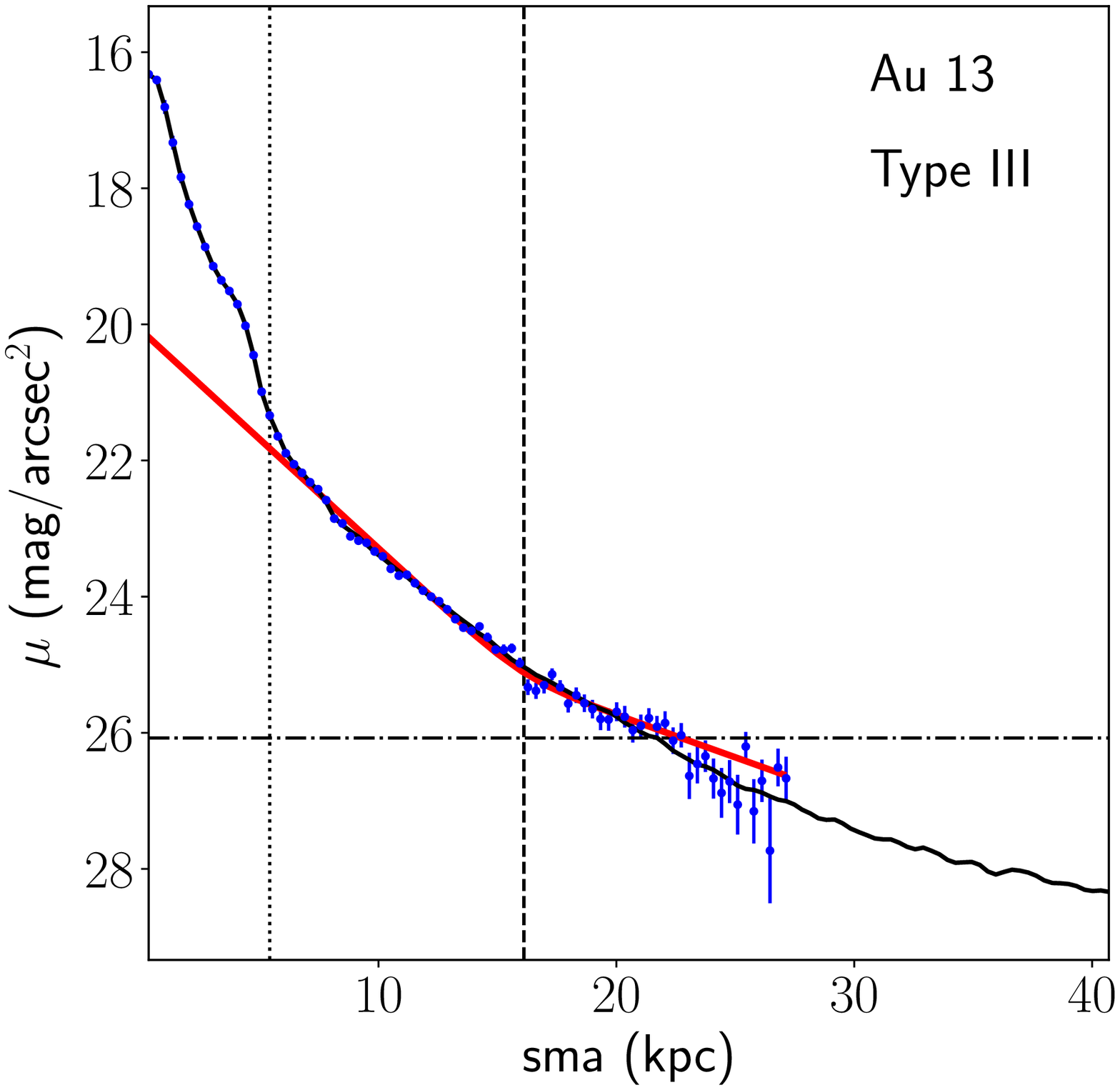} & \includegraphics[width=25em]{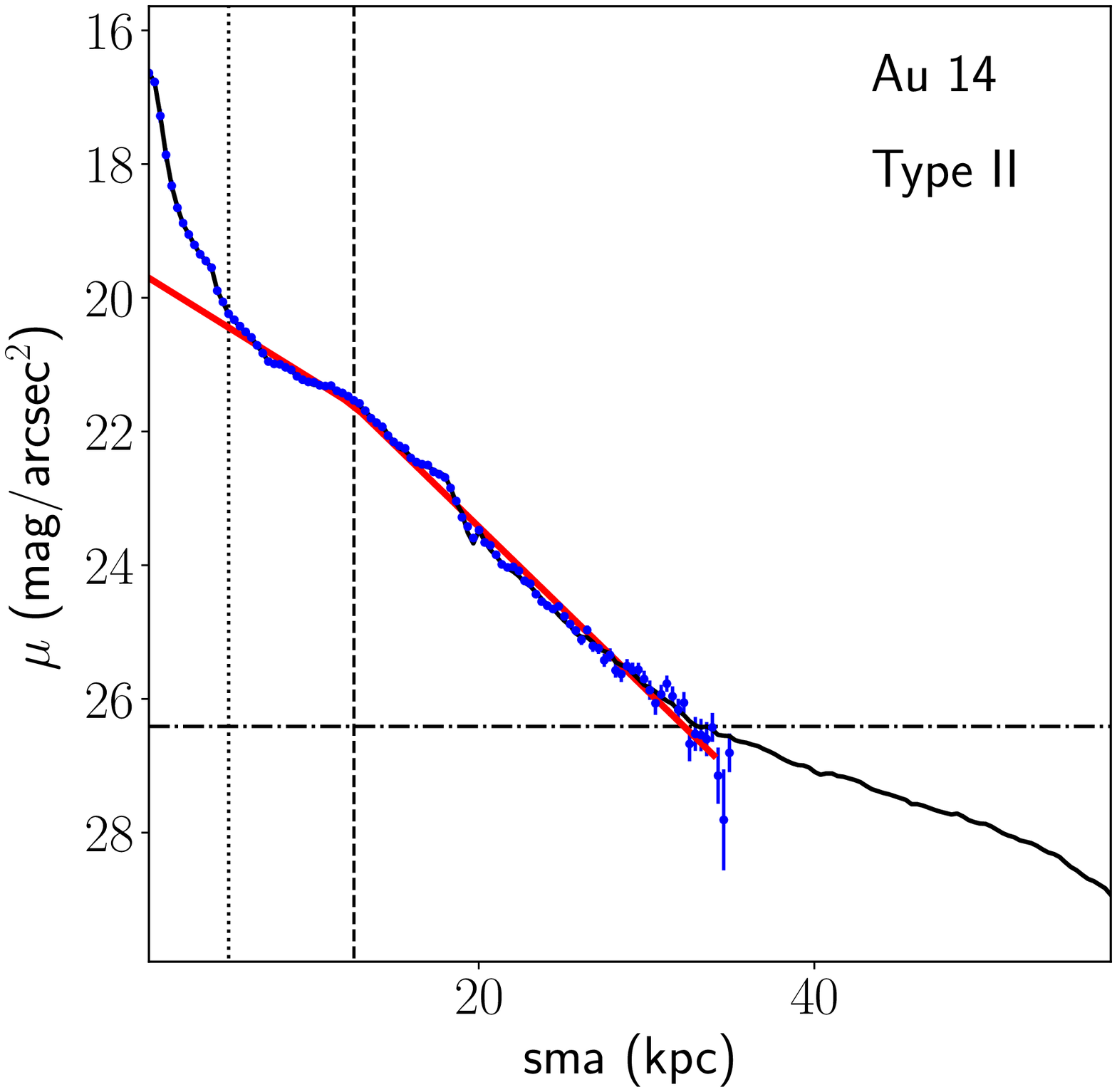} \\
\includegraphics[width=25em]{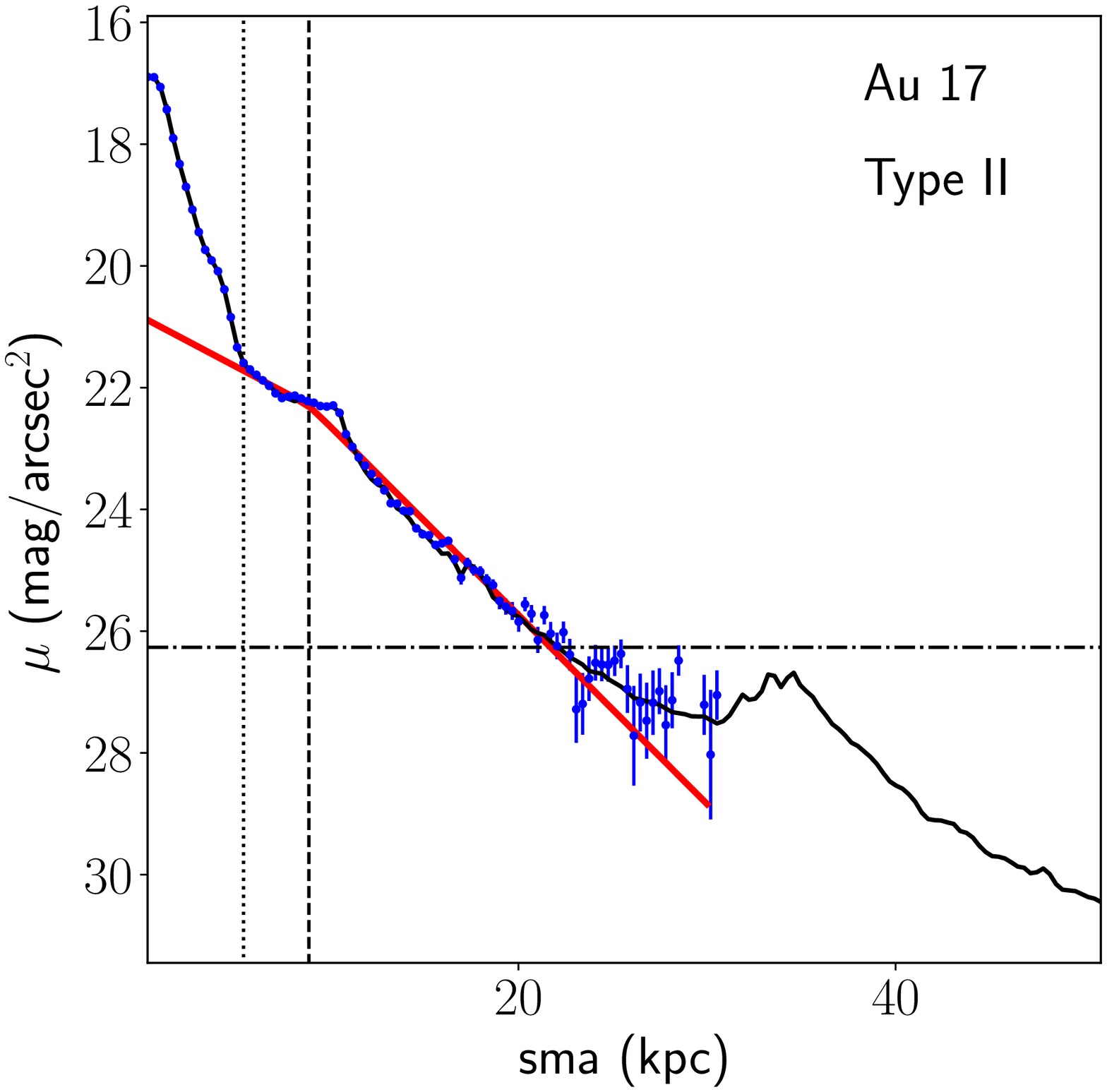} & \includegraphics[width=25em]{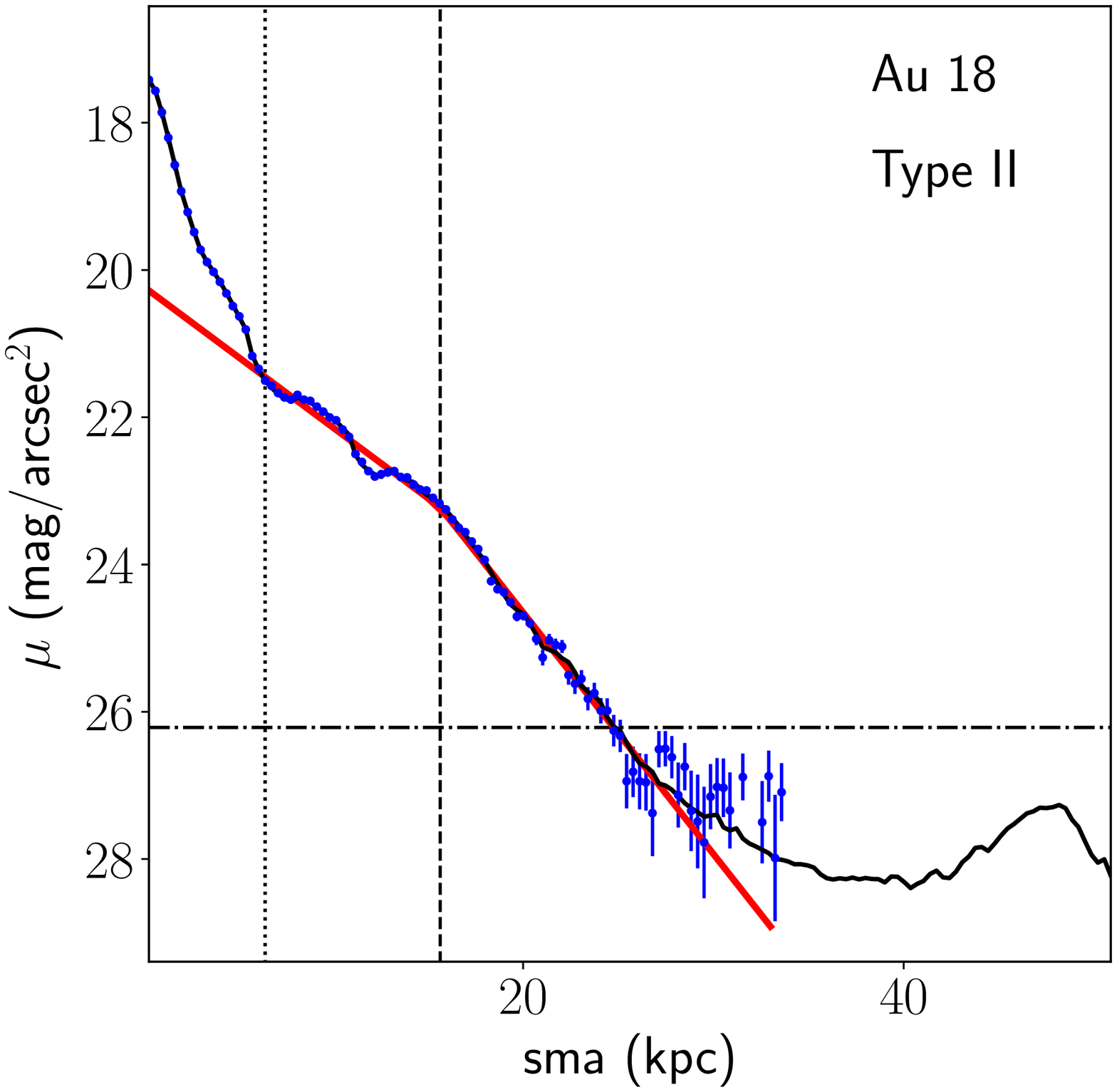} \\
\end{tabular}
\contcaption{}
\end{figure}

\begin{figure}
\centering
\begin{tabular}{ccc}
\includegraphics[width=25em]{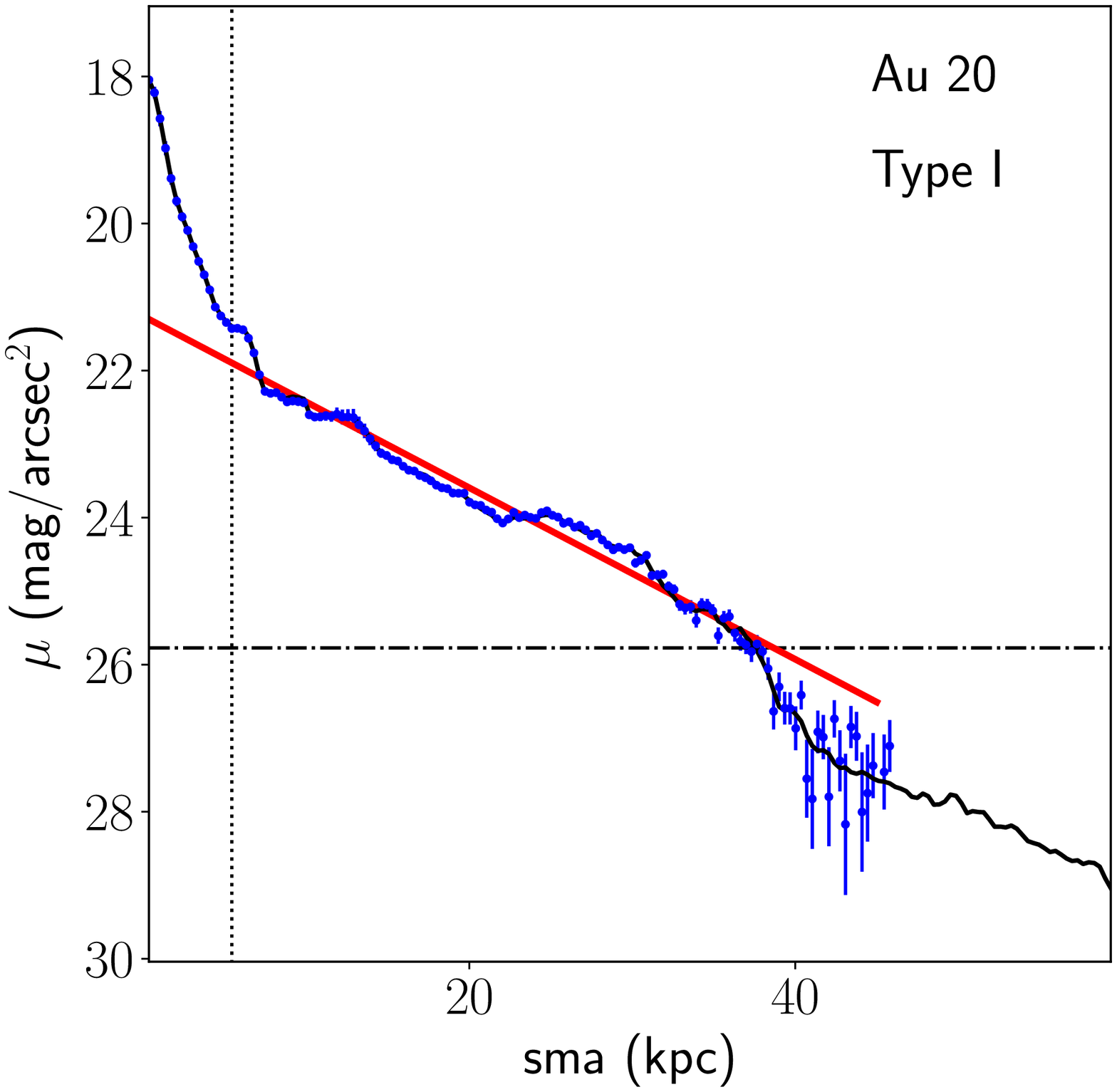} & \includegraphics[width=25em]{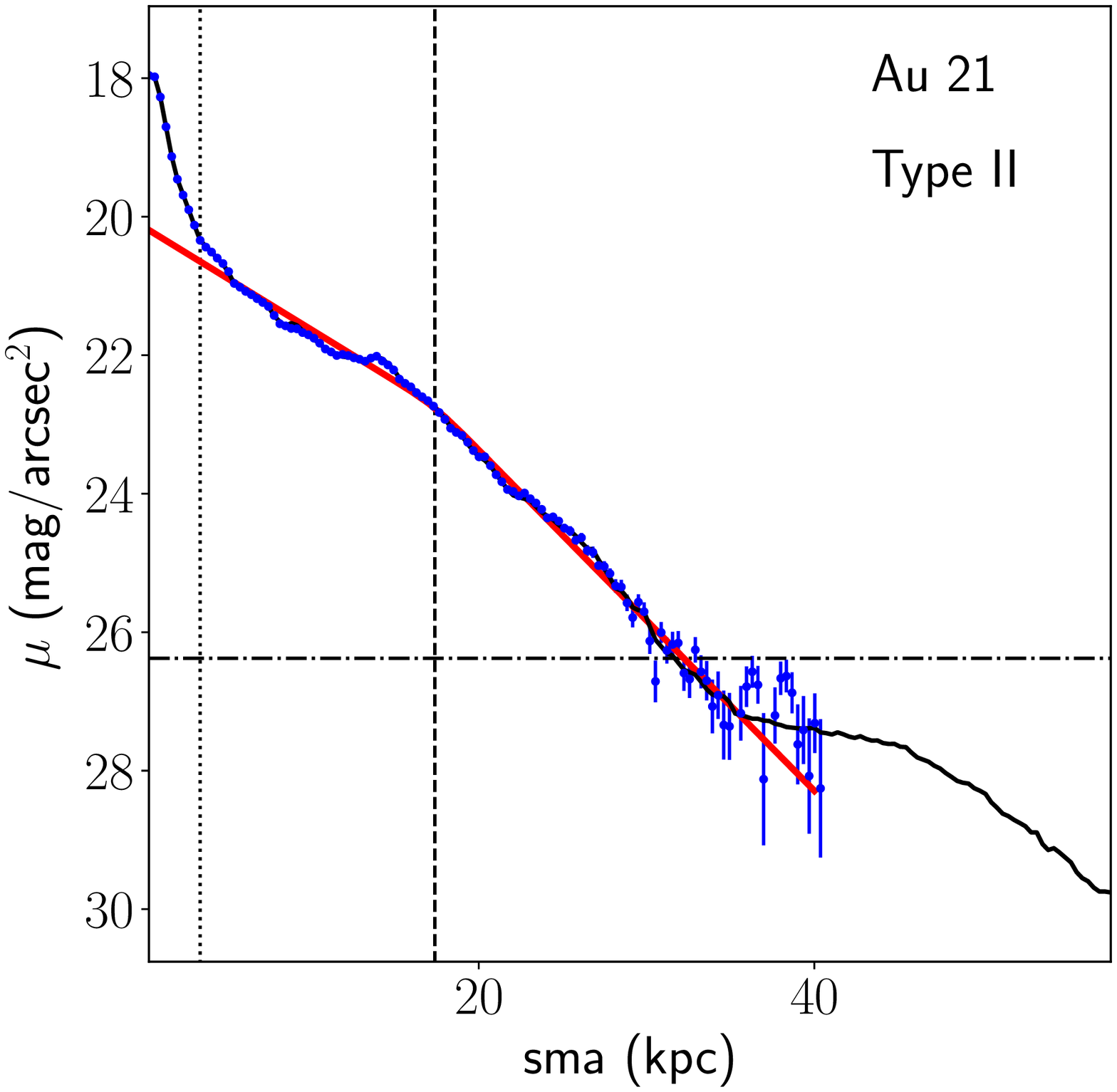} \\
\includegraphics[width=25em]{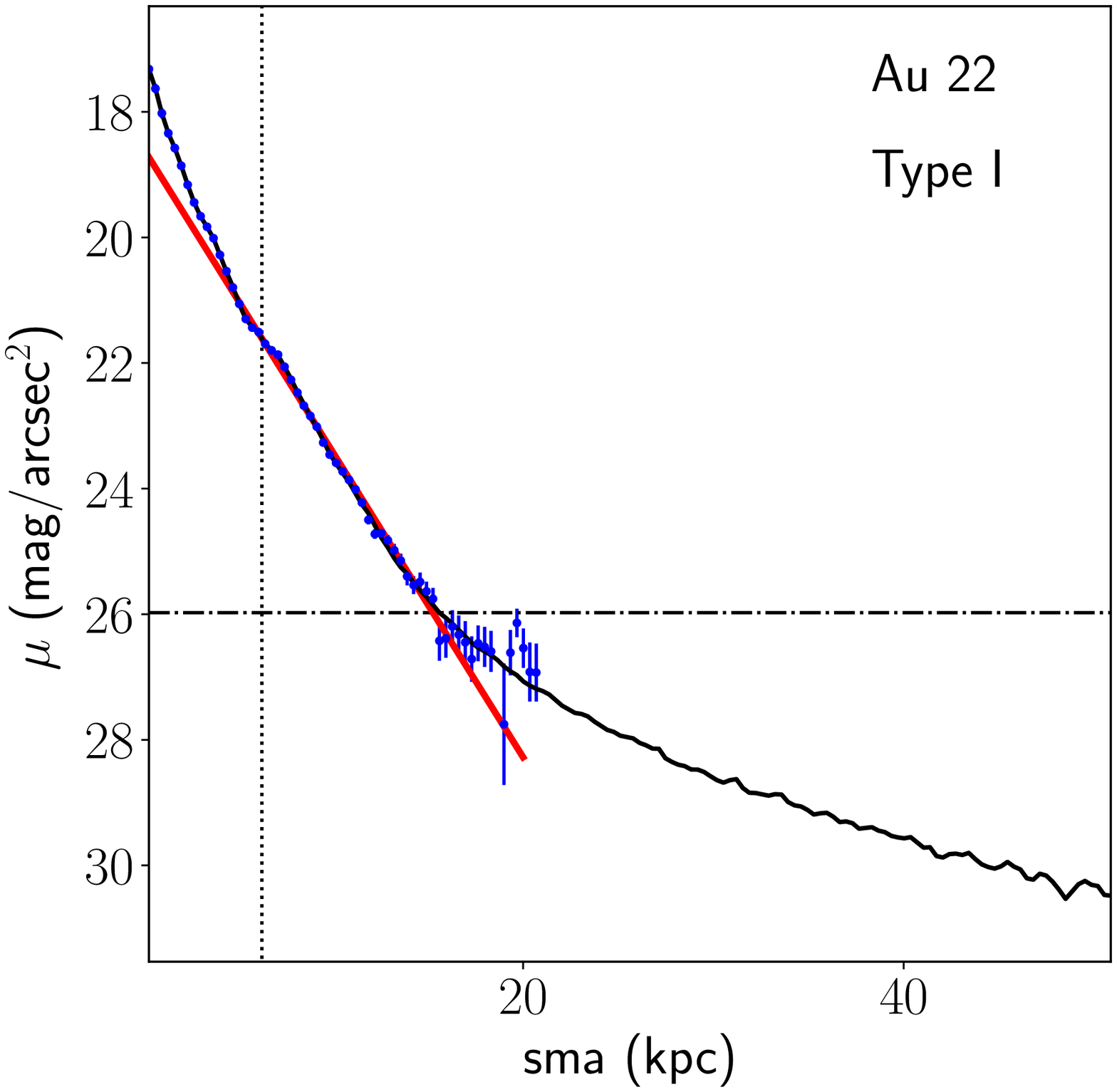} & \includegraphics[width=25em]{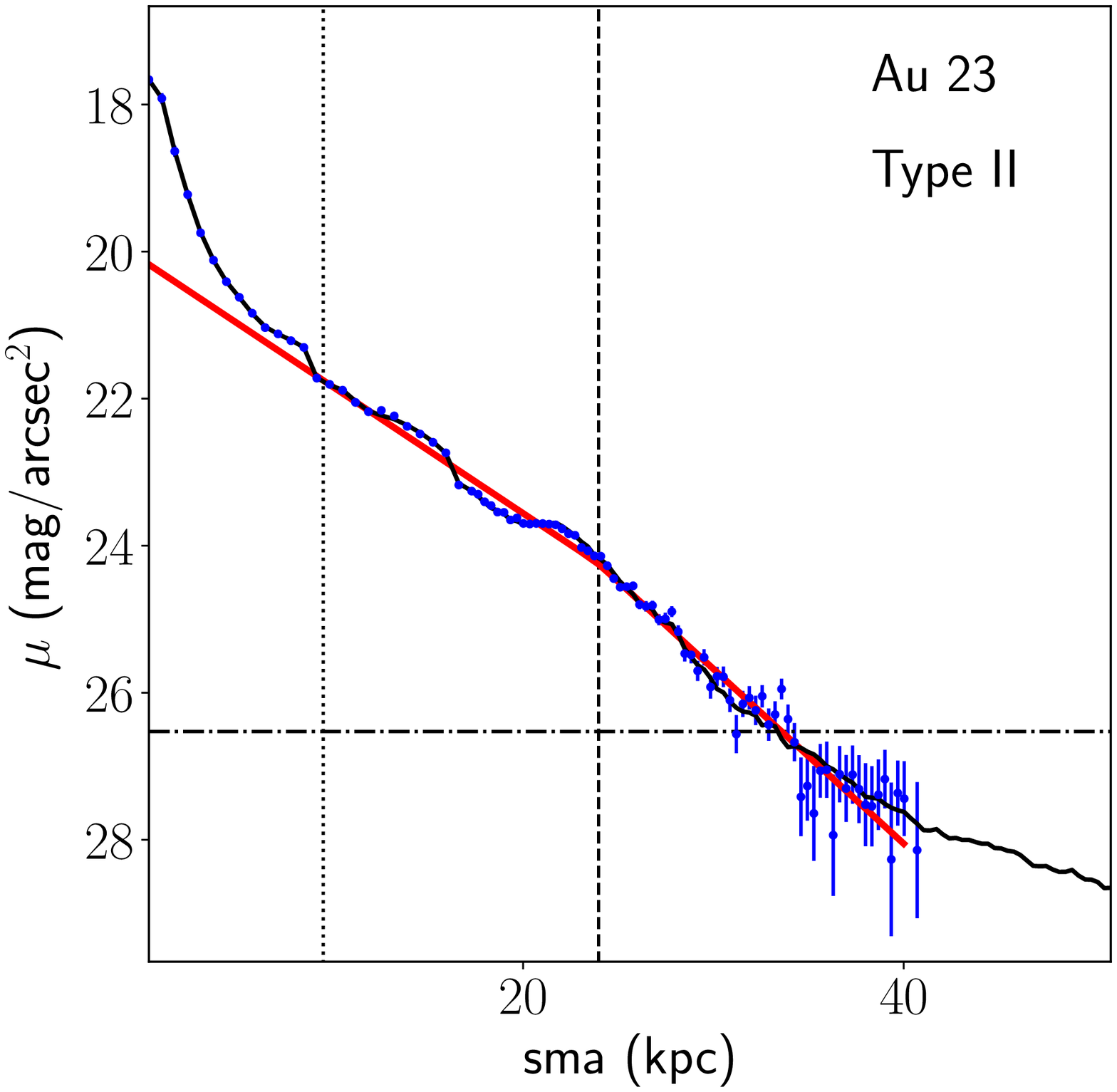} \\
\includegraphics[width=25em]{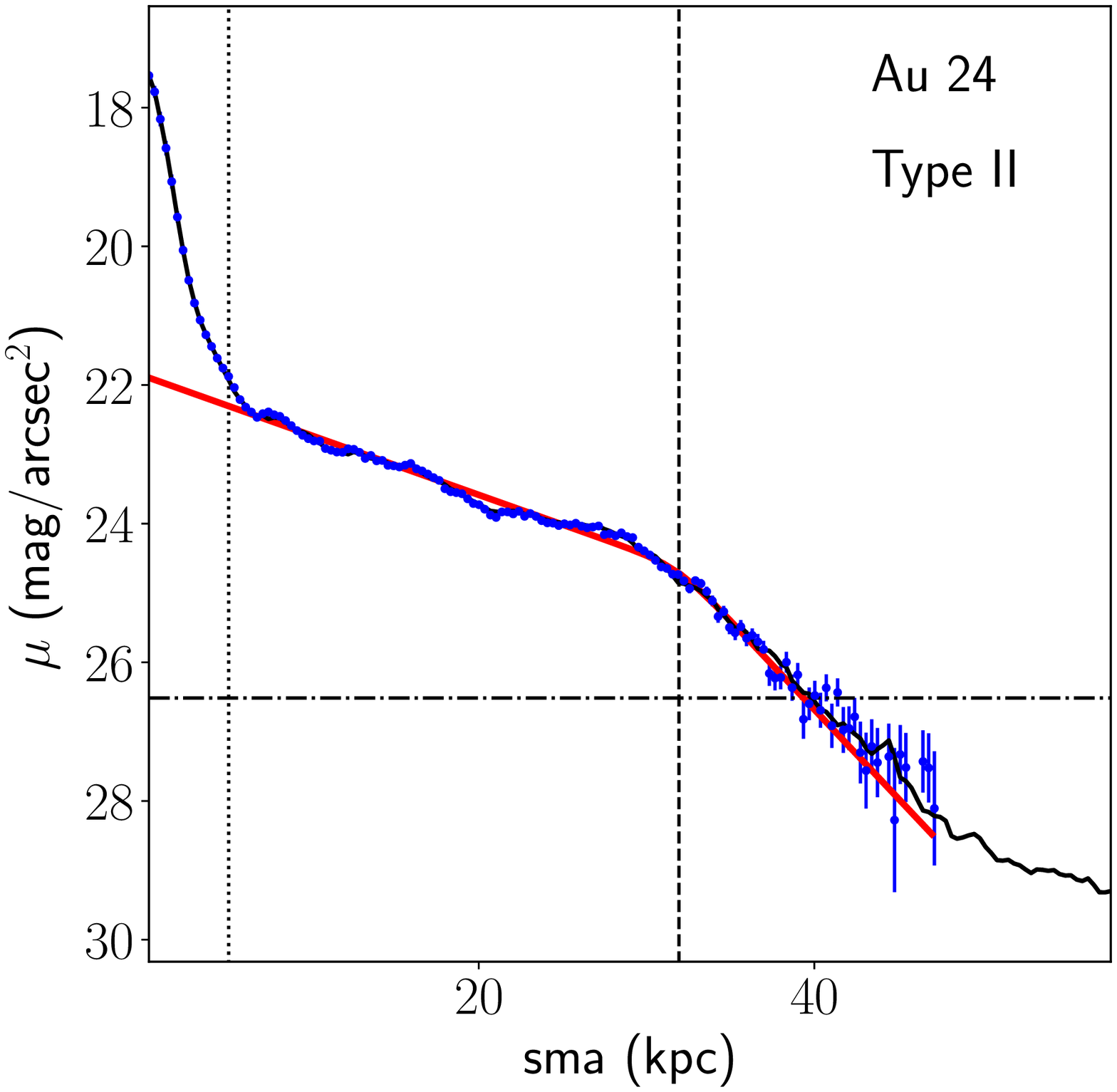} & \includegraphics[width=25em]{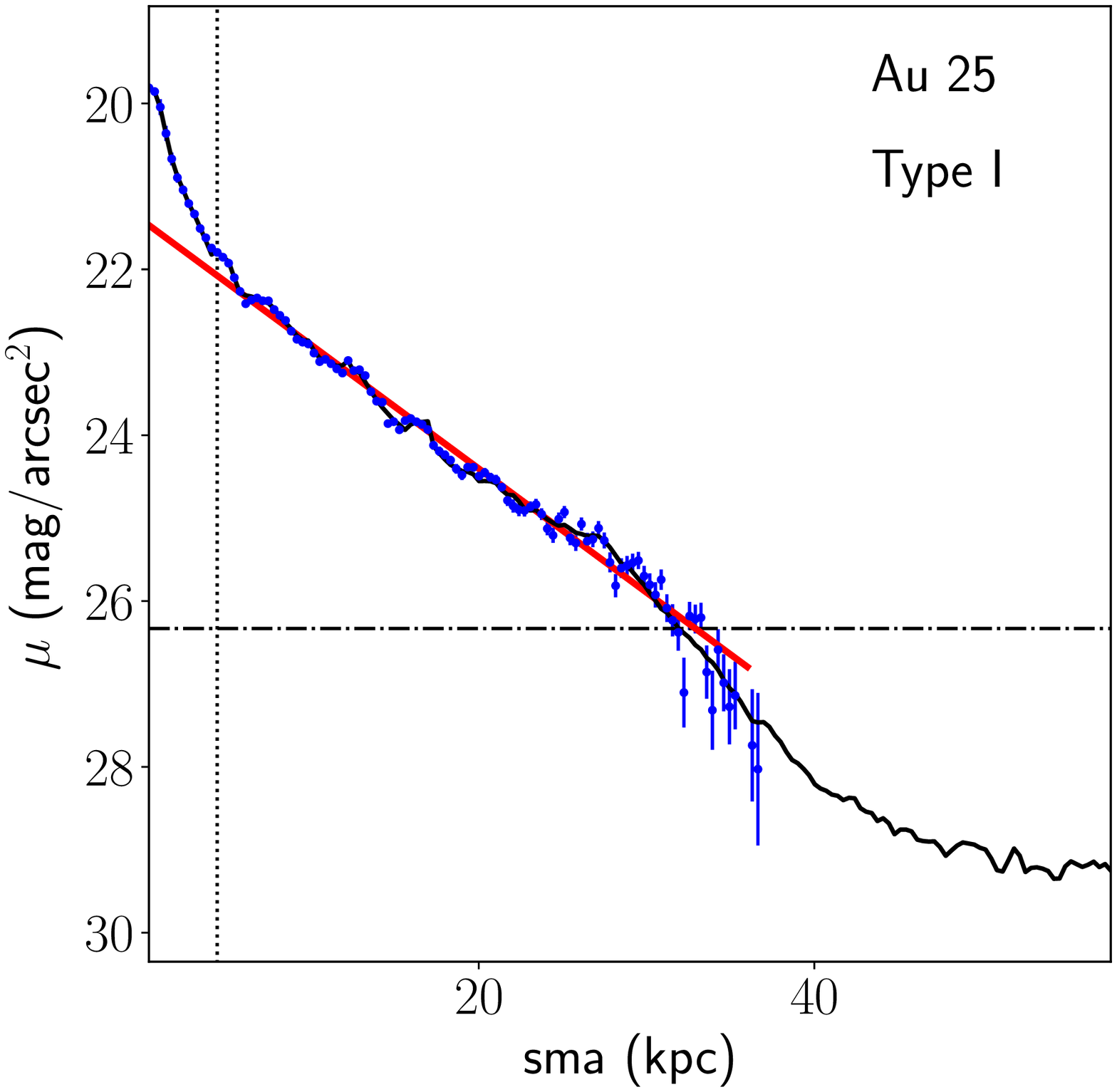} \\

\end{tabular}
\contcaption{}
\end{figure}

\begin{figure}
\centering
\begin{tabular}{ccc}
\includegraphics[width=25em]{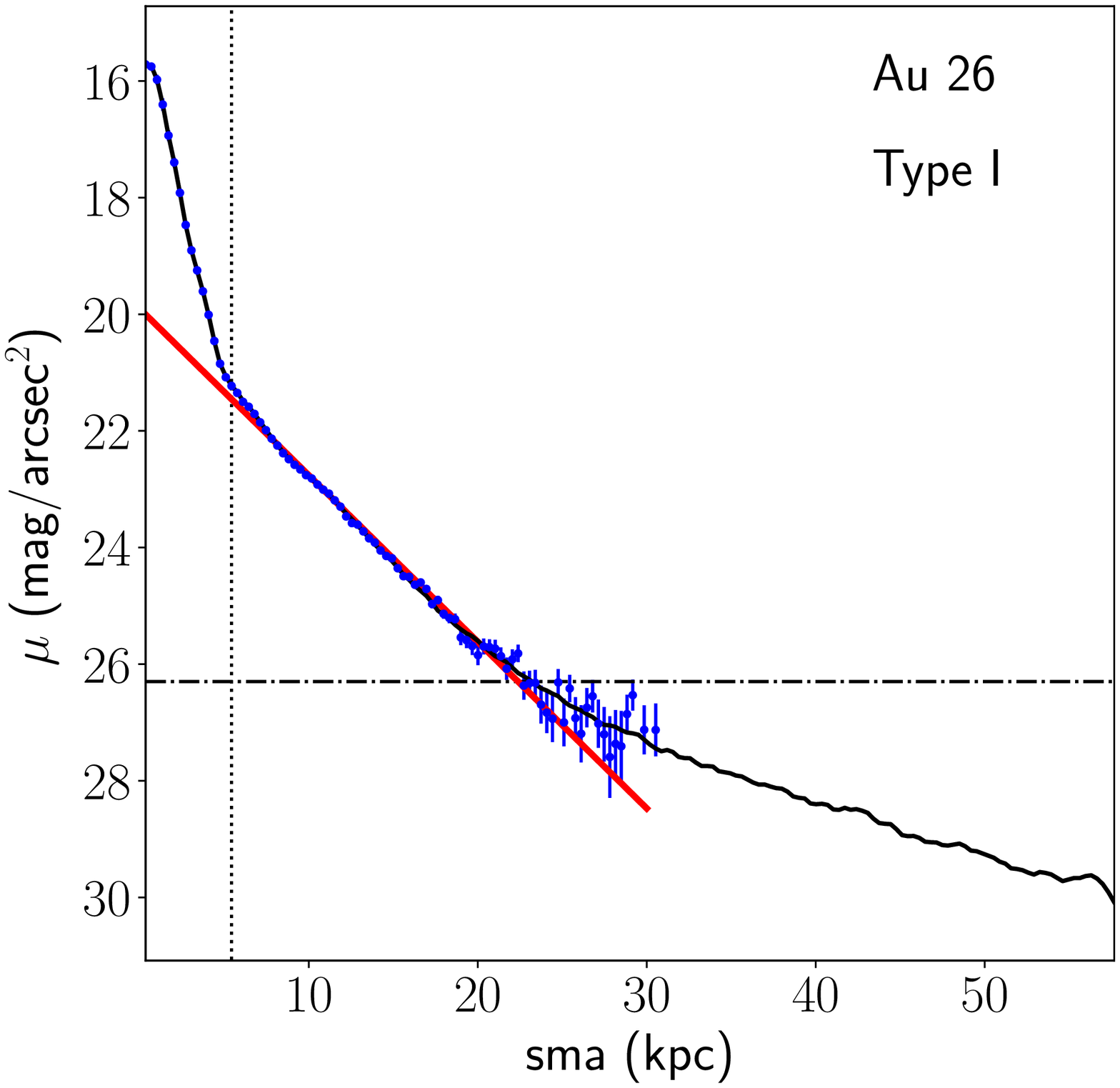} & \includegraphics[width=25em]{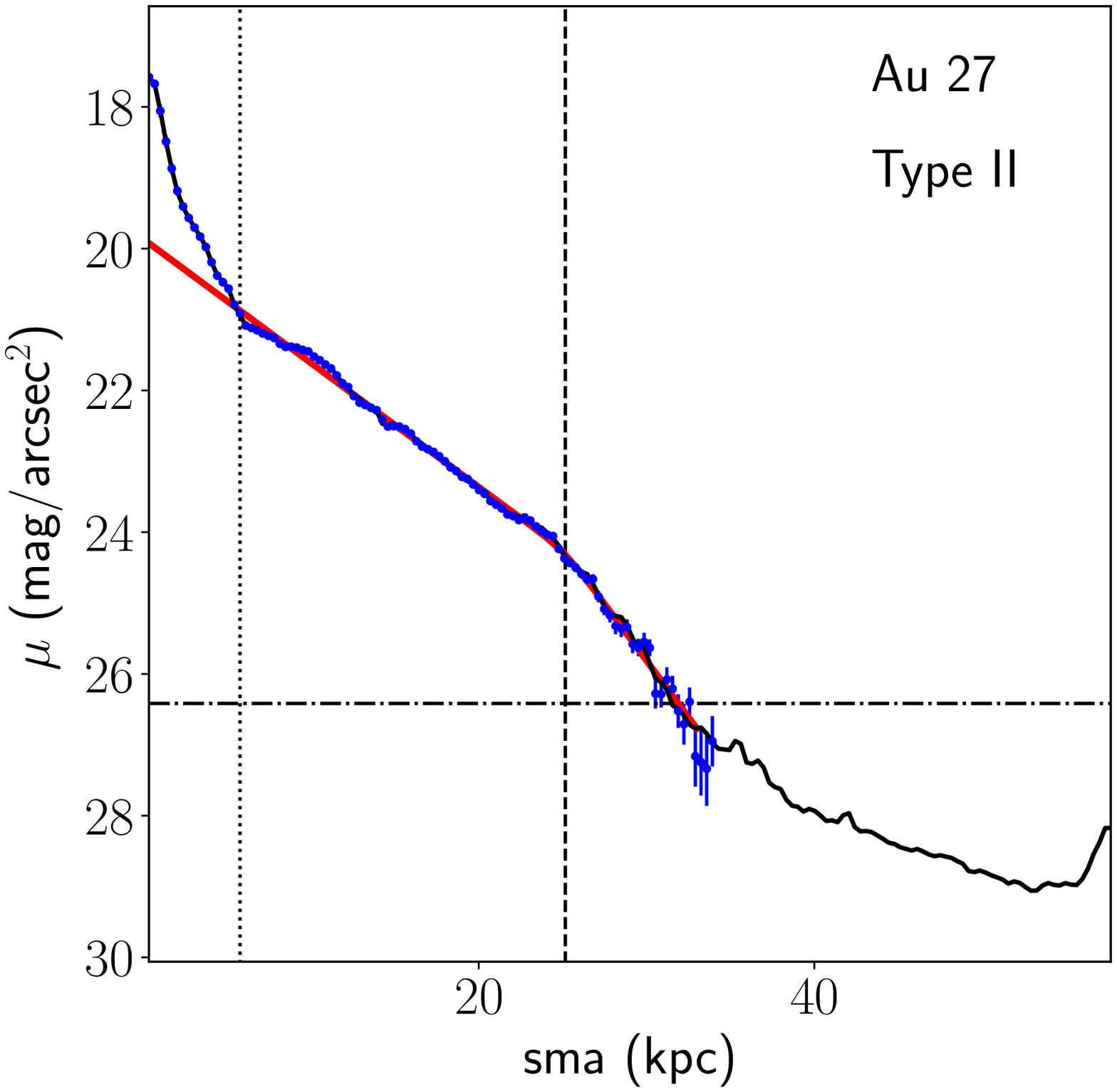} \\
\multicolumn{2}{c}{\includegraphics[width=25em]{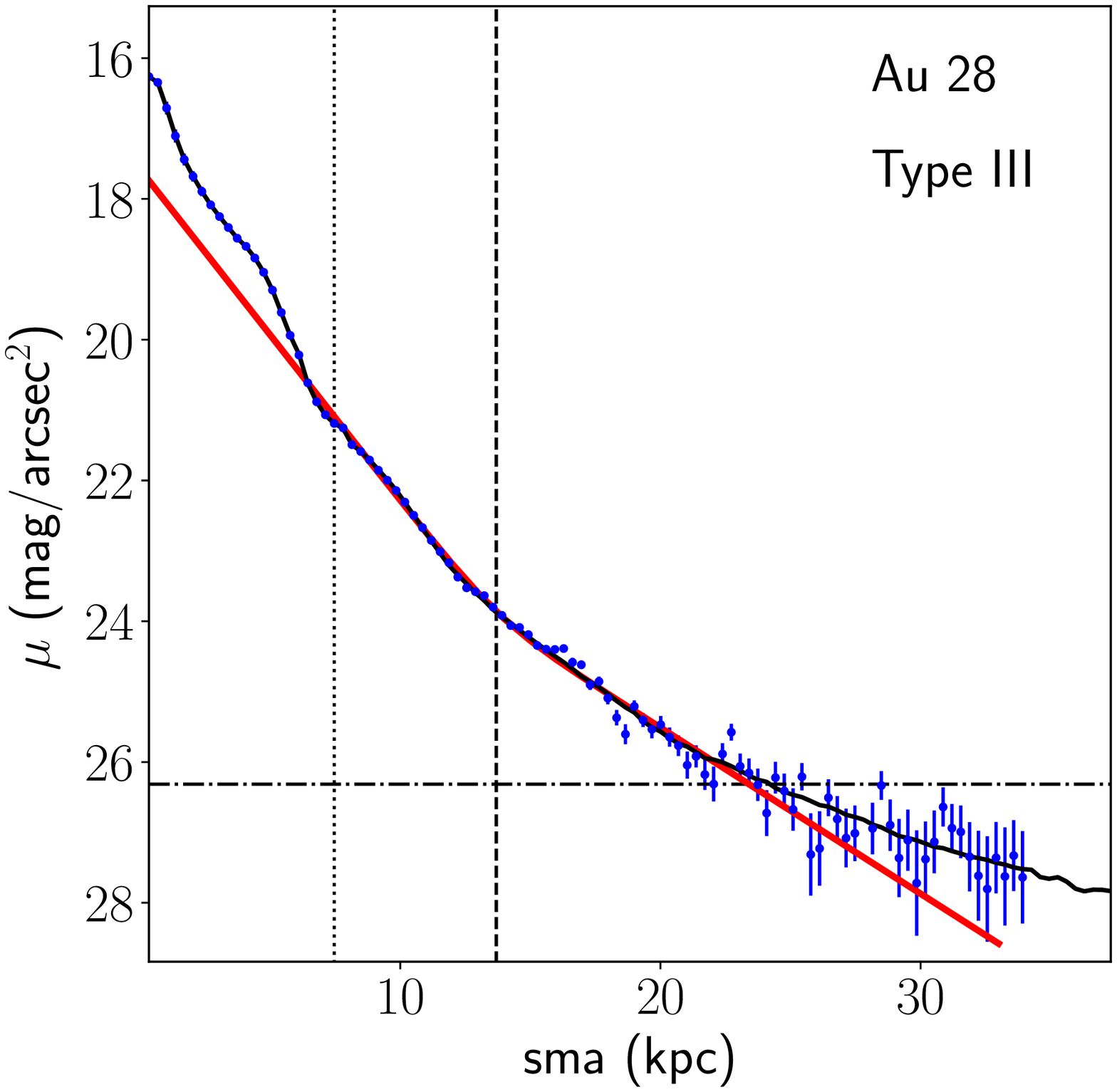} }\\
\end{tabular}
\contcaption{}
\end{figure}

\section{Ellipse fitting and 2D decomposition for noiseless synthetic images}\label{sec:sbp}

\begin{figure}
\begin{minipage}{0.45\textwidth}
\begin{table}
	\centering
	\caption{Bar and disc morphological parameters obtained from ellipse fits, using {\it r} band synthetic images without noise (analogue to Table 1 in the paper). The columns show: (1) the galaxy simulation name; (2) and (3), the lower and upper limit of bar lengths from ellipse fits, respectively; (4) the maximum ellipticity that defines $a_\text{max}$; (5) the bar strength.}
	\label{tab:ellipse_results_nn}
	\begin{tabular}{l c c c c c}
		\hline
		Galaxy & $a_\text{max}$ & $L_\text{bar}$ & $\epsilon_\text{max}$ & $f_\text{bar}$  \\
		~ & (kpc) & (kpc) &  &  \\
		(1) & (2) & (3) & (4) & (5) \\
		\hline
		Au1  & 3.05 & 4.07 & 0.56 & 0.57 \\

		Au2  & 5.09 & 8.82 & 0.58 & 0.61 \\

		Au5  & 3.39 & 4.41 & 0.50 & 0.51 \\

		Au6  & 3.73 & 5.43 & 0.42 & 0.42 \\

		Au7  & 3.05 & 5.43 & 0.47 & 0.46 \\

		Au9  & 4.07 & 6.11 & 0.66 & 0.76 \\

		Au10  & 3.39 & 6.45 & 0.62 & 0.67 \\

		Au12  & 3.05 & 3.73 & 0.57 & 0.60 \\

		Au13$^*$  & 3.73 & 5.43 & 0.57 & 0.59 \\

		Au14  & 4.07 & 5.09 & 0.61 & 0.66 \\

		Au17$^\dagger$  & 4.07 & 5.43 & 0.57 & 0.59 \\

		Au18$^*$  & 4.07 & 6.45 & 0.59 & 0.62 \\

		Au20  & 3.39 & 5.43 & 0.61 & 0.66 \\

		Au21  & 2.71 & 3.39 & 0.41 & 0.41 \\

		Au22$^*$  & 3.73 & 6.45 & 0.55 & 0.56 \\

		Au23$^*$  & 7.80 & 9.16 & 0.63 & 0.69 \\

		Au24  & 3.05 & 5.09 & 0.52 & 0.52 \\

		Au25  & 3.39 & 4.07 & 0.43 & 0.43 \\

		Au26$^\dagger$  & 3.39 & 5.43 & 0.48 & 0.48 \\

		Au27  & 3.73 & 5.77 & 0.63 & 0.69 \\

		Au28  & 4.75 & 7.46 & 0.69 & 0.81 \\
   		\hline
	\end{tabular}
\end{table}
\end{minipage}
\begin{minipage}{0.45\textwidth}

	\centering
	\includegraphics[width=\textwidth]{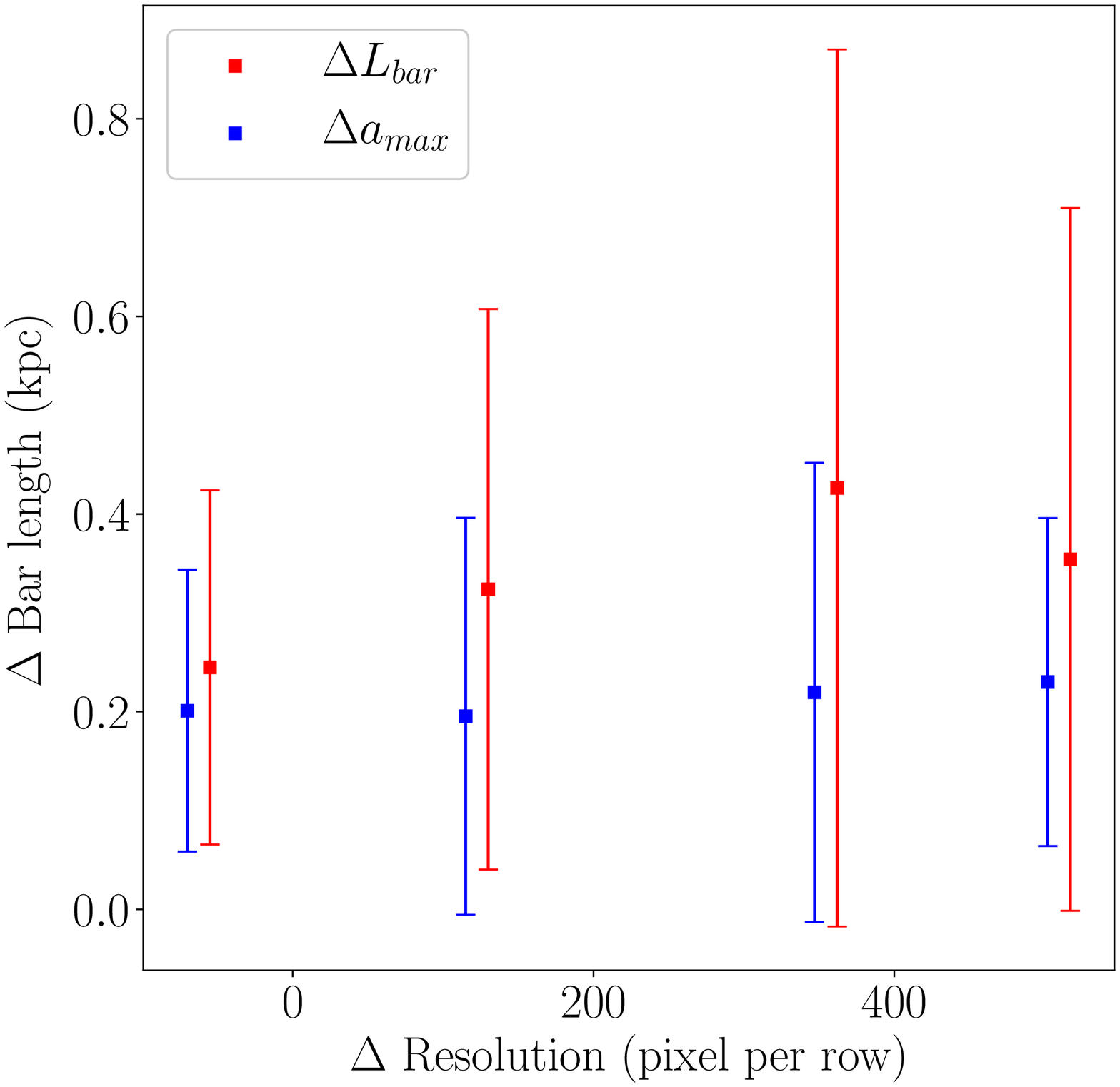}
    \caption{Mean bar length differences as a function of the variation in image resolution with respect to the results showed in Table 1 in the paper, where synthetic images without noise and with a 348 pixel per row were used. The error bar represents the standard deviation of the different estimates. Differences in $L_\text{bar}$ are larger and more scattered than differences in the $a_\text{max}$ measurements.}
\end{minipage}
\begin{minipage}{\textwidth}
\begin{table}
	\centering
	\caption{{\sc galfit} 2D multicomponent decomposition results for the face-on {\it r} band synthetic images without noise of the 21 Auriga barred galaxies, similar to the results from noisy images presented in Table 2 in the paper. Disc, bulge, and bar components are fitted with an exponential, S{\'e}rsic, and modified Ferrer profile, respectively. The columns show: (1) the galaxy simulation name; (2)-(4) the central surface brightness for each component; (5) the bulge effective radius; (6) the disc scalelength; (7) the bar radius; (8) the bulge S{\'e}rsic index; (9)-(10) the Ferrer function $\alpha$ and $\beta$ parameters, respectively; (11)-(12) the axis ratio for the outermost ellipse for bulge and bar component; and (13)-(15) the luminosity ratio for bulge, disc, and bar, respectively. Galaxies with B/P bulges are indicated by $^*$, and buckling bars by $^\dagger$.}
	\label{tab:decomposition_results_nn}
	\begin{tabular}{lcccccccccccccc} 
		\hline
		Galaxy & $\mu_\text{e,B}$ & $\mu_{0,\text{D}}$ & $\mu_{0,\text{bar}}$ & $r_\text{e,B}$ & $r_\text{s,D}$ & $r_\text{bar}$ & $n$ & $\alpha$ & $\beta$ & $q_\text{B}$ & $q_\text{bar}$ & $B/T$ & $D/T$ & ${\rm Bar}/T$\\
		~ & $\left(\frac{\text{mag}}{\text{arcsec}^2}\right)$ & $\left(\frac{\text{mag}}{\text{arcsec}^2}\right)$ & $\left(\frac{\text{mag}}{\text{arcsec}^2}\right)$ & (kpc) & (kpc) & (kpc) &  &  &  &  &  &  &  & \\
		(1) & (2) & (3) & (4) & (5) & (6) & {(7)} & (8) & {(9)} & (10) & (11) & (12) & (13) & (14) & (15)\\
		\hline
		Au1  & 19.66 & 20.60 & 17.63 & 1.60 & 4.14 & 4.21 & 0.94 & 1.50 & 1.62 & 0.38 & 0.30 & 0.17 & 0.69 & 0.13 \\
		Au2  & 19.28 & 21.60 & 19.66 & 1.19 & 10.40 & 8.03 & 0.54 & 1.50 & 1.53 & 0.49 & 0.38 & 0.08 & 0.86 & 0.07 \\
		Au5  & 18.26 & 19.62 & 18.48 & 0.76 & 4.11 & 4.32 & 0.45 & 1.50 & 1.43 & 0.70 & 0.36 & 0.10 & 0.84 & 0.06 \\
		Au6  & 20.15 & 20.76 & 20.21 & 0.96 & 5.53 & 5.43 & 0.41 & 1.50 & 1.45 & 0.63 & 0.42 & 0.04 & 0.92 & 0.04 \\
		Au7  & 18.52 & 19.86 & 19.62 & 0.96 & 4.88 & 5.61 & 0.44 & 1.50 & 1.23 & 0.52 & 0.43 & 0.09 & 0.86 & 0.06 \\
		Au9  & 18.21 & 20.11 & 18.98 & 0.98 & 4.03 & 6.03 & 0.50 & 1.50 & 1.02 & 0.41 & 0.27 & 0.15 & 0.71 & 0.14 \\
		Au10  & 17.19 & 19.40 & 16.46 & 0.84 & 2.92 & 5.60 & 0.28 & 1.50 & 1.60 & 0.47 & 0.31 & 0.17 & 0.48 & 0.35 \\
		Au12  & 18.22 & 19.18 & 19.21 & 0.85 & 3.70 & 3.78 & 0.27 & 1.50 & 0.01 & 0.46 & 0.34 & 0.06 & 0.87 & 0.06 \\
		Au13$^*$  & 16.93 & 19.90 & 18.06 & 0.97 & 3.13 & 5.43 & 0.47 & 1.50 & 1.05 & 0.43 & 0.47 & 0.32 & 0.36 & 0.31 \\
		Au14  & 17.26 & 19.34 & 17.49 & 0.81 & 4.80 & 5.09 & 0.34 & 1.50 & 1.53 & 0.53 & 0.33 & 0.11 & 0.81 & 0.09 \\
		Au17$^\dagger$  & 17.53 & 20.41 & 19.28 & 1.26 & 4.08 & 5.43 & 0.49 & 1.50 & 0.01 & 0.41 & 0.40 & 0.37 & 0.44 & 0.19 \\
		Au18$^*$  & 18.23 & 19.94 & 19.15 & 1.12 & 4.33 & 6.45 & 0.47 & 1.50 & 1.14 & 0.47 & 0.33 & 0.17 & 0.72 & 0.12 \\
		Au20  & 18.79 & 20.98 & 19.31 & 0.97 & 6.92 & 5.43 & 0.49 & 1.50 & 1.24 & 0.43 & 0.38 & 0.09 & 0.82 & 0.09 \\
		Au21  & 18.68 & 19.73 & 19.80 & 1.00 & 5.19 & 4.40 & 0.33 & 1.50 & 0.82 & 0.56 & 0.48 & 0.07 & 0.89 & 0.04 \\
		Au22$^*$  & 18.55 & 19.28 & 17.30 & 0.81 & 2.48 & 5.85 & 0.61 & 1.50 & 1.66 & 0.69 & 0.33 & 0.14 & 0.62 & 0.24 \\
		Au23$^*$  & 18.59 & 20.26 & 18.13 & 1.28 & 5.50 & 9.16 & 0.55 & 1.50 & 1.75 & 0.43 & 0.27 & 0.14 & 0.76 & 0.10 \\
		Au24  & 18.19 & 21.23 & 19.53 & 1.07 & 7.19 & 5.09 & 0.55 & 1.50 & 1.46 & 0.43 & 0.42 & 0.18 & 0.77 & 0.05 \\
		Au25  & 20.77 & 20.96 & 21.12 & 1.07 & 5.15 & 4.07 & 0.38 & 1.50 & 0.85 & 0.40 & 0.34 & 0.03 & 0.94 & 0.03 \\
		Au26$^\dagger$  & 16.33 & 19.72 & 17.75 & 1.20 & 3.45 & 5.32 & 0.51 & 1.50 & 1.22 & 0.41 & 0.52 & 0.49 & 0.30 & 0.21 \\
		Au27  & 18.41 & 19.70 & 18.40 & 0.87 & 5.09 & 5.32 & 0.37 & 1.50 & 1.35 & 0.49 & 0.29 & 0.06 & 0.87 & 0.07 \\
		Au28  & 17.21 & 19.04 & 16.43 & 0.90 & 3.15 & 7.46 & 0.37 & 1.50 & 1.59 & 0.45 & 0.27 & 0.13 & 0.49 & 0.38 \\
        \hline
	\end{tabular}
\end{table}
\end{minipage}
\end{figure}



\label{lastpage}